\documentclass[sigconf]{acmart}
\pdfoutput=1 

\renewcommand\footnotetextcopyrightpermission[1]{}
\usepackage[english]{babel}
\usepackage{blindtext}

    \usepackage{indentfirst}   
	\usepackage{graphicx} 
	\usepackage{balance}
	 \usepackage{amssymb} 
	\usepackage{amsmath}
	\usepackage{multirow, rotating, wasysym, url}
	\usepackage[tight]{subfigure}
	\usepackage[ruled,linesnumbered,vlined]{algorithm2e}
	\usepackage{hyperref} %
	\usepackage{color}
	\usepackage[utf8]{inputenc}
	\usepackage[english]{babel}
	\usepackage{comment}
    \usepackage[normalem]{ulem}
    \usepackage{enumitem}
    \usepackage{appendix}
	\usepackage{makecell}

	\newcommand{\presec}{\vspace{-0.1cm}}
	\newcommand{\postsec}{\vspace{-0cm}}
	\newcommand{\presub}{\vspace{-0.1cm}}
	\newcommand{\postsub}{\vspace{-0cm}}

    \newcommand{\vvv}{\vspace{-0.0in}}
	
	\newcommand{\postsubfig}{\vspace{-0.0in}}
	
	\newcommand{\ie}{\textit{i.e.}}
    \newcommand{\eg}{\textit{e.g.}}

    \newcommand{\etc}{\textit{etc.}}

	\newcommand{\bbb}{\noindent\textbf}

    \newcommand{\leftshift}{$< \kern -.2em <$}
    \newcommand{\rightshift}{$> \kern -.2em >$}
    
\graphicspath{{./GraphPPT/},{./EvaluateFigures/},{./figure/}}

\thanks{\textsuperscript\textdagger School of Computer Science,  and National Engineering Laboratory for Big Data Analysis Technology and Application, Peking University, China.}
\thanks{\textsuperscript\textdaggerdbl Peng Cheng Laboratory, Shenzhen, China.}
\thanks{\textsuperscript\textsection Department of Computer Science and Engineering, Hong Kong University of Science and Technology, Hong Kong.}
\thanks{\textsuperscript\textparagraph Huawei, Theory Lab, China}

\thanks{This work is supported by Key-Area Research and Development Program of Guangdong Province 2020B0101390001, National Natural Science Foundation of China (NSFC) (No. U20A20179).}

\author{Tong Yang}
\authornotemark[2]
\authornotemark[3]
\affiliation{Peking University}

\author{Jizhou Li}
\authornotemark[2]
\affiliation{Peking University}

\author{Yikai Zhao}
\authornotemark[2]
\affiliation{Peking University}

\author{Kaicheng Yang}
\authornotemark[2]
\affiliation{Peking University}

\author{Hao Wang}
\authornotemark[4]
\affiliation{Hong Kong University of Science and Technology}

\author{Jie	Jiang}
\authornotemark[2]
\affiliation{Peking University}

\author{Yinda Zhang}
\authornotemark[2]
\affiliation{Peking University}

\author{Nicholas Zhang}
\authornotemark[5]
\affiliation{Huawei}



\copyrightyear{2022}
\acmYear{2022}
\setcopyright{acmcopyright}
\acmConference[WWW '22] {Proceedings of the ACM Web Conference 2022}{April 25--29, 2022}{Virtual Event, Lyon, France.}
\acmBooktitle{Proceedings of the ACM Web Conference 2022 (WWW '22), April 25--29, 2022, Virtual Event, Lyon, France}
\acmPrice{15.00}
\acmISBN{978-1-4503-9096-5/22/04}
\acmDOI{10.1145/3485447.3511980}

\title{QCluster: Clustering Packets for Flow Scheduling}

\begin{document}
    \begin{abstract}

\noindent Flow scheduling is crucial in data centers, as it directly influences user experience of applications.
According to different assumptions and design goals, there are four typical flow scheduling problems/solutions: SRPT, LAS, Fair Queueing, and Deadline-Aware scheduling.
When implementing these solutions in commodity switches with limited number of queues, they need to set static parameters by measuring traffic in advance, while optimal parameters vary across time and space.
This paper proposes a generic framework, namely QCluster, to adapt all scheduling problems for limited number of queues. 
The key idea of QCluster is to cluster packets with similar weights/properties into the same queue.
QCluster is implemented in Tofino switches, and can cluster packets at a speed of 3.2 Tbps. To the best of our knowledge, QCluster is the fastest clustering algorithm. 
Experimental results in testbed with programmable switches and ns-2 show that QCluster reduces the average flow completion time (FCT) for short flows up to 56.6\%, and reduces the overall average FCT up to 21.7\% over state-of-the-art.
All the source code in ns-2 is available in Github \cite{opensource}. 

\vspace{-0.25cm}
\end{abstract}

    \maketitle
    \sloppy

	\presec
\section{Introduction} 
\label{sec:introduction}

\subsection{Background and Motivation} 

Given some flows/packets in a node, flow scheduling is to decide the forwarding sequences of packets for some optimization goals, such as flow completion time, fairness, or meeting deadlines.
Flow scheduling has been a hot topic in data centers (
\cite{pias,homa,pfabric,auto,coflow,hula,universal-pkt-sched,narayan2018restructuring,timely,rc3,remy,sivaraman2014experimental,lee2015accurate} 
), because it directly determines bandwidth usage, latency, and Quality of Service of applications.

According to different optimization goals and assumptions, there are typically four kinds of scheduling problems/solutions: 
SRPT (Shortest Remaining Processing Time first, \eg, pFabric \cite{pfabric}), 
LAS (Least Attained Service, \eg, PIAS \cite{pias}, Auto \cite{auto}), 
Fair Queueing (\eg, Nagle \cite{nagle}, BR\cite{BR}, AFQ \cite{afq}), and Deadline-Aware Scheduling (\eg, pFabric \cite{pfabric} with Earliest-Deadline-First (pFabric-EDF), $D^3$ \cite{d3}, PDQ \cite{pdq} and Karuna \cite{karuna}).
These works have made great contributions, and can achieve near-optimal or excellent performance when there are a great number of queues in each egress port.
For example, pFabric \cite{pfabric} achieving near optimal performance, assumes there are infinite number of queues.
Another example is Fair Queueing. The earlier work, Nagle \cite{nagle}, assigns each flow one queue to achieve the theoretical Fair Queueing, requiring a great many queues. Instead, Bit by bit round robin \cite{BR} uses one preemptive queue to conduct the scheduling. However, BR is still far from practice, and the preemptive queue can only be approximately implemented with multiple queues, which is done by approximate Fair Queueing (AFQ) \cite{afq}. However, AFQ needs to rotate the queue priorities, which has not been achieved in current switches.

The number of queues in commodity switches is very limited, \eg, $k=8$ queues, and therefore the above works must be adapted to a limited queue version for practical use. 
To adapt for $k$ queues, one commonly used approach is to measure traffic in advance.
Specifically, one first builds a measurement system to collect traffic from switches or end hosts, analyzes the statistics, and makes many tests to find appropriate parameters.
However, when traffic changes and mismatches the parameters, the performance could degrade a lot.
For example, the authors of PIAS show that when thresholds mismatch traffic, the flow completion time (FCT) could be degraded by 38\% \cite{pias,auto}; For another example, when implementing pFabric in $k=8$ queues, the FCT of pFabric could be degraded by $30\%$ \cite{pfabric,auto}.
However, the optimal parameters often vary across time and space, and the traffic distribution in network changes frequently and quickly. 
Measuring traffic in advance cannot adapt to the quick change of traffic.
Therefore, it is desirable but challenging to adapt existing solutions for limited number of queues without measuring in advance.
The design goal is to devise a framework to address this challenge for all existing scheduling problems. 

\subsection{Our Solution} 
\label{sec:intro:solution}

Our insight is that adapting existing solutions for $k$ queues is actually a clustering problem: clustering packets into queues.
The $k$ queues in current commodity switches are first-in-first-out (FIFO).
If packets with drastically different weights or properties are placed into one queue, the performance will be poor because of FIFO.
For example, if a small flow and a large flow are placed into the same queue, the small flow will be blocked by the large flow and the overall FCT will be large. 
For another example, if a common flow and a flow with a deadline are placed in one queue, the deadline may be missed.
Therefore, our insight is that packets in the same queue should have similar weights, and this is actually a clustering problem, and we name it the \textit{\textbf{Queue Clustering}} problem.

Queue Clustering has the following two requirements that existing clustering problems often do not have.
1) The clustering speed needs to catch up with the line rate, \eg, 3.2 Tbps, and no existing clustering algorithm can achieve this speed;
2) Packet disorder should be avoided. 
Due to the above special requirements, existing clustering algorithms \cite{kmeans, cluster-1, cluster-2} cannot be directly used, and this paper proposes the \textit{\textbf{QCluster}} to cluster packets with similar weights/priorities into the same queue.

QCluster uses packet weight and property as features. 
For flows with special properties, such as deadline or time sensitiveness, they should be in different clusters/queues from other flows.
For flows with the same property, we cluster them according to the packet weight.
Inspired by $k$-means \cite{kmeans}, we maintain the average packet weight for each queue, and call it \textit{queue weight}.
Given an incoming packet, we compare the packet weight with the $k$ queue weights to choose a queue.
The packet weight is recorded and updated in the Scheduling Count-min sketch (see details in Section \ref{subsec:timesketch}), which also records timestamp and last queue ID.
For different scheduling problems, packet weight has different definitions, and different dequeuing policy should be chosen.
For LAS, we define packet weight as the number of bytes sent, and use strict priority to dequeue packets.
We also propose an adaptive method (see detail in \S \ref{subsec:cluster}) to make high-priority queue have fewer packets so that the probability that small flows are blocked by large flows will be reduced.
For Fair Queueing, we also define packet weight as the number of bytes sent, but use round robin to dequeue packets.
For SRPT and Deadline-Aware, we implement them in ns-2, but not in our testbed because these two policies need to know the remaining flow size that the current TCP protocol does not support.

During the clustering process, the latter packets of a flow could be placed into a higher-priority queue, and thus packet disorder could happen.
Existing solutions (\eg, pFabric) also have this problem.
To avoid packet disorder, we propose the PDA algorithm. 
Our key idea is that given an incoming packet $a_{now}$ of flow $a$, only if all previous packets of $a$ are not in any queue of this switch, we can place $a_{now}$ to any queue and packet disorder will not happen; Otherwise, we need to schedule this packet according to the state of the previous packet and the scheduling policy.

\bbb{Key Contributions:}

\noindent 1) We propose the QCluster to adapt all existing scheduling algorithms for limited number of queues (\S 3). In QCluster, we propose the Scheduling Count-Min sketch to record all necessary information, and propose the PDA algorithm to avoid packet disorder.

\noindent 2) We apply QCluster to four typical scheduling policies (SRPT, LAS, Fair Queueing, and Deadline-Aware Scheduling) as case studies. 

\noindent 3) We implement QCluster in Tofino switch and build a testbed (\S 5). We also conduct large-scale simulations using ns-2. 
 
\noindent 4) Extensive experimental results on a testbed and simulators show that QCluster can well adapt existing scheduling solutions to limited number of queues, achieving similar or better performance (\S 6).

	\presec
\section{Background and Related Work} 
\postsec
\label{sec:background}

Due to the significance of flow scheduling, there are a great many works in the literature, and we introduce them briefly.

\bbb{1) SRPT:} 
SRPT (shortest remaining processing time first) assumes that the remaining size of each flow is known, and lets the flow with the smallest remaining size go first to minimize FCT.
Typical solutions include pFabric \cite{pfabric}, Homa \cite{homa}, and more \cite{phost}.
In pFabric \cite{pfabric}, the smaller the remaining flow size is, the higher priority a packet gets.
And for dequeuing, the switch should find the earliest packet from the flow with the highest priority. 
However, this is hard to be deployed in commodity switches.
A recent work, Homa \cite{homa}, also belongs to this kind.

\bbb{2) LAS:} 
%
LAS (least attained service first) lets the flow with the smallest bytes sent go first.
Typical solutions include PIAS \cite{pias} and AuTO \cite{auto}.
PIAS \cite{pias} can achieve small FCT by carefully setting the thresholds for each queue according to the flow size distribution and network load.
Using the handcrafted thresholds for fixed distribution, PIAS achieves very small FCT.
However, if the distribution changes, the FCT will drop significantly \cite{auto}.
%
%

\bbb{3) Fair Queueing:} 
Fair Queueing was first introduced by Nagle \cite{nagle}, which owns some good characteristics compared to FCFS (Fisrt Come First Serve).
To achieve per-flow fairness, all active flows in the switch should have the same priority to use the bandwidth.
Typical solutions include Nagle \cite{nagle}, RFQ \cite{RFQ}, BR(bit-by-bit round robin) \cite{BR}, Gearbox \cite{Gearbox}, and AFQ \cite{afq}. 

\bbb{4) Deadline-Aware Scheduling:}
In data centers, some flows could have deadlines, and are called \textit{deadline flows}.
The goal of Deadline-Aware scheduling is to meet the deadlines of deadline flows first, and then minimize the FCT of non-deadline flows. 
Typical solutions include pFabric-EDF \cite{pfabric}, $D^3$ \cite{d3}, PDQ \cite{pdq}, D2TCP \cite{d2tcp}, MCP \cite{MCP} and Karuna \cite{karuna}.
pFabric \cite{pfabric} with Earliest-Deadline-First (pFabric-EDF) assigns priorities of packets for the deadline flows to be the flow’s deadline. 
And it assigns priorities of packets for the non-deadline flows based on remaining flow size. 
$D^3$ \cite{d3} proposes a deadline-aware control protocol, which uses explicit rate control to apportion bandwidth to meet the deadlines of flows. 
%
%

\bbb{5) Recent Hardware Solutions:}
Recent works also leverage the emerging new hardware in networking. \cite{packet-transactions,kim2015band,sivaraman2015towards,drmt,narayana2017language,hashpipe,pifo,sivaraman2015dc,narayana2016compiling,shrivastav2019fast,alcoz20sppifo,sunflow}.
PIFO \cite{pifo} designs a priority queue. 
In this design, each incoming packet can be enqueued into an arbitrary position of the queues, while PIFO dequeues packets from the head. 
PIEO \cite{shrivastav2019fast} is a generalization of the PIFO primitive allowing dequeue from arbitrary positions.
SP-PIFO \cite{alcoz20sppifo} uses strict-priority queues to achieve similar behavior of an ideal PIFO. 
PIFO is indeed very flexible and generic to a great many traffic optimization problems.
However, PIFO cannot implement pFabric when dealing with starvation. 

	\presec
\section{The QCluster Framework} 
\postsec
\label{sec:qcluster}

In this section, we first propose a generic framework, \textbf{\textit{QCluster}}, to address the \textit{queue clustering} problem. 
Second, we show how to use the SCM sketch to record and update flow information.
Third, we show how to control cluster sizes.
Last, we propose an algorithm to avoid packet disorder.
    \presub
\subsection{The QCluster Framework} 
\postsub
\label{subsec:frame}

\bbb{Queue Clustering:}
Given a switch with $k$ queues per port, there are incoming packets belonging to different flows.
The problem is how to cluster packets with similar weights/properties into the same queue without measuring traffic in advance.
Note that the queue clustering problem has four requirements that most existing clustering problems do not have (see Section \ref{sec:intro:solution}).

\begin{figure}[h]
\setlength\abovecaptionskip{-0.0cm}
\setlength\belowcaptionskip{-0.4cm}
	\centering
	\includegraphics[width=1\linewidth]{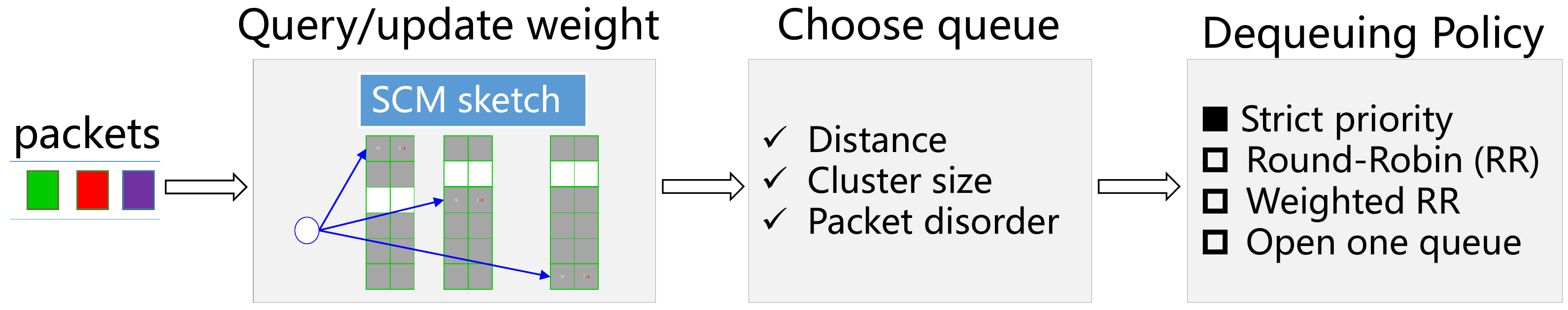}
	\caption{The QCluster framework.}
	\label{draw:QCluster:framework}
\end{figure}

\bbb{The QCluster Framework:}
Our framework is inspired by $k$-means, and can be applied to all flow scheduling problems.
QCluster works as follows (see Figure \ref{draw:QCluster:framework}). 
1) For packets with the same special property, \eg, with a deadline, or time-sensitive, QCluster clusters them into queues different from other flows; 
for flows with the same property, QCluster clusters them according to the packet weight.
2) Packet weight has different definitions for different scheduling problems. 
For example, the packet weight in LAS is the number of bytes sent.
Packet weight is recorded and updated by the SCM sketch which is detailed in Section \ref{subsec:timesketch}. 
QCluster maintains the average packet weight for each queue, namely \textit{queue weight}. 
3) When choosing the queue, we consider three factors: distance, cluster size, and packet disorder.
Given an incoming packet, we compare the packet weight with the $k$ queue weights to choose two adjacent queues, and then choose one of them according to our requirement for cluster size.
For example, for Fair Queueing, we need to let all clusters have the same size; for LAS, we need to let higher priority queues have fewer packets. 
The reason behind is shown in Section \ref{subsec:cluster}.
Section \ref{subsec:pda} shows how to avoid packet disorder.
4) For different scheduling problems, we need to choose the corresponding dequeuing policy. 
For example, we should choose strict priority \cite{pias} for minimizing FCT, and weighted round-robin \cite{weightedroundrobin} for fairness.

	\presub
\subsection{The Scheduling Count-Min Sketch} \postsub
\label{subsec:timesketch}

In QCluster, given an incoming packet $a_{now}$ with flow ID $a$, we need to know three kinds of information: 
1) The number of bytes sent;
2) the arriving time of the last packet of the incoming flow;
3) the queue that the last packet was sent into.
To record these kinds of information with extremely limited memory in switches, we propose an enhanced CM sketch \cite{cmsketch}, namely the Scheduling Count-Min sketch (SCM).
Compared to CM, SCM has three additional functions:  
1) SCM can automatically delete the information of aged flows; 
2) SCM records the queue ID of the previous packet of each flow goes; 
3) SCM can distinguish messages\footnote{In data centers, applications often establish persistent TCP connections. Communications can reuse the opened connections. Such communications are known as ``messages''. Packets that belong to different messages will be scheduled individually.}
and Flowlets.

\begin{figure}[htbp]
\setlength\abovecaptionskip{-0.2cm}
\setlength\belowcaptionskip{-0.4cm}
\centering
\includegraphics[width=0.85\linewidth]{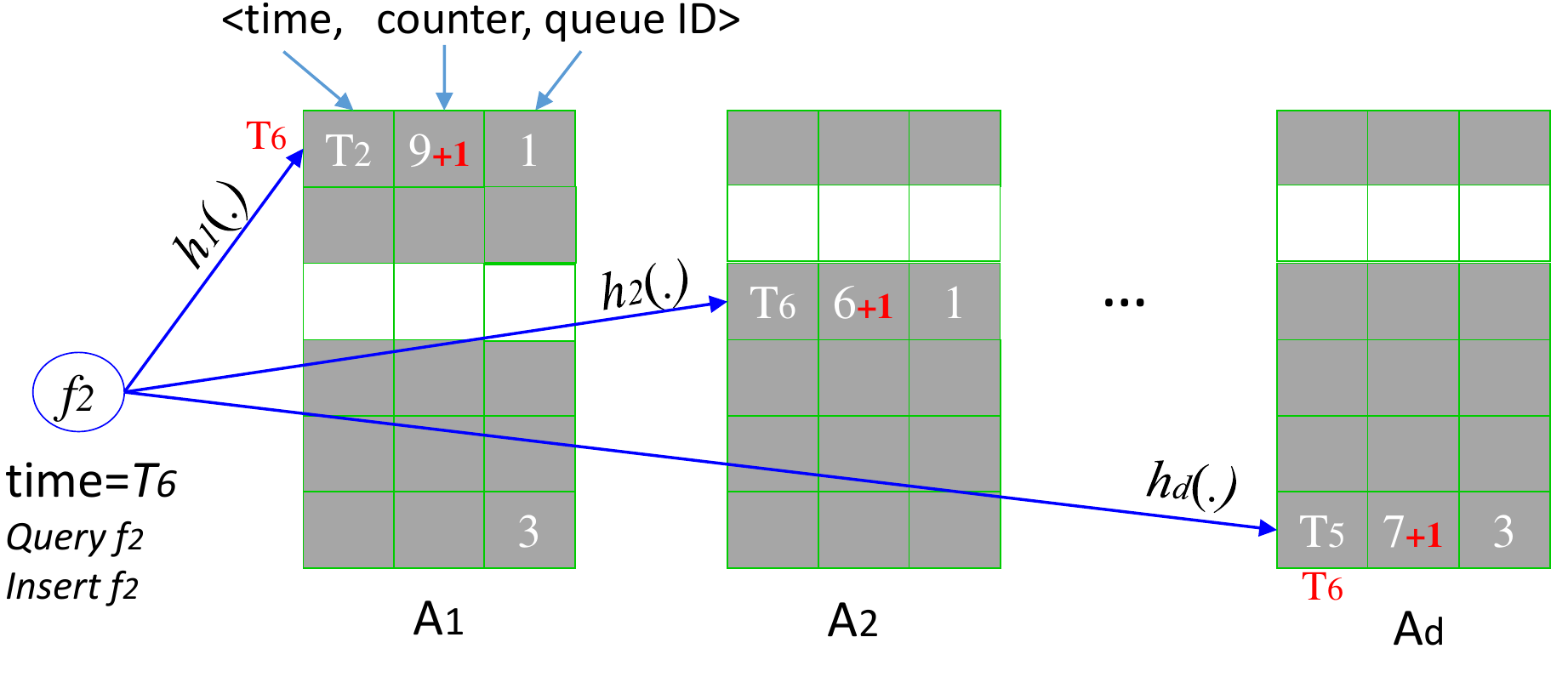}
\caption{The SCM. To query the packet with flow ID $f_2$, we get $d$ hashed buckets, and then report the oldest time $T_2$, and the smallest counter 6.
To insert a packet of flow $f_2$, for each hashed buckets, we update its timestamp to the current time $T_6$ and increment the counter by 1.
}
\label{draw:active:sketch}
\end{figure}

\bbb{Data Structure (Figure \ref{draw:active:sketch}):}
The SCM sketch consists of $d$ arrays, $A_1$, $\cdots$, $A_d$. Each array has $l$ buckets. Each bucket has three fields: a \textit{timestamp} recording the last access time, a \textit{counter} recording the number of bytes (or packets), and a \textit{queue ID} recording the queue that the last packet hashed into this bucket was sent into. 
Each array is associated with a hash function $h_i(.)$.

\bbb{Insert:} To insert a packet $a_{now}$ with flow ID $a$, we calculate $d$ hash functions, and get $d$ \textit{hashed buckets}.
We first define a threshold $\Delta T_{message}$. Suppose that the time now is $t_{now}$.
For each of the $d$ hashed buckets, if its timestamp $t_{bucket}$ is older than $t_{now} - \Delta T_{message}$, we consider the incoming packet as the first packet of a new message.
We clear this bucket, set the counter field to the number of bytes in the packet, and set the timestamp field to the current time.
Otherwise, we add the number of bytes in the packet to the counter, and update timestamp to the current time.

\bbb{Query:} To query a packet $a_{now}$ with flow ID $a$, similarly, we get the $d$ hashed buckets.
We choose the smallest size $\hat{w}(a_{now})$ as the weight of packet $a$ and the oldest timestamp $\hat{t}(a)$ as its timestamp. 
If $\hat{t}(a)$ is older than $t_{now} - \Delta T_{message}$, it means that no packets arrived in recent $\Delta T_{message}$ time.

Distinguishing flowlets is very similar to distinguishing message.
The recorded queue ID is only used for packet disorder avoidance.
The insertion and query complexity of the SCM sketch are $O(1)$.
	\presub
\subsection{Adjusting Cluster Size} 
\postsub
\label{subsec:cluster}

QCluster clusters all the packets into $k$ clusters, and each cluster corresponds to a queue.
The mean of each cluster corresponds to the queue weight.
Let $q_i$ be the $i^{th}$ queue, and let $m_i$ be the queue weight of $q_i$.
Given an incoming packet with weight $w(a_{now})$, we compare $w(a_{now})$ with $m_1$, $\cdots$, $m_k$.
Suppose $m_i\leqslant w(a_{now})<m_{i+1}$, it means we should choose $q_i$ or $q_{i+1}$.
Different from traditional approaches which choose the nearest one, we may choose differently in order to control the cluster size.
We have two strategies: Same-Cluster-Size and Proportional-Cluster-Size.

\bbb{Same-Cluster-Size.}
For Fair Queueing, we need to let all clusters have the similar size.
In practice, when all clusters have the similar size, the queue weight will be inversely proportional to the number of flows in the queue.
In dequeuing, we need to perform weighted round robin, and the weight is inversely proportional to the queue weight. 
We propose a technique, namely \textit{Adaptive Threshold}.
We define the threshold between $q_i$ and $q_{i+1}$ as: $thres_i=m_i*\beta+m_{i+1}*(1-\beta)$, where $\beta=(\frac{p_i}{p_i+p_{i+1}})^\alpha$ and $p_i$ is the number of packets in $q_i$.
When increasing/decreasing $\alpha$, the number of packets in $q_i$ will increase/decrease. 
In implementation, $\alpha$ will increase or decrease automatically according to the cluster size.

\bbb{Proportional-Cluster-Size.}
For policies of LAS, SRTP, Deadline-Aware, we need to let the size of each cluster be proportional to the queue weight.
It is well known that a large number of flows only contain a small number of packets \cite{elastic,UnivMon}. 
If we still let all clusters have the same size, small flows will be blocked by large flows in the first queue.
Therefore, Proportional-Cluster-Size can reduce FCT for small flows.
Similarly, we can also use the above adaptive-threshold technique.
The only difference is that we need to change $p_i$ to $\frac{p_i}{m_i}$.
We also try three other methods (arithmetic mean, geometric mean, harmonic mean) to achieve similar performance.
According to our experimental results (\see Figure \ref{fig:Mean:W4}), we recommend using adaptive-threshold or geometric mean.

	\presub
\subsection{Packet Disorder Avoidance}
\postsub
\label{subsec:pda}

\begin{table*} [t]
\setlength\abovecaptionskip{0.1cm}
\setlength\belowcaptionskip{0.0cm}
\centering
\caption{Applying QCluster to Scheduling Algorithms. We have applied QCluster to the first four scheduling problem. According to pFabric \cite{pfabric}, 
Deadline-aware scheduling means Deadline-first-then-SRPT in default: after meeting the dealines, we should use SRPT for other non-deadline flows. Similarly, we have Deadline-first-then-LAS, Deadline-first-then-SJF.
}
\begin{tabular}{|c|l|l|l|l|l|}
\hline												
\textbf{	Scheduling	}&\makecell[c]{\textbf{	Strategy	}}&\makecell[c]{\textbf{	Special	}}&\makecell[c]{\textbf{	Packet	}}&\makecell[c]{\textbf{	PDA (in a flowlet)	}}&\makecell[c]{\textbf{	Dequeuing	}}\\
\textbf{	problem	}&		&\makecell[c]{\textbf{	property	}}&\makecell[c]{\textbf{	weight	}}&		&\makecell[c]{\textbf{	policy	}}\\
\hline												
\footnotesize{	SRPT \cite{pfabric,homa,phost}	}&\footnotesize{	Smallest weight first	}&\footnotesize{	--	}&\footnotesize{	\# bytes remained	}&\footnotesize{	Priority cannot ascend	}&\footnotesize{	Strict priorty	}\\
\hline												
\footnotesize{	LAS \cite{auto,L2DCT,pias}	}&\footnotesize{	Smallest weight first	}&\footnotesize{	--	}&\footnotesize{	\# bytes sent	}&\footnotesize{	Priority cannot ascend	}&\footnotesize{	Strict priorty	}\\
\hline												
\footnotesize{	Fair Queueing \cite{afq,nagle,BR}	}&\footnotesize{	Fair scheduling	}&\footnotesize{	--	}&\footnotesize{	\# bytes sent	}&\footnotesize{	Queue cannot change	}&\footnotesize{    Weighted round robin   }\\
\hline												
\footnotesize{	Deadline-aware-SRPT \cite{pfabric,pdq,d3}   }&\footnotesize{	Deadline flow first, then ...	}&\footnotesize{	Deadline flows	}&\footnotesize{	Hybrid	}&\footnotesize{	Priority cannot ascend	}&\footnotesize{	Strict priorty	}\\
\hline												
\footnotesize{	Shortest Job First (SJF)	}&\footnotesize{	Smallest weight first	}&\footnotesize{	--	}&\footnotesize{	Total flow size	}&\footnotesize{	Priority cannot ascend	}&\footnotesize{	Strict priorty	}\\
\hline												
\footnotesize{	Deadline-first-then-	}&\footnotesize{	Deadline flow first, then ...	}&\footnotesize{	Deadline flows	}&\footnotesize{	Hybrid	}&\footnotesize{	--	}&\footnotesize{	Hybrid	}\\
\footnotesize{	LAS/SJF/Fairness	}&		&		&		&		&		\\
\hline												
\footnotesize{	Coflow scheduling \cite{coflow, coda, coflowbest18}	}&\footnotesize{	Weighted fair scheduling	}&\footnotesize{	--	}&\footnotesize{	\# bytes sent	}&\footnotesize{	Priority cannot ascend	}&\footnotesize{	Weighted sharing	}\\
\hline												
\footnotesize{	Minimum rate guarantees \cite{tcpminrate}	}&\footnotesize{	Flow below its	minimum }&\footnotesize{	Flows below their	}&\footnotesize{	Hybrid	}&\footnotesize{	--	}&\footnotesize{	Hybrid	}\\
		&\footnotesize{	 rate first	}&\footnotesize{    minimum rate	}&		&		&		\\
\hline												
\footnotesize{	TSN flow scheduling \cite{TSN,802.1Qbv}	}&\footnotesize{	TSN flow first	}&\footnotesize{	TSN flows	}&\footnotesize{	Hybrid	}&\footnotesize{	--	}&\footnotesize{	Hybrid	}\\
\hline												
\footnotesize{	Hybrid scheduling scenarios	}&\footnotesize{	Hybrid	}&\footnotesize{	--	}&\footnotesize{	Hybrid	}&\footnotesize{	--	}&\footnotesize{	Hybrid	}\\
\hline												
\end{tabular}
\vspace{-0.4cm}
\label{tab:application}
\end{table*}

When deployed in switches, QCluster will automatically adjust queue thresholds across time and space. 
On the one hand, dynamic thresholds can achieve better performance; on the other hand, latter packets could be sent to higher-priority queues while the former packets of the same flow are still in a congested low-priority queue, thus packet disorder may happen.
While existing solutions \cite{presto,juggler,kandula2007dynamic, conga, hula,drill} can significantly reduce the probability of packet disorder, we aim to avoid disorder.

We propose an algorithm named Packet Disorder Avoidance (PDA).
In the PDA algorithm, we need to know whether the previous packet of the current flow is in the switch, which currently cannot be implemented in commodity switches, and thus we use the SCM sketch and \textit{Flowlet} for approximate implementation.
Flowlet is first proposed by Erico Vanini \textit{et al.} \cite{flowlet}, and we change the definition a little: given an incoming packet $a_{now}$ with flow ID $a$, if all the packets in all queues do not belong to flow $a$, we consider $a_{now}$ as the beginning of a Flowlet.
In this case, the last packet $a_{last}$ of flow $a$ has already been sent to the next-hop switches, and $a_{now}$ can go to any queue. 
In other words, different Flowlets can be scheduled individually and packet disorder will not happen.

The SCM sketch reports the last arriving time of flow $a$. If the last packet of flow $a$ was sent to a queue more than $\Delta T_{Flowlet}$ ago, we consider $a_{now}$ the first packet of a new Flowlet.
If $a_{now}$ is not the first packet of a new Flowlet, we query the SCM sketch to get which queue the previous packet stays.
PDA works slightly differently for LAS/SRPT/Deadline-Aware and Fair Queueing.
For LAS/SRPT/Deadline-Aware, given an incoming packet $a_{now}$ of flow $a$, if $a_{now}$ is not the first packet of a Flowlet, and the previous packet of flow $a$ is in queue $q_i$, we do not allow $a_{now}$ to go to \textit{any higher-priority queue} than $q_i$; Otherwise, we can choose the queue according to the clustering algorithm, and packet disorder will not happen.
For Fair Queueing, if the previous packet of flow $a$ is still in one queue ($q_i$), we do not allow $a_{now}$ to change the queue, \ie, $a_{now}$ can only go to $q_i$; Otherwise, we choose the queue according to the clustering algorithm.

	\presec
\section{Applications}
\postsec
\label{sec:application}

We have applied QCluster to four scheduling problems: SRPT, LAS, Fair Queueing, and Deadline-Aware scheduling.
This section also shows how to apply QCluster to other 6 flow scheduling problems.
We list all the scheduling problems and applications in Table \ref{tab:application}.
When applying QCluster to different scheduling problems, the differences lie in the following aspects:

\bbb{1) Strategy:} The strategies for scheduling flows. If there is no special property, most scheduling problems let the packet with the smallest weight go first.

\bbb{2) Special property:} Special requirements that the algorithms must meet. For example, for Deadline-Aware Scheduling, deadline flows must complete before their deadlines; for TSN flows \cite{TSN}, they have the highest priority when they arrive at a switch.
Flows with special property are clustered into high-priority queues, and the other flows are clustered into low-priority queues. When all high-priority queues have no packet, we can dequeue packets from low-priority queues using strict priority or round robin.

\bbb{3) Packet weight:} The packet weight for scheduling. Packet weight can be the number of bytes already sent, total flow size, or the remaining flow size. Hybrid means for flows without special property, the weight is different for SRPT, LAS, Fair Queueing, \etc

\bbb{4) PDA (in a flowlet):} The requirements for packet disorder avoidance. For Fair Queueing, we do not allow the incoming packet to change its egress queue in a flowlet. For other scheduling policies, we do not allow the incoming packet to go to a queue with a higher priority than the previous packets in a flowlet.

\bbb{5) Dequeuing policy:} Most scheduling problems use strict priority. Fair Queueing uses weighted round robin. Weighted sharing is proposed by Aalo \cite{coflow}. 

In practice, applications may need hybrid policies, as shown in the end of Table \ref{tab:application}. As QCluster can implement all basic scheduling policies, and thus can also be adapted for hybrid scheduling scenarios.
For example, there are many deadline flows, and users may also want to maximize fairness for deadline flows. In this case, we can cluster deadline flows into the first several queues, and dequeue with weighted round robin.

    \presec
\section{Testbed and Implementations} 
\postsec
\label{sec:implementations}

We build a testbed to evaluate QCluster, and deploy it in a Tofino switch.
In our testbed, we focus on LAS and Fair Queueing. For other two policies, we show the large-scale simulations in ns-2. 

\presub
\subsection{Testbed Setup} 
\postsub

Our testbed consists of 7 servers and an Edgecore Wedge 100BF-32X switch (with Tofino ASIC). 
We use another 1000 Mbps switch to manage the servers and the Tofino switch. 
Each server runs Ubuntu 16.04-64bit with Linux kernel 4.13, and is connected to the Tofino switch via a Mellanox ConnectX-3 40GbE NIC.
We are able to achieve about 37Gbit/s throughput between each pair of servers.
To improve the FCT, we set the RTO-min in Linux kernel to 10ms.
In the switch, we use the per-port ECN marking, and set the marking threshold to 300KBytes (about 200 packets).

\presub
\subsection{Implementation in P4} 
\postsub
\label{sec:imple:p4}

We have fully implemented a P4 version of QCluster with 500 lines of P4 code, including all the \textit{registers} and \textit{metadata} for QCluster in the data plane, and compiled it to the Tofino switch~\cite{tofino}.

\bbb{Using Registers and SALUs.}
In the Tofino switch, we use \textit{registers} to implement the SCM sketch, where registers are a kind of stateful objects.
We leverage the SALU (Stateful ALU) in each stage to look up and update the entries in the registers. 
In the current Tofino switches, a SALU can at most update a pair of up to 32-bit registers, while one bucket of our SCM sketch has three fields. 
To address this problem, we divided our SCM sketch into two sketches.
The two sketches have the same number of buckets and the same $k$ hash functions.
The difference is that each bucket of the first sketch includes the timestamp and counter, while that of the second sketch includes the timestamp and Queue ID.
The timestamps of two sketches are the same, and thus redundant, but inevitable.

\bbb{Update of Queue Weights.}
For each queue, we maintain two variables: \textit{weight sum} (the number of total weight) and \textit{packet number}.
The queue weight is the weight sum divided by packet number.
After an incoming packet is sent to the chosen queue, we add its weight to the weight sum and increment the packet number by 1. 
When one packet dequeues, we do not update the queue weights because the ingress/egress pipelines in Tofino switches do not share memory, so it is tricky to implement the update.

\vspace{-0.5cm}
\begin{table} [!h]
\setlength\abovecaptionskip{0.1cm}
\setlength\belowcaptionskip{0.1cm}
\centering\caption{\small H/W resources used in P4 by QCluster.
}
{\centering
\small
\begin{tabular}{|l|r| r|c|}
\hline
\textbf{Resource} & \textbf{Usage} & \textbf{Percentage}\\ \hline
SRAM & 61 & 6.35\% \\ 
TCAM & 3 & 1.04\% \\  
Hash Bits & 94 & 1.88\% \\ 
Stateful ALU & 5 & 10.42\% \\ 
\hline 
\end{tabular}
}
\label{tab:overhead}
\end{table}
\vspace{-0.4cm}

\bbb{The Problem of Division.}
The current Tofino switch does not support the division operation in the data plane, and thus we cannot directly calculate the queue weight.
However, we noticed that these queue weights can tolerate errors, which allows some delay before updating the average values.
Our key idea is to use the control plane to calculate and update the queue weights periodically. 
Every packet goes through a \textit{range match-action table} to determine which queue to send, increments the \textit{packet number}, and accumulates the \textit{weight sum} for this queue.
Using range match-action tables is because comparing one weight with the $k$ queue weights one by one needs too many stages.
The thresholds of the range match entries are inserted and updated by the control plane periodically.

\bbb{Resource Usage.}
In this way, we only need 6 stages: one stage to get the timestamp, two stages for lookup and insertion of SCM sketch, one stage for calculating $\hat{w}$, one stage for range matching, and one stage for the increment of weight sum and packet number.
Table~\ref{tab:overhead} shows the resource usage.
As a result, we can fit the QCluster into the switch ASIC for packet processing at line-rate.

    \presec
\section{Experimental Results} 
\postsec

In this section, we conduct extensive experiments in a testbed and ns-2, and compare our QCluster with the state-of-the-art solutions.
In all experimental figures below, QCluster for different problems is abbreviated as follows.
\bbb{QC-SRPT:} QCluster for SRPT.
\bbb{QC-LAS:} QCluster for LAS.
\bbb{QC-FQ:} QCluster for Fair Queueing.
\bbb{QC-DDL:} QCluster for Deadline-Aware Scheduling.
    \presub
\subsection{Experimental Setup} 
\postsub

\begin{figure}[htbp]
\setlength\abovecaptionskip{-0.1cm}
\setlength\belowcaptionskip{-0.3cm}
\setlength\subfigcapskip{-0.5cm}
\subfigure[Cumulative of Messages]{
\begin{minipage}[b]{0.185\textwidth}
\includegraphics[width=\textwidth]{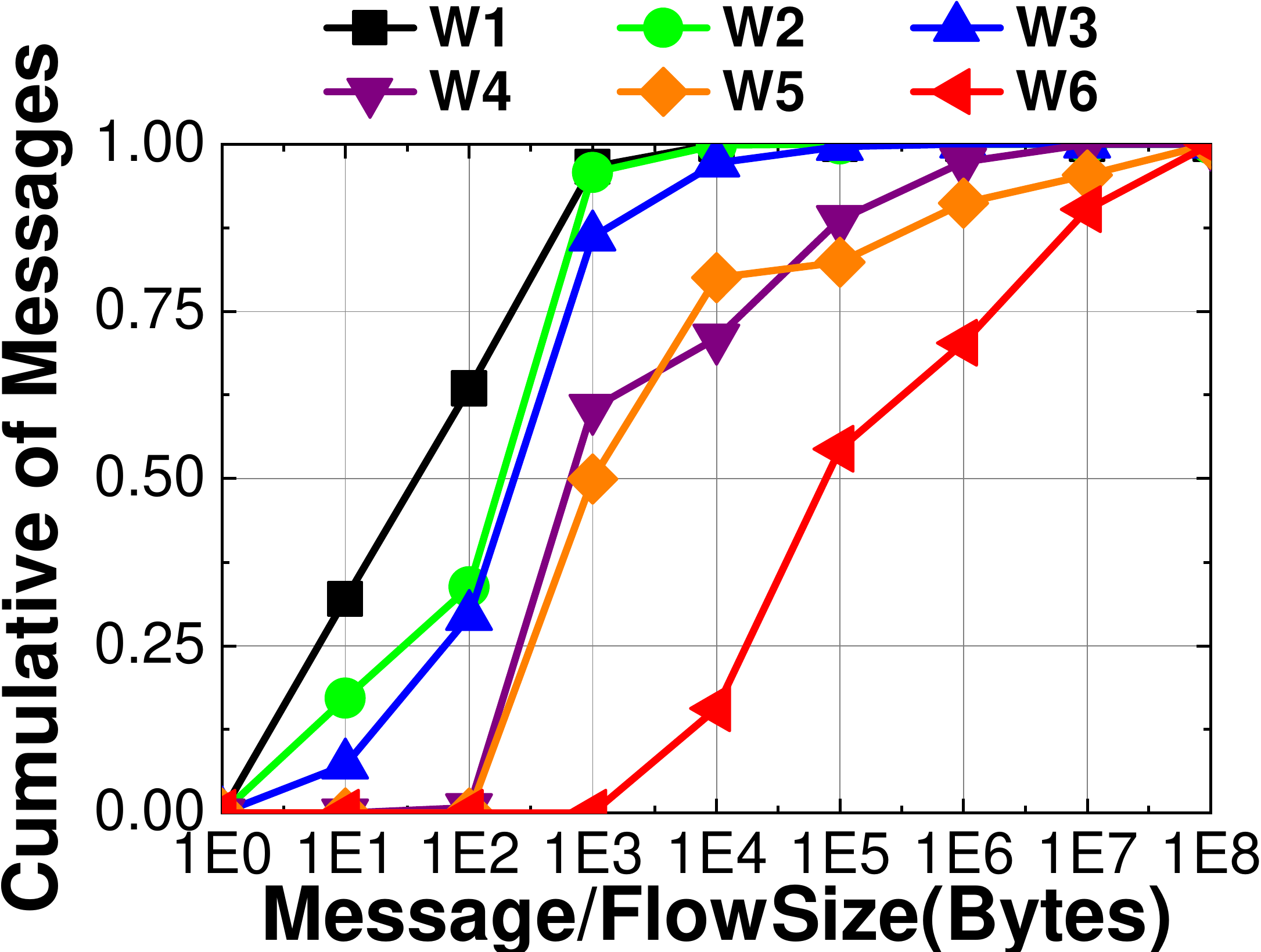}
\postsubfig
\label{fig:thres:W4}
\end{minipage}
}
\subfigure[Cumulative of Bytes]{
\begin{minipage}[b]{0.185\textwidth}
\includegraphics[width=\textwidth]{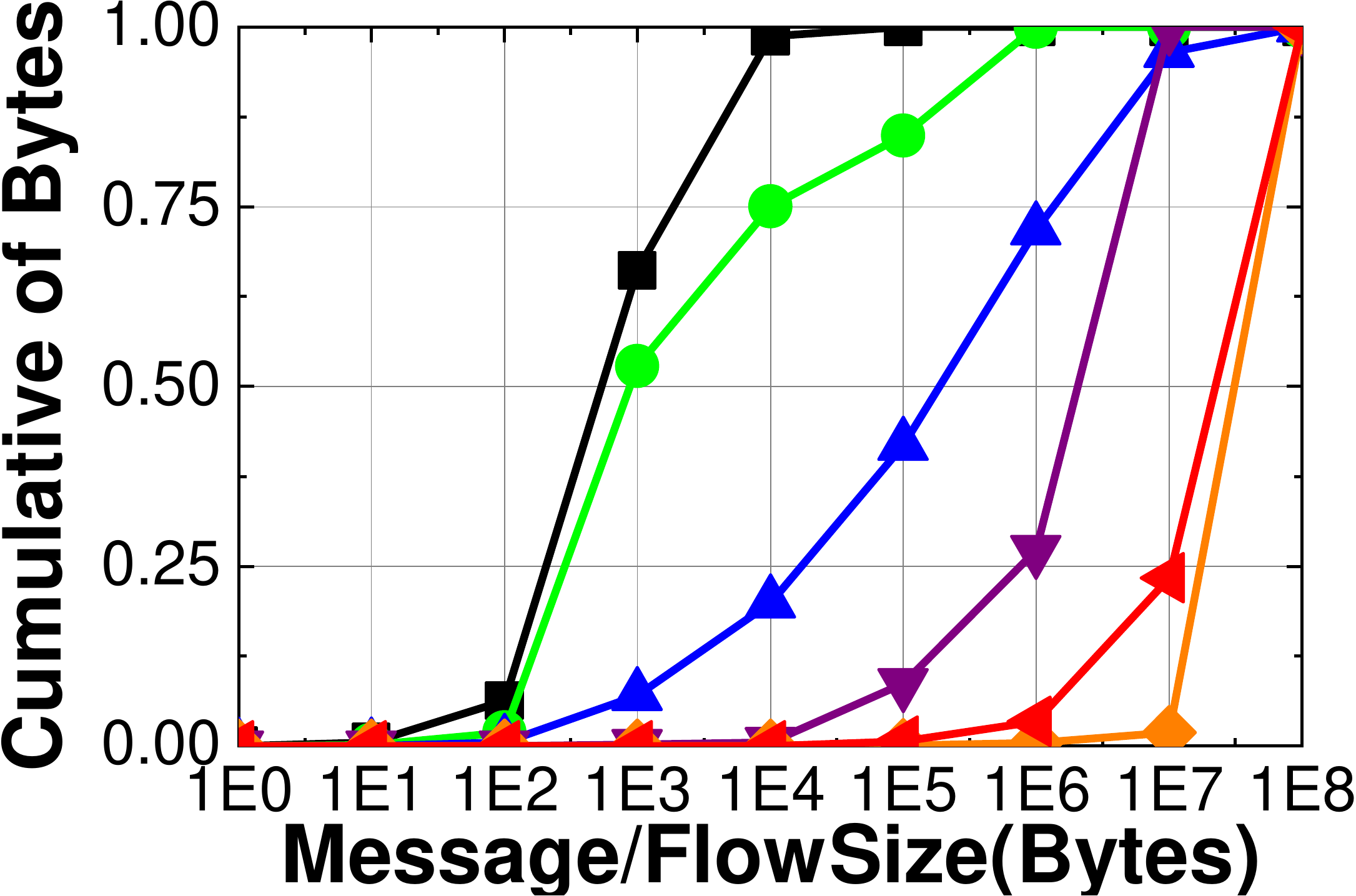}
\postsubfig
\label{fig:thres:W6}
\end{minipage}
}
\postsubfig
\caption{The workloads used to evaluate. The distributions are based on measurements from real production data centers. \cite{keyvalue,hadoop,googledata,vl2,dctcp}}
\label{fig:distribution}
\end{figure}

\bbb{Workloads (Figure~\ref{fig:distribution}):}
We use six workloads as previous works did \cite{homa,pias,pfabric}. Their distributions are shown in Figure \ref{fig:distribution}.
We use W1-W4 which were used to evaluate Homa. Besides, we also use the Data Mining workload(W5) and the Web Search workload(W6) which were used to evaluate DCTCP\cite{dctcp}, pFabric\cite{pfabric} and PIAS\cite{pias}.
We consider flows smaller than 1KB as small flows, and flows larger than 10KB as large flows.
Because there is no flow smaller than 1KB in W6, we consider flows less than 10KB and flows larger than 100KB as small flows and large flows in W6. 
In the following experiments, we mainly use W4 and W6 for comparisons because these two workloads are frequently used by existing solutions.
We show the performances on the remaining workloads in supplementary material due to space limitation.

\bbb{Comparison with state-of-the-art:}

\noindent 1) \textit{For SRPT and LAS}, we compare QC-LAS and QC-SRPT with simulations of pFabric\cite{pfabric}, PIAS\cite{pias}, and DCTCP\cite{dctcp}. 
%
We do not compare with Homa \cite{homa} because of two reasons: 
(1) Homa does not provide simulation codes in ns-2.
(2) Homa assumes congestion often happens in the downlinks of edge switches, while we do not.

\noindent\textit{2) For Fair Queueing,} we compare QC-FQ with ideal fair queueing, ideal fair queueing with ECN, and AFQ.
We use the BR\cite{BR} algorithm as the ideal fair queueing.
AFQ is chosen because it is approximately the practical version of BR.

\noindent\textit{3) For Deadline-Aware Scheduling}, we compare our QC-DDL with pFabric-EDF\cite{pfabric} and DCTCP\cite{dctcp}. Like pFabric-EDF, QC-DDL aims to minimize the FCT of non-deadline flows and maximize the throughput of deadline flows. 
However, other algorithms, like MCP\cite{MCP} and Karuna\cite{karuna}, only address one of these situations in their evaluation.

\bbb{Metrics: }

\bbb{Flow completion times (FCT): }
FCT is generally used in measuring the performance of scheduling algorithms.
We measure the average FCT across all flows, and separately for different flow sizes. 
We also consider the 99th percentile flow completion time. 

\bbb{Jain's fairness index \cite{jain}: }
As AFQ \cite{afq} does, we use Jain's Fairness index to measure the fairness. 
It is defined as $J(x_1,\cdots,x_n) = (\sum_{i=1}^n x_i)^2/(n\cdot \sum_{i=1}^n x_i^2)$, where $x_i$ is the average throughput of flows with the same order of magnitude. 

\bbb{Application throughput: }
For deadline traffic, we measure the application throughput which is defined as the fraction of flows that meet their deadlines.

    \presub
\subsection{Experiments in Testbed} \postsub
\label{exp:tb_reorder}

\begin{figure}[htbp]
\setlength\abovecaptionskip{-0.1cm}
\setlength\belowcaptionskip{-0.6cm}
\setlength\subfigcapskip{-0.5cm}
\subfigure[W4]{
\begin{minipage}[b]{0.185\textwidth}
\includegraphics[width=\textwidth]{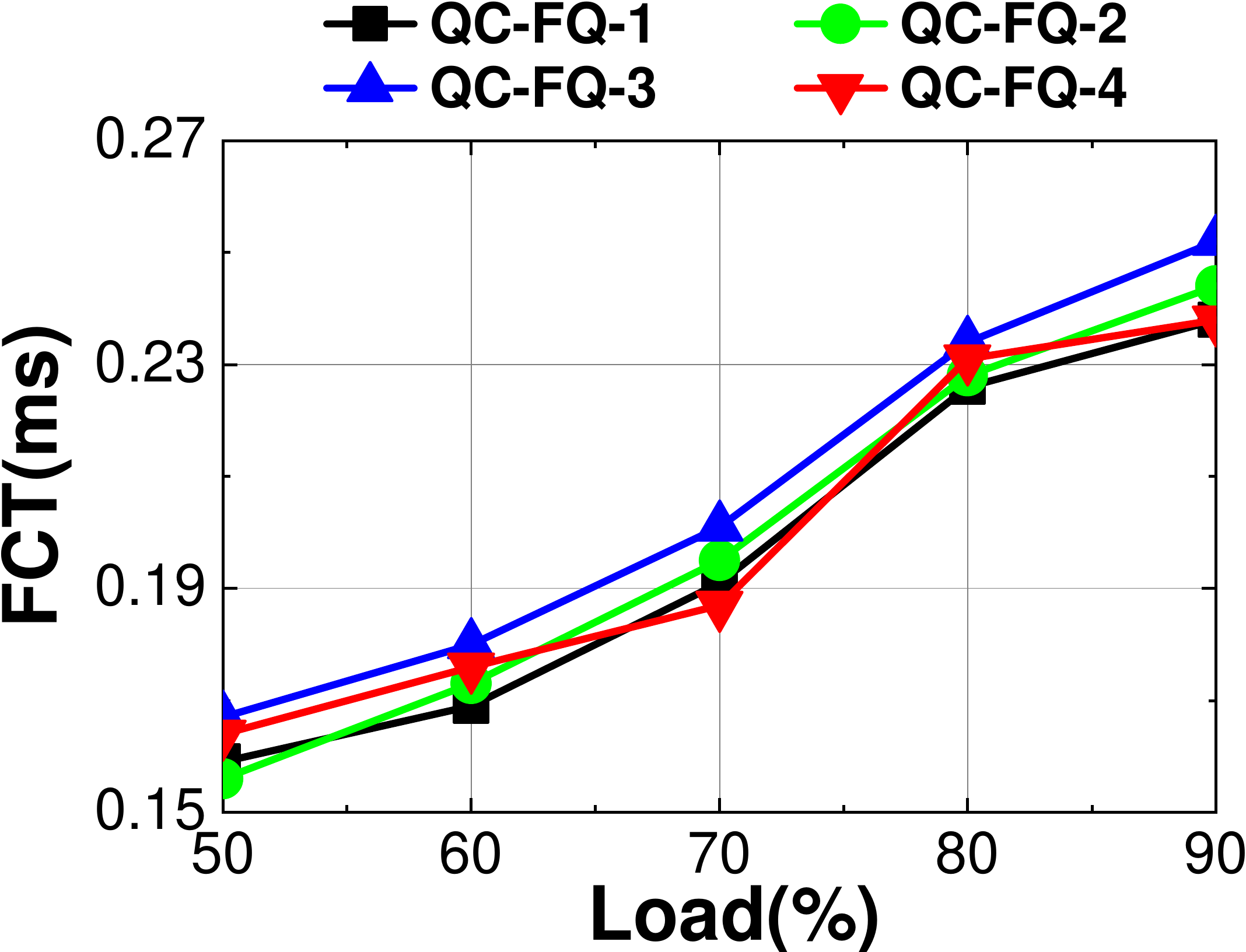}
\postsubfig
\label{fig:thres:W4}
\end{minipage}
}
\subfigure[W6]{
\begin{minipage}[b]{0.18\textwidth}
\includegraphics[width=\textwidth]{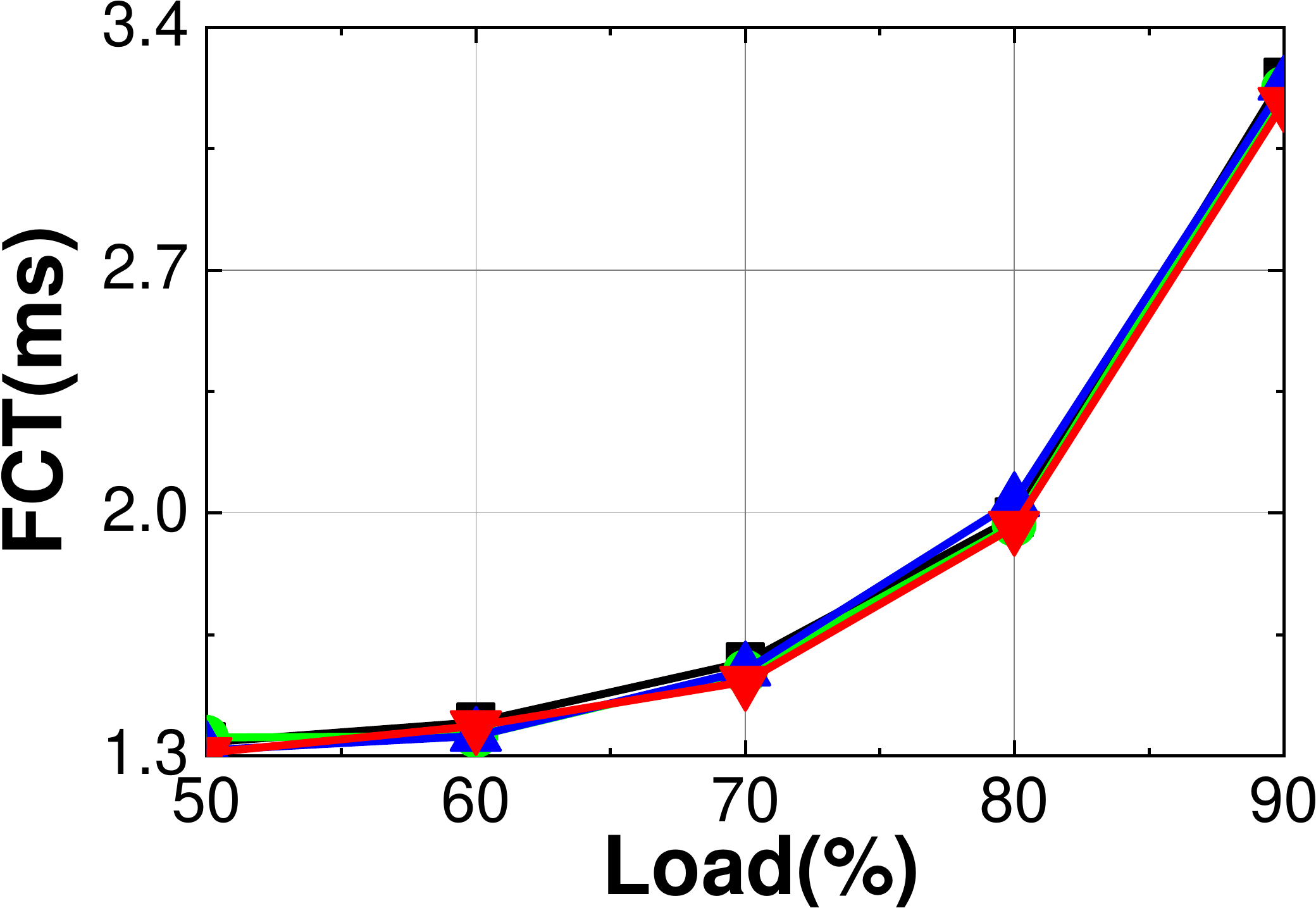}
\postsubfig
\label{fig:thres:W6}
\end{minipage}
}
\postsubfig
\caption{Overall average FCT of different initial thresholds for QC-FQ on different workloads.}
\label{fig:thres}
\end{figure}

\begin{figure}[htbp]
\setlength\abovecaptionskip{-0.1cm}
\setlength\belowcaptionskip{-0.5cm}
\setlength\subfigcapskip{-0.4cm}
\centering
\subfigure[QC-FQ-1]{
\begin{minipage}[b]{0.185\textwidth}
\centering
\includegraphics[width=\textwidth]{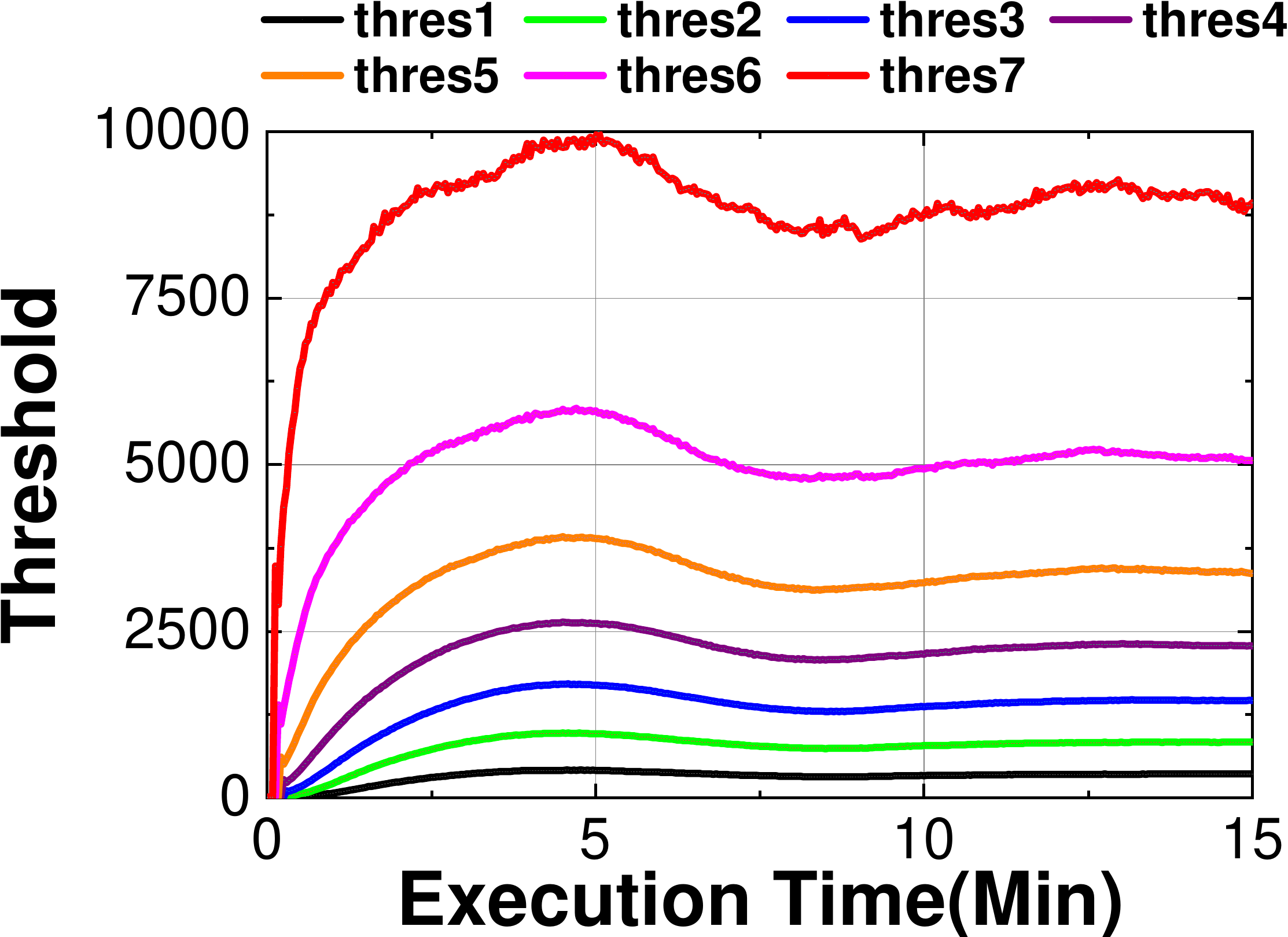}
\postsubfig
\label{fig:tb_thres:vary:1}
\end{minipage}
}
\subfigure[QC-FQ-3]{
\begin{minipage}[b]{0.185\textwidth}
\centering
\includegraphics[width=\textwidth]{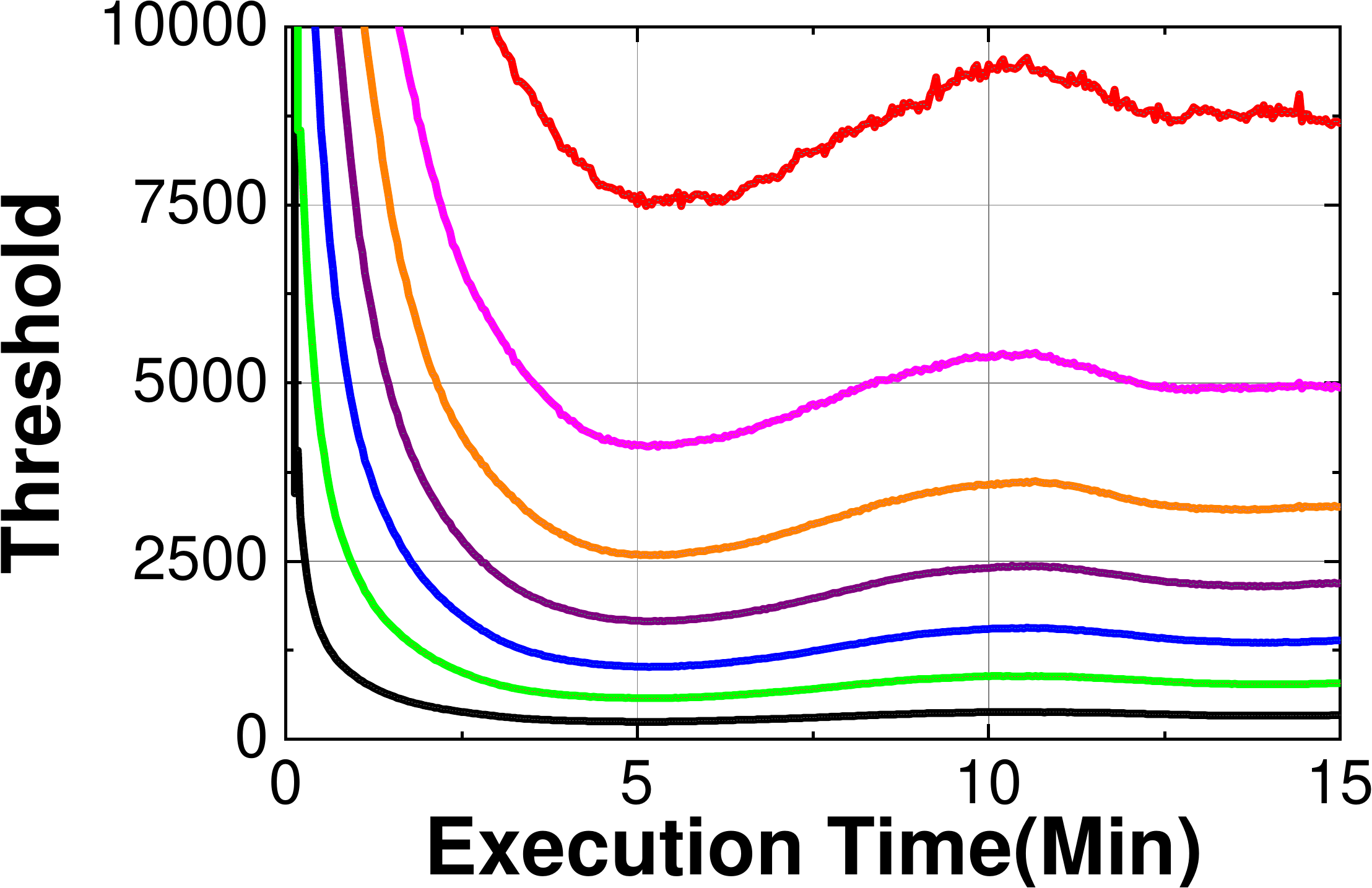}
\postsubfig
\label{fig:tb_thres:vary:2}
\end{minipage}
}
\postsubfig
\caption{Change of thresholds over time for QC-FQ on W6.}
\label{fig:tb_thres:vary}
\end{figure}

\begin{figure}[htbp]
\setlength\abovecaptionskip{-0.1cm}
\setlength\belowcaptionskip{-0.4cm}
\setlength\subfigcapskip{-0.5cm}
\centering
\subfigure[QC-FQ-1]{
\begin{minipage}[b]{0.185\textwidth}
\centering
\includegraphics[width=\textwidth]{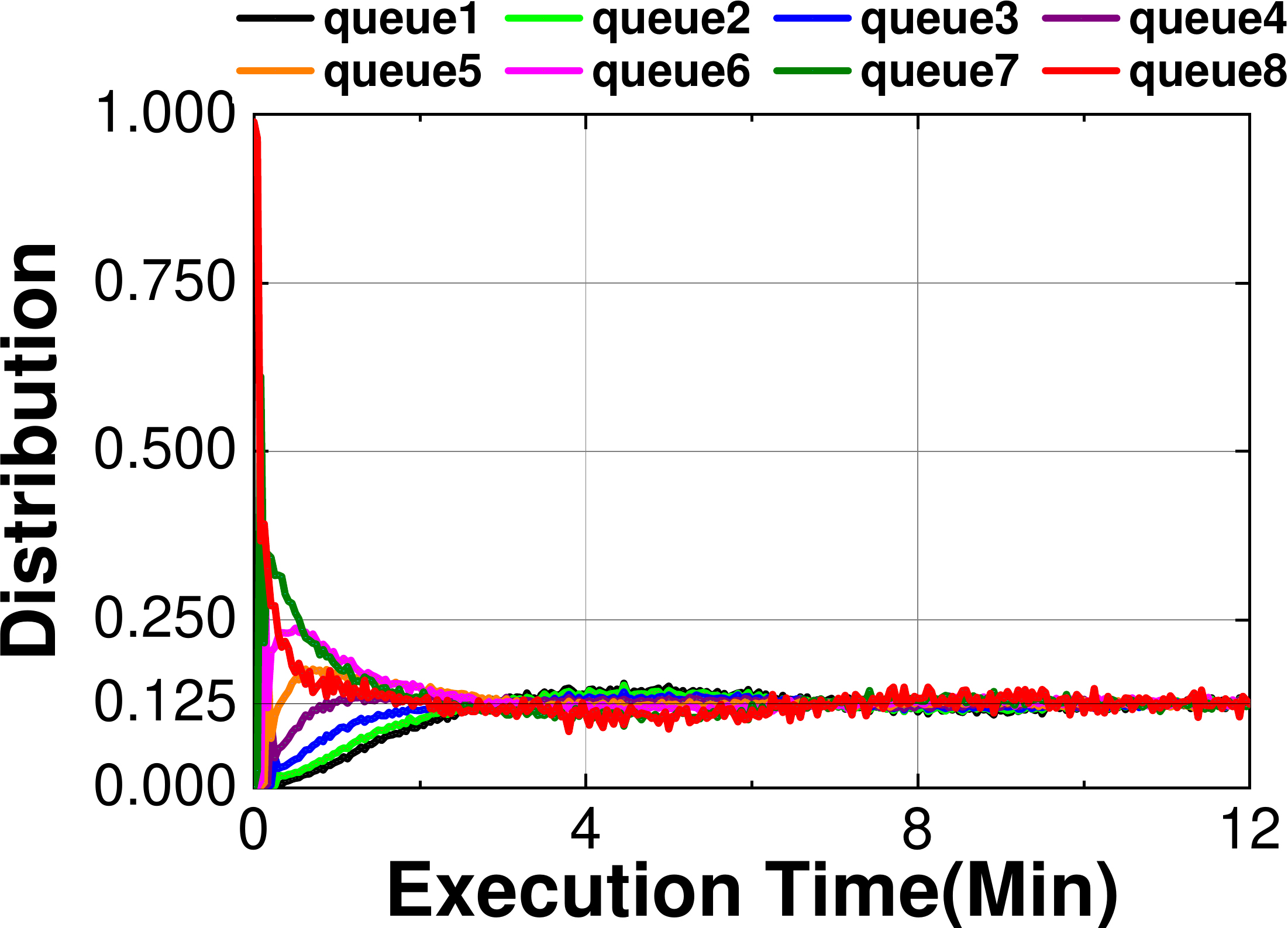}
\postsubfig
\label{fig:tb_dist:vary:1}
\end{minipage}
}
\subfigure[QC-FQ-3]{
\begin{minipage}[b]{0.185\textwidth}
\centering
\includegraphics[width=\textwidth]{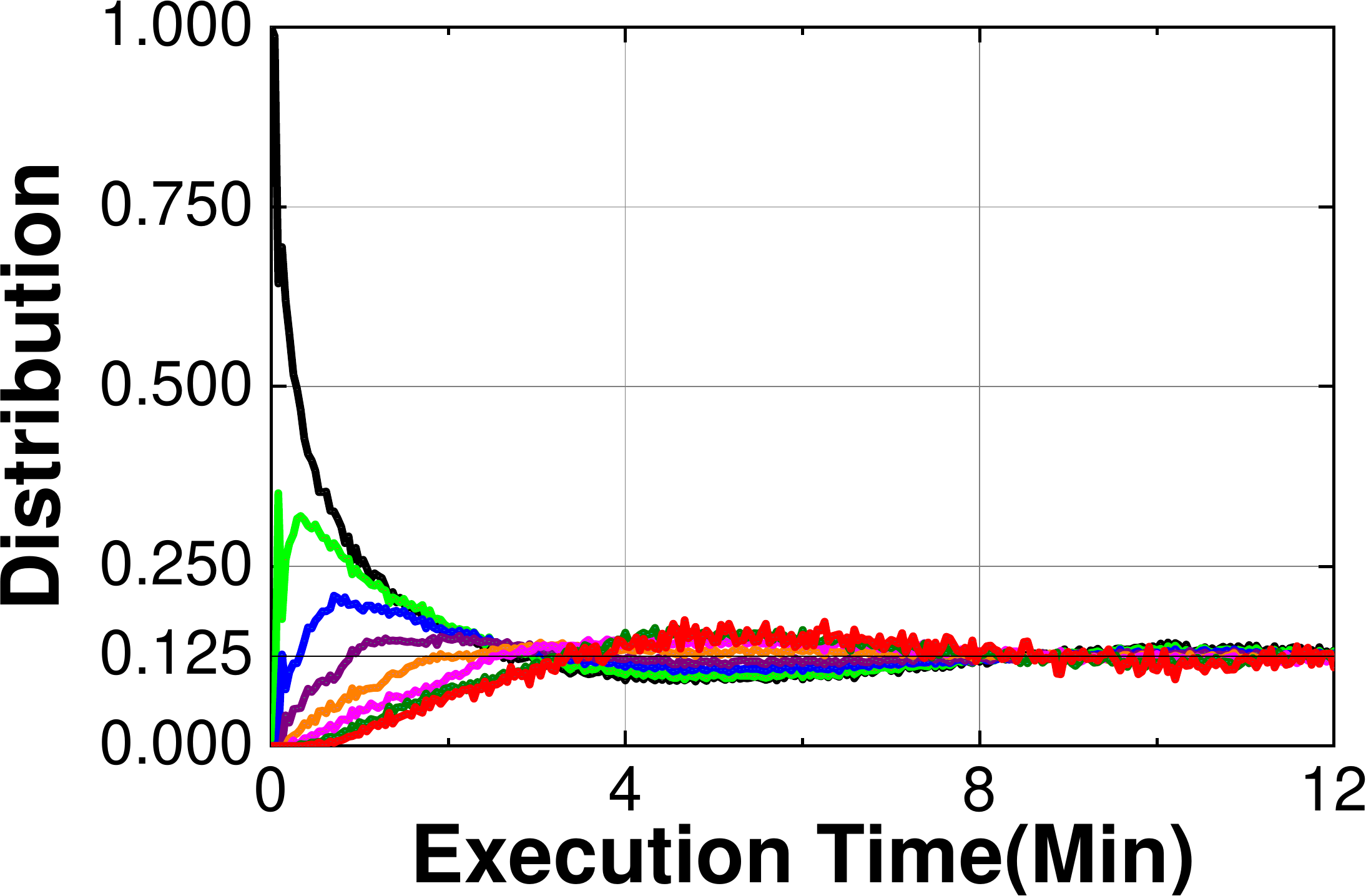}
\postsubfig
\label{fig:tb_dist:vary:2}
\end{minipage}
}
\postsubfig
\caption{Change of packet distributions over time for QC-FQ on W6. The packet distribution is the proportion of packets in one queue among all packets.}
\label{fig:tb_dist:vary}
\end{figure}

In this section, we show the experimental results in our testbed.
We first take QC-FQ as the example to show the influence of different initial thresholds, and then compare QC-LAS and QC-FQ with DCTCP and PIAS. 
We use four different initial threshold settings for QC-FQ ( QC-FQ-1 to QC-FQ-4). In the first three settings, all flows will go to the queue with the lowest, middle or highest priority at the beginning. In QC-FQ-4, all flows will go to queues with either the highest or the lowest priority.
The performance of PIAS heavily depends on the thresholds. 
We implement PIAS in P4 switch, and use two sets of thresholds for PIAS. 
In our testbed, the flow size distribution is known, and thus we can obtain the optimal static thresholds, ``PIAS-OPT''.
We can also find static thresholds that result in poor performance, ``PIAS-WST''. 
A simple method producing ``PIAS-WST'' is to let 60\% packets go to the first queue, 30\% packets go to the last. 

\bbb{QC-FQ performance on different initial thresholds: 
(Figure~\ref{fig:thres}-\ref{fig:tb_dist:vary}): }
The overall FCT using different initial thresholds are close to each other in both W4 and W6.
For different initial thresholds in 90\% workload of W6, the thresholds of same queues converge to the similar values respectively, and the proportion of packets in each queue becomes close to each other.
It is because that QCluster will update and optimize the thresholds in a short period of time, and the poor performance at the beginning has little impact on performance in the long term.

\vspace{-0.1cm}
\begin{figure}[htbp]
\setlength\abovecaptionskip{-0.1cm}
\setlength\belowcaptionskip{-0.3cm}
\setlength\subfigcapskip{-0.5cm}
\subfigure[W4]{
\begin{minipage}[b]{0.185\textwidth}
\includegraphics[width=\textwidth]{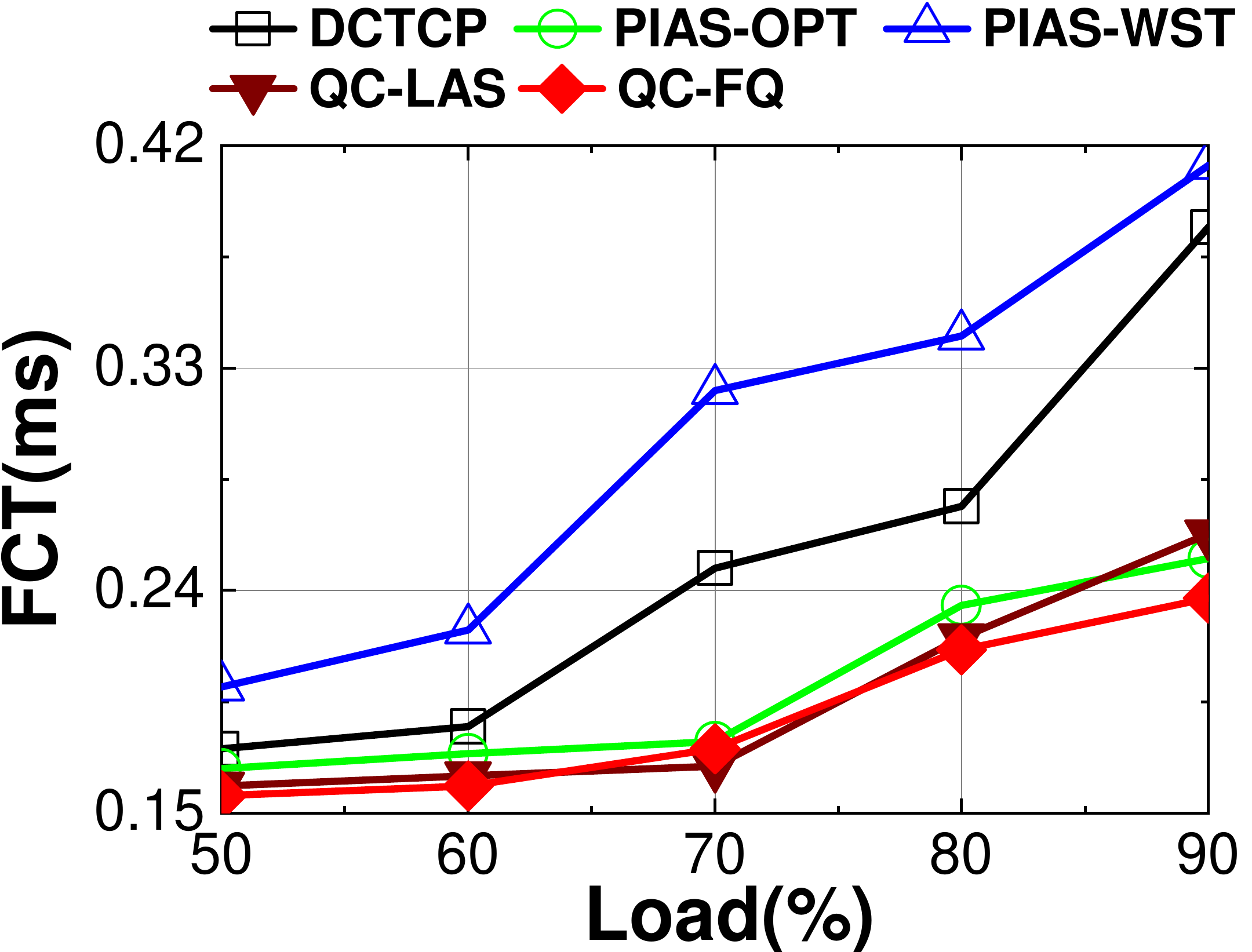}
\postsubfig
\label{fig:end:W4}
\end{minipage}
}
\subfigure[W6]{
\begin{minipage}[b]{0.177\textwidth}
\includegraphics[width=\textwidth]{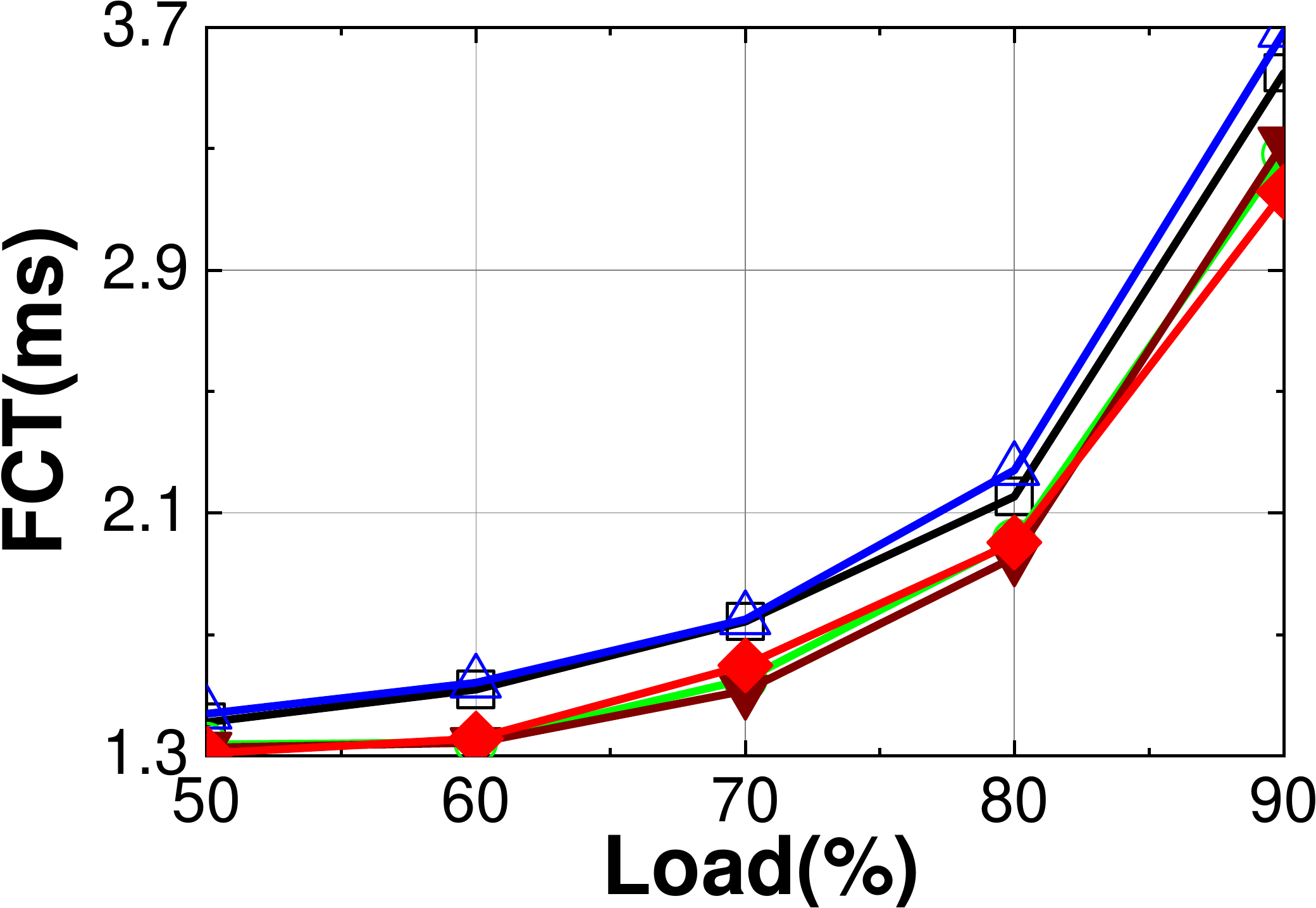}
\postsubfig
\label{fig:end:W6}
\end{minipage}
}
\postsubfig
\caption{Overall average FCT on different workloads.}
\label{fig:end}
\end{figure}

\bbb{Overall performance for different scheduling policies: 
(Figure~\ref{fig:end}): }
The overall FCT of our QC-FQ is about 42.4\% and 14.3\% lower than that of PIAS-WST, and about 38.8\% and 11.0\% lower than that of DCTCP on W4 and W6.
Compared to PIAS-OPT, the overall FCT of QC-FQ is about 6.3\% and 3.7\% lower on W4 and W6.

\begin{figure*}[htbp]
\setlength\abovecaptionskip{-0.1cm}
\setlength\belowcaptionskip{-0.4cm}
\setlength\subfigcapskip{-0.5cm}
\centering
\subfigure[(0,1KB): Avg]{
\begin{minipage}[b]{0.202\textwidth}
\centering
\includegraphics[width=\textwidth]{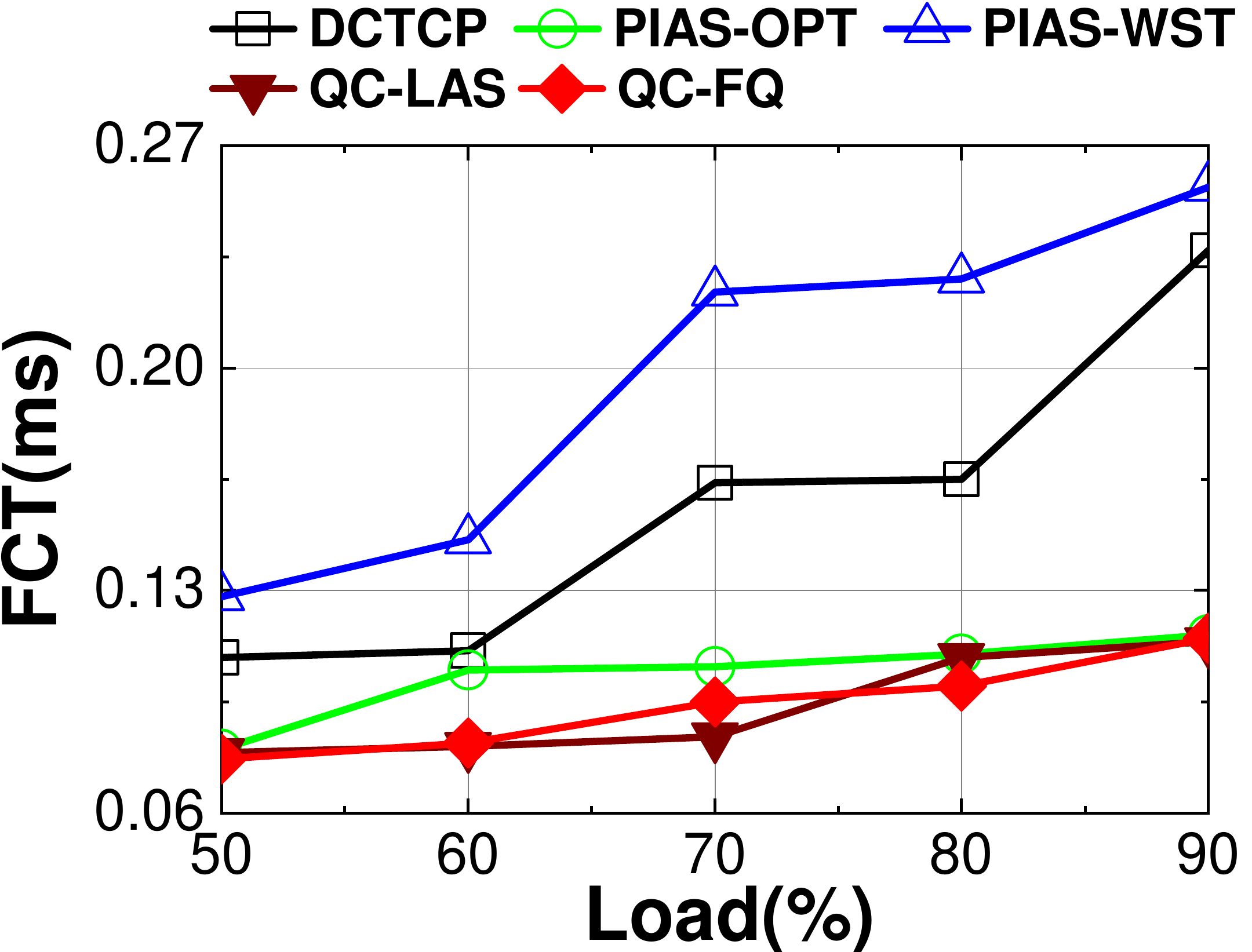}
\postsubfig
\label{fig:tb_end:W4:1K}
\end{minipage}
}
\subfigure[(0,1KB): 99th Percentile]{
\begin{minipage}[b]{0.181\textwidth}
\centering
\includegraphics[width=\textwidth]{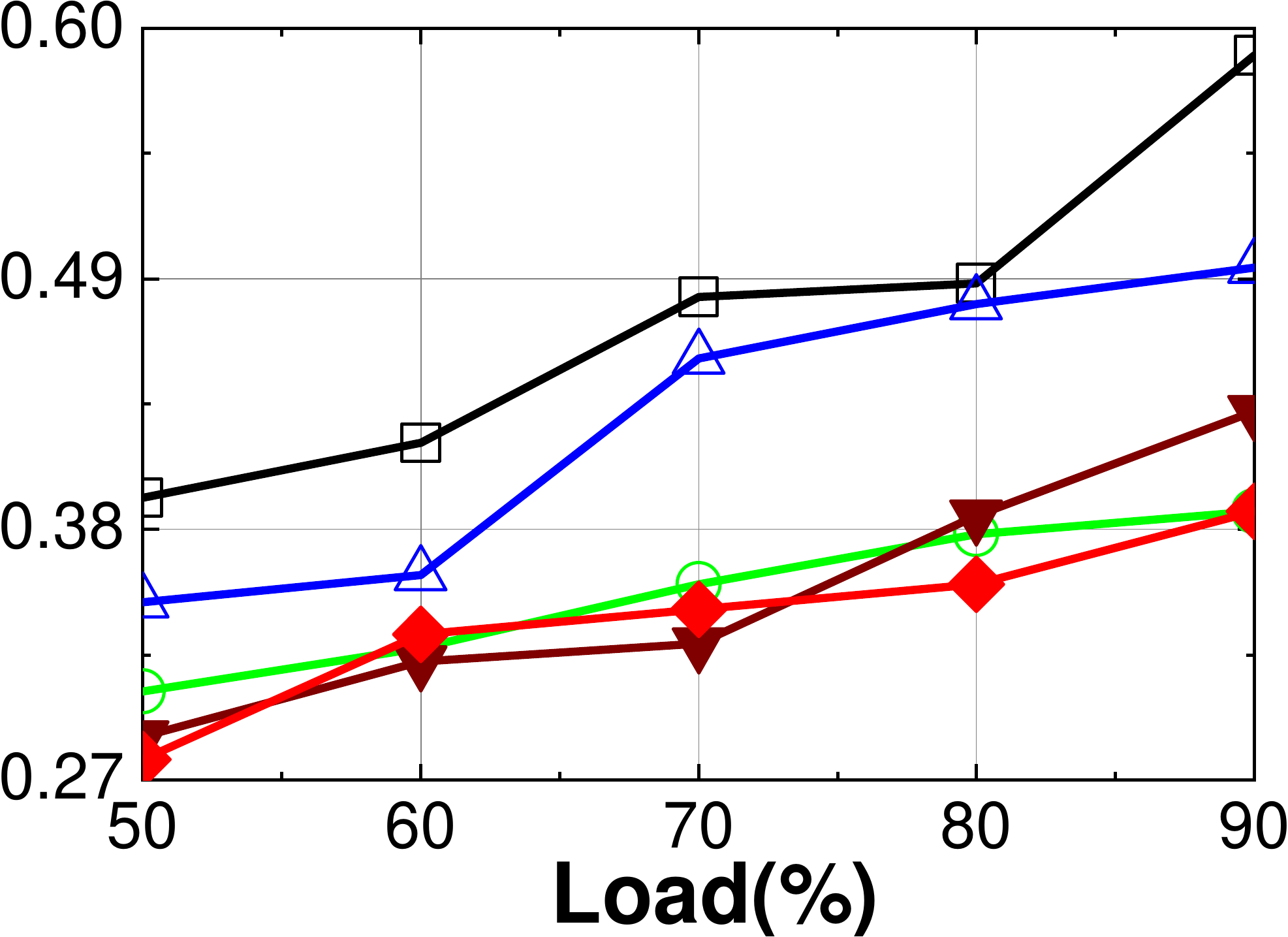}
\postsubfig
\label{fig:tb_end:W4:99th}
\end{minipage}
}
\subfigure[(1KB,10KB): Avg]{
\begin{minipage}[b]{0.182\textwidth}
\centering
\includegraphics[width=\textwidth]{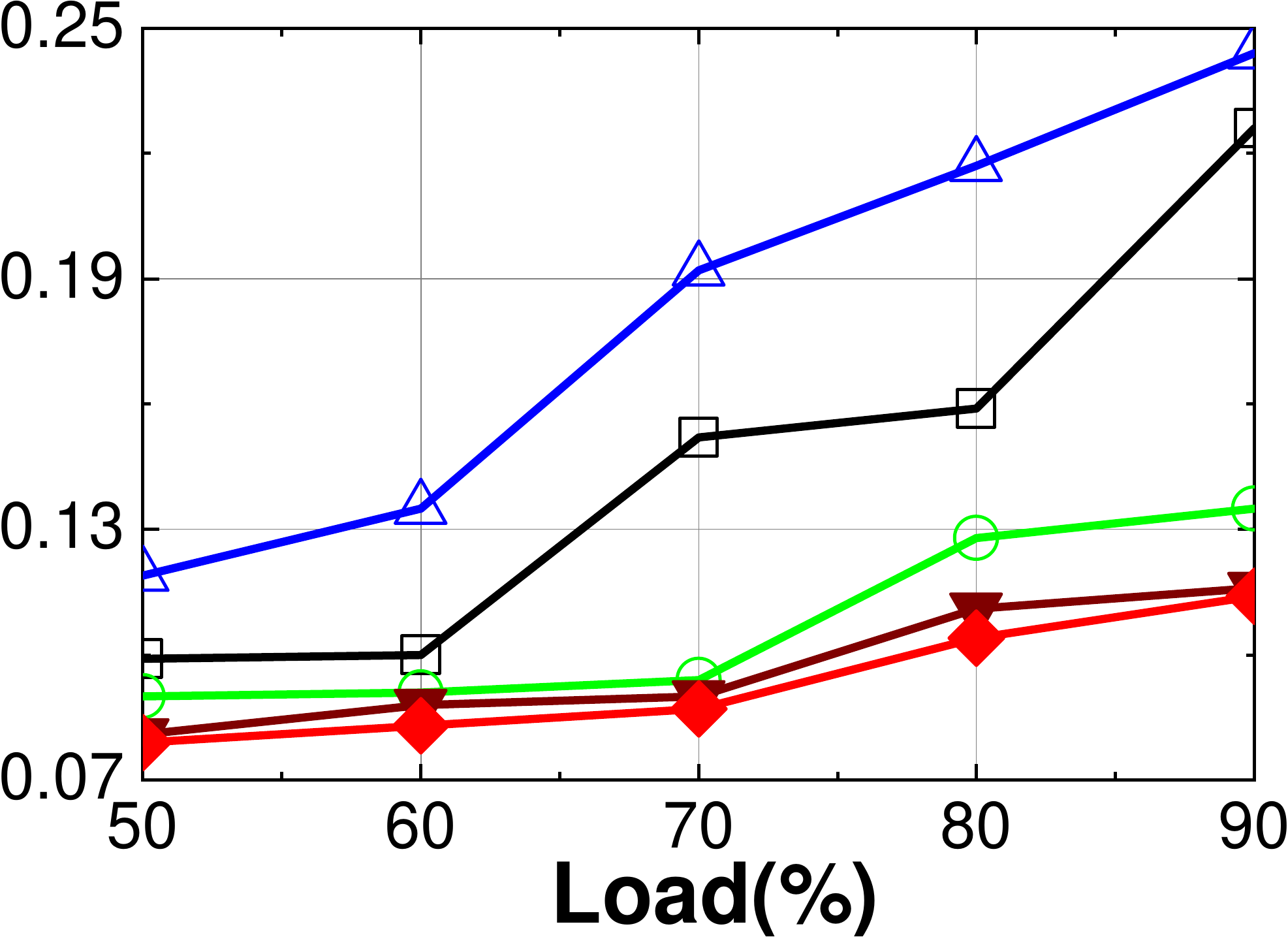}
\postsubfig
\label{fig:tb_end:W4:10K}
\end{minipage}
}
\subfigure[(10KB,$\infty$): Avg]{
\begin{minipage}[b]{0.183\textwidth}
\centering
\includegraphics[width=\textwidth]{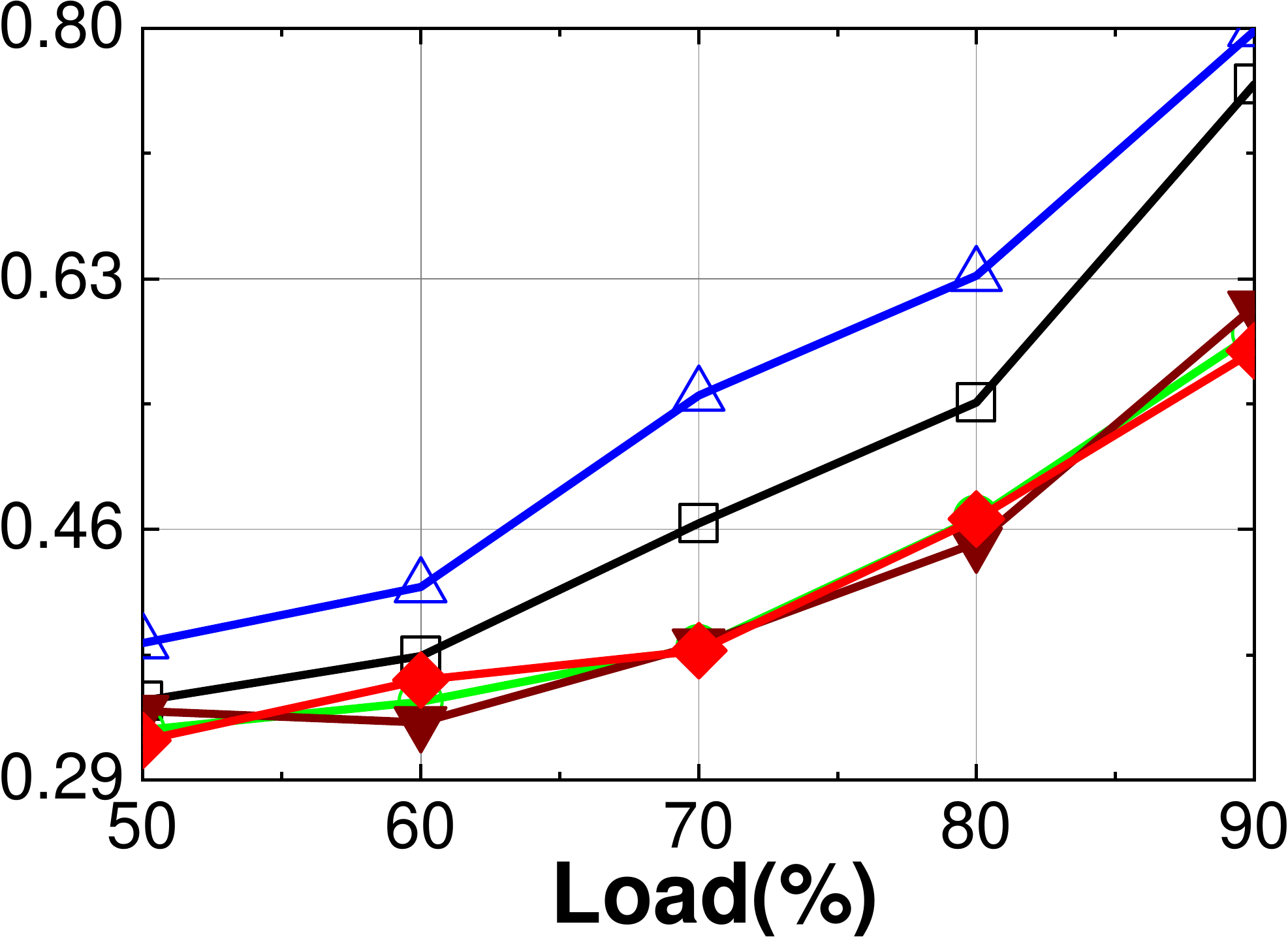}
\postsubfig
\label{fig:tb_end:W4:infty}
\end{minipage}
}
\postsubfig
\caption{FCT across different flow sizes on W4.}
\label{fig:tb_end_all_W4}
\end{figure*}

\bbb{Performance on W4
(Figure~\ref{fig:tb_end_all_W4}):}
The FCT of QC-FQ is about 55.3\% lower than that of PIAS-WST and about 51.5\% lower than that of DCTCP for small flows in (0, 1KB).
For the 99th percentile flow of small flows, the FCT of QC-FQ is about 21.6\% lower than that of PIAS-WST.
For middle flows in (1KB, 10KB) and large flows in (10KB, $\infty$), QC-FQ reduces the FCT by about 53.3\% and 27.2\% compared to PIAS-WST, respectively.

    \presub
\subsection{Experiments in ns-2} 
\postsub


\bbb{ns-2 settings:}
We use the leaf-spine topology, which consists of 4 spine switches and 9 leaf (ToR) switches. Each leaf switch is connected to 16 hosts via 10Gbps links, and connected to each spine switch via a 40Gbps link.   
The round-trip time between hosts under different leaf switches is 40.8$\mu$s.
We use the packet spraying \cite{spray} for load balancing and disable dupACKs.
QCluster is deployed in all switches.
For each output port, We use a SCM sketch with 83KB memory.

\subsubsection{Evaluation on SRPT and LAS} \postsub~

\begin{figure}[htbp]
\setlength\abovecaptionskip{-0.1cm}
\setlength\belowcaptionskip{-0.4cm}
\setlength\subfigcapskip{-0.5cm}
\centering

\subfigure[W4]{
\begin{minipage}[b]{0.190\textwidth}
\includegraphics[width=1.0\textwidth]{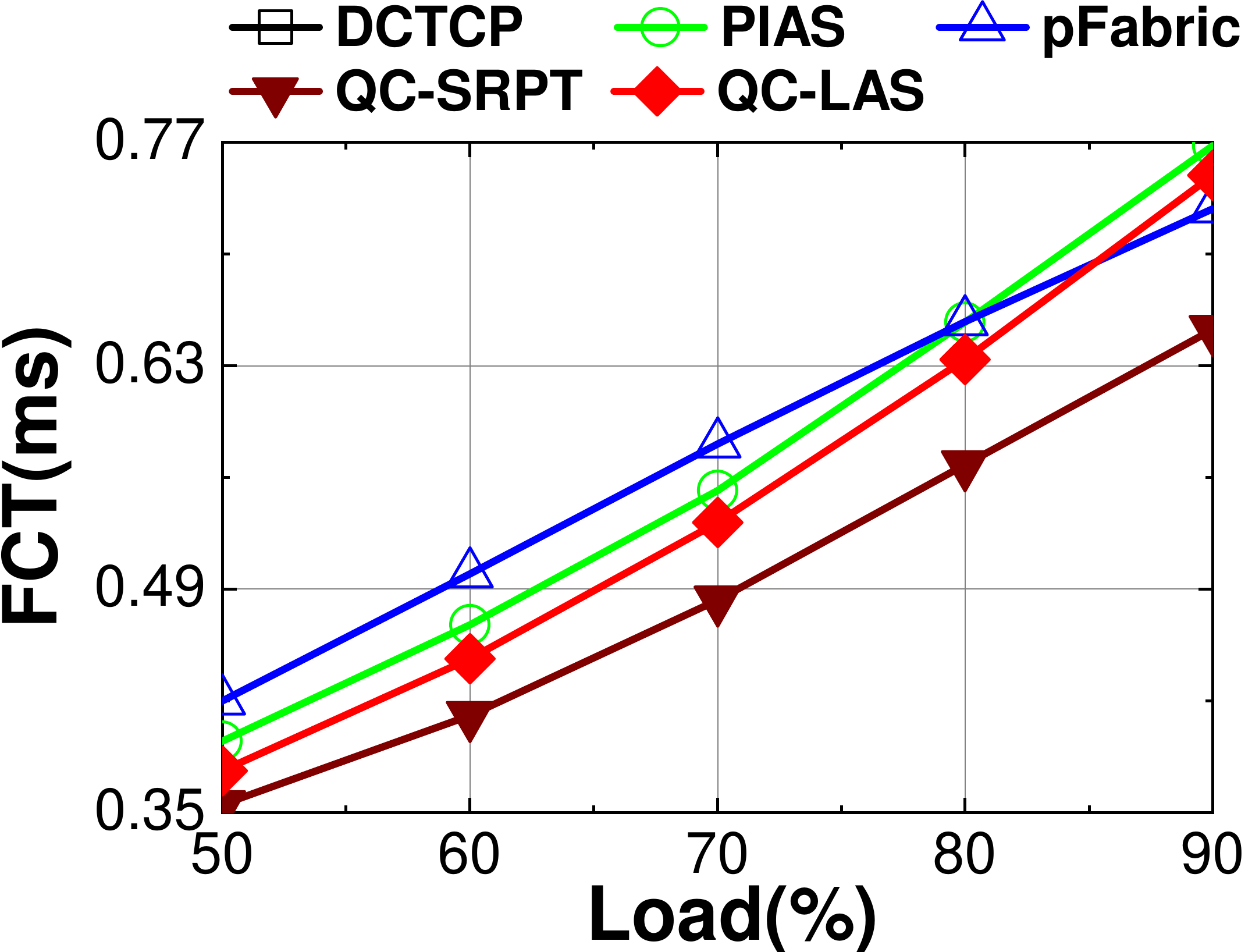}
\postsubfig
\label{fig:AVG:W4}
\end{minipage}
}
\subfigure[W6]{
\begin{minipage}[b]{0.177\textwidth}
\includegraphics[width=1.0\textwidth]{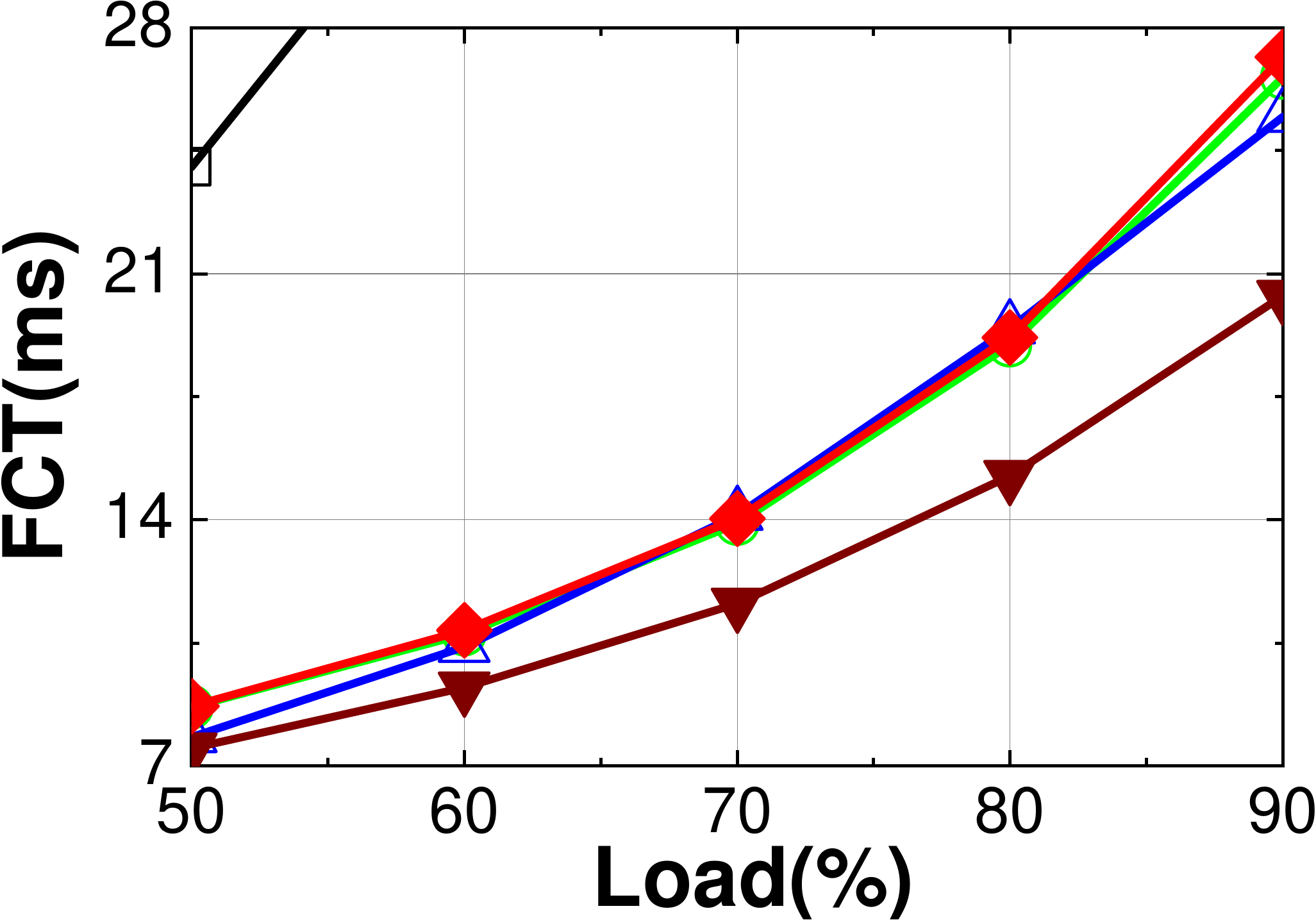}
\postsubfig
\label{fig:AVG:W6}
\end{minipage}
}
\postsubfig
\caption{Overall average FCT on different workloads for SRPT and LAS.}
\label{fig:AVG}
\end{figure}

We show the performances on W4 and W6 in Figure~\ref{fig:AVG}.

\bbb{Overall performance (Figure~\ref{fig:AVG}): }
Compared to pFabric, the average FCT of QC-SPRT is about 4.3\% lower. 
Especially on W4 and W6, the average FCT of QC-SPRT is about 11.3\% and 21.7\% lower.
Besides, for LAS, the average FCT of QC-LAS is about 4.73\% lower than the average FCT of PIAS.
For SRPT, with ECN for congestion control, QC-SPRT can perform better on most workloads.
Next, we show the detailed analysis on W4.

\begin{figure}[htbp]
\setlength\abovecaptionskip{-0.1cm}
\setlength\belowcaptionskip{-0.4cm}
\setlength\subfigcapskip{-0.5cm}
\subfigure[W4]{
\begin{minipage}[b]{0.19\textwidth}
\includegraphics[width=\textwidth]{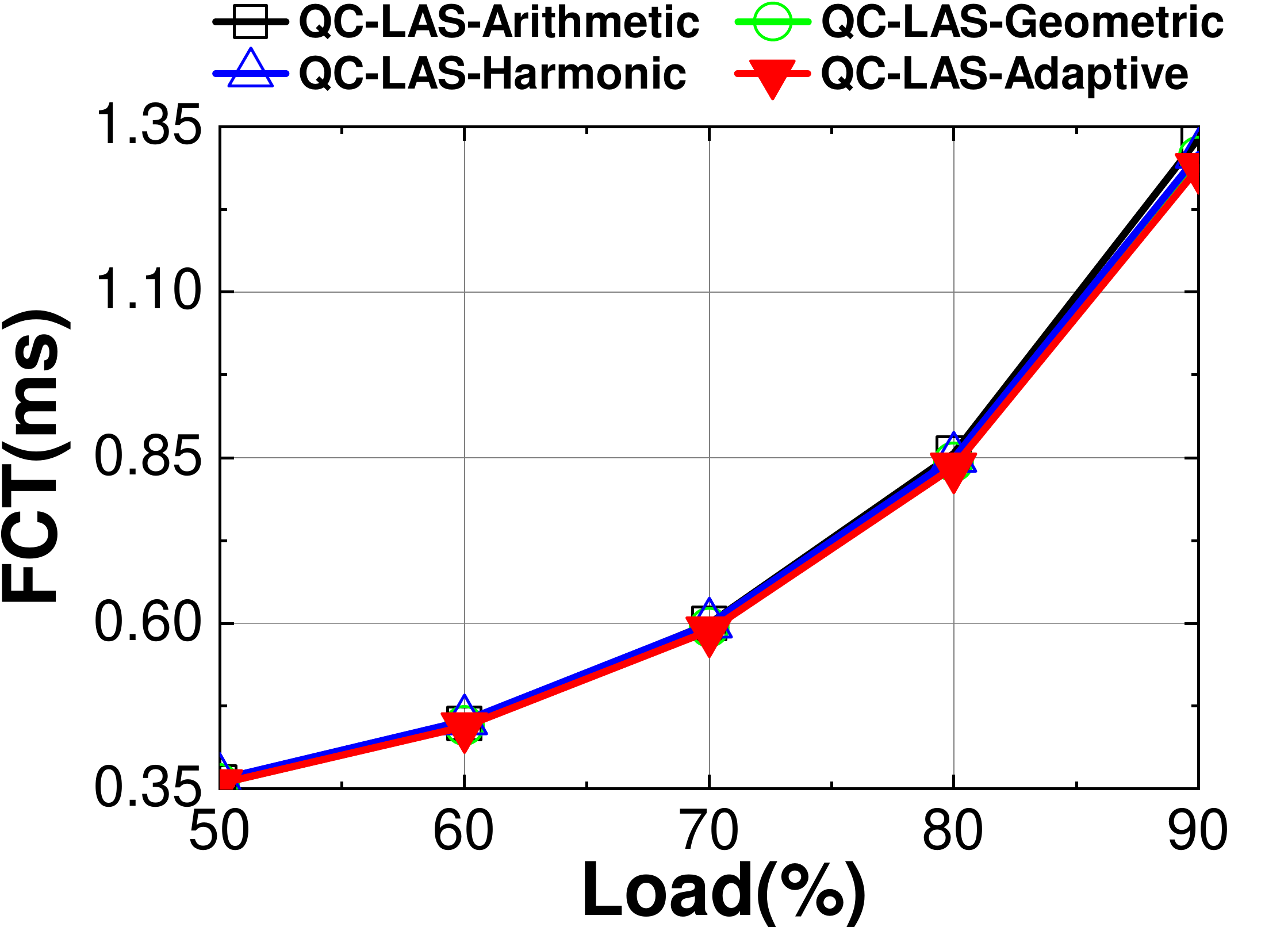}
\postsubfig
\label{fig:Mean:W4}
\end{minipage}
}
\subfigure[W6]{
\begin{minipage}[b]{0.171\textwidth}
\includegraphics[width=\textwidth]{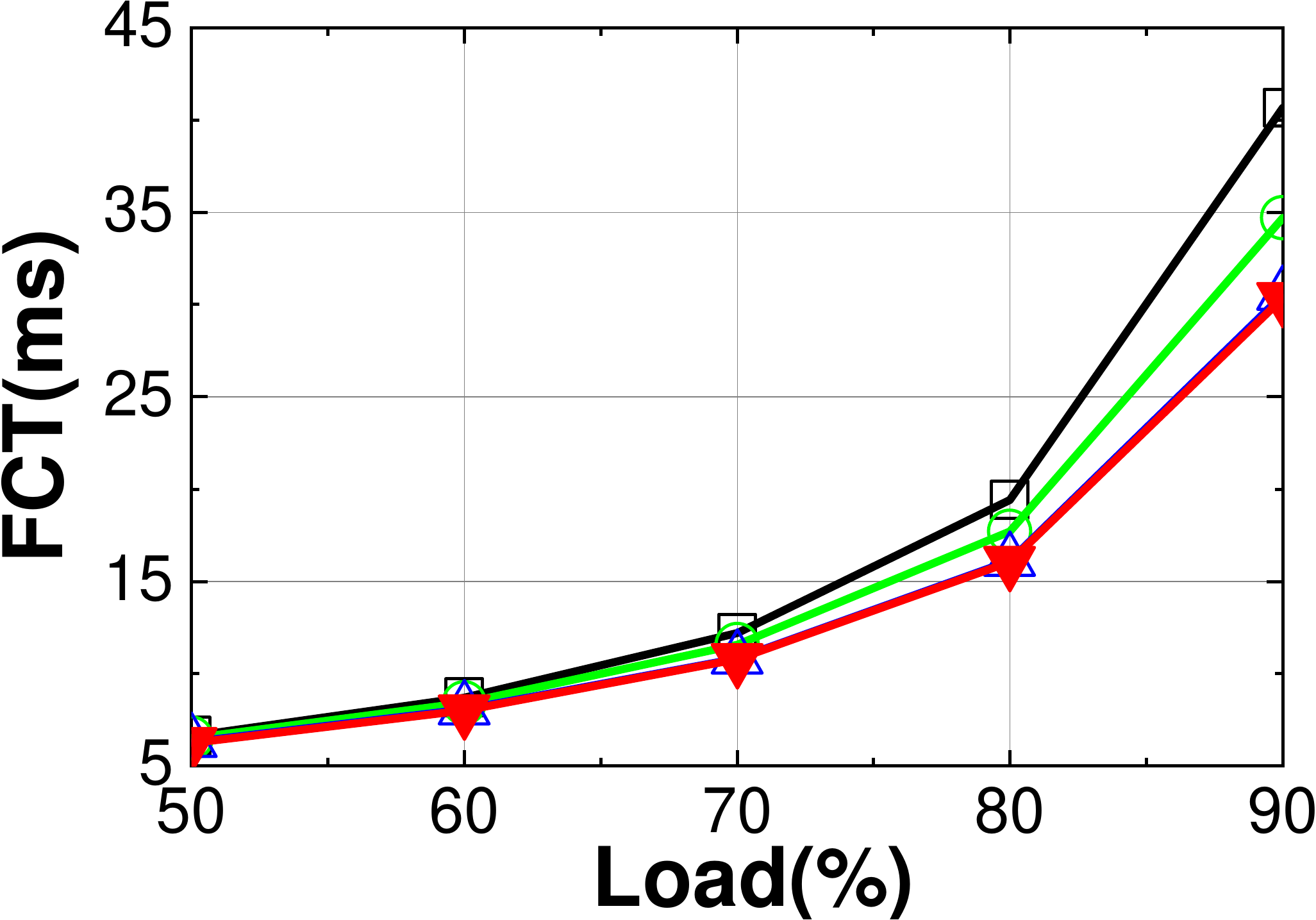}
\postsubfig
\label{fig:Mean:W6}
\end{minipage}
}
\postsubfig
\caption{FCT comparisons using different methods to achieve Proportional-Cluster-Size for LAS.}
\label{fig:Mean}
\end{figure}

\begin{figure}[htbp]
\setlength\abovecaptionskip{-0.1cm}
\setlength\belowcaptionskip{-0.4cm}
\setlength\subfigcapskip{-0.5cm}
\subfigure[Packet Ordered Ratio]{
\begin{minipage}[b]{0.185\textwidth}
\includegraphics[width=\textwidth]{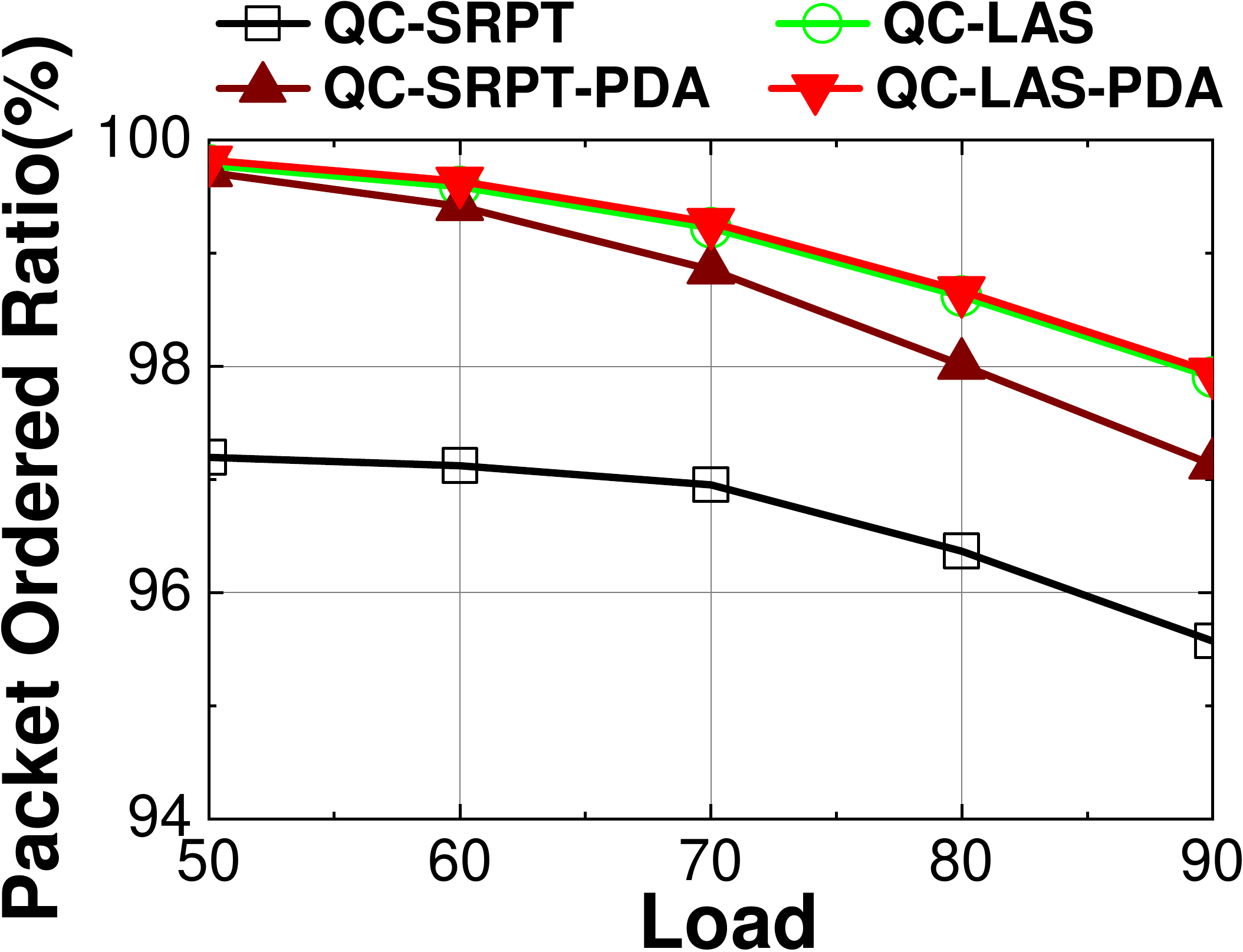}  
\postsubfig
\label{fig:PDA:order}
\end{minipage}
}
\subfigure[FCT]{
\begin{minipage}[b]{0.177\textwidth}
\includegraphics[width=\textwidth]{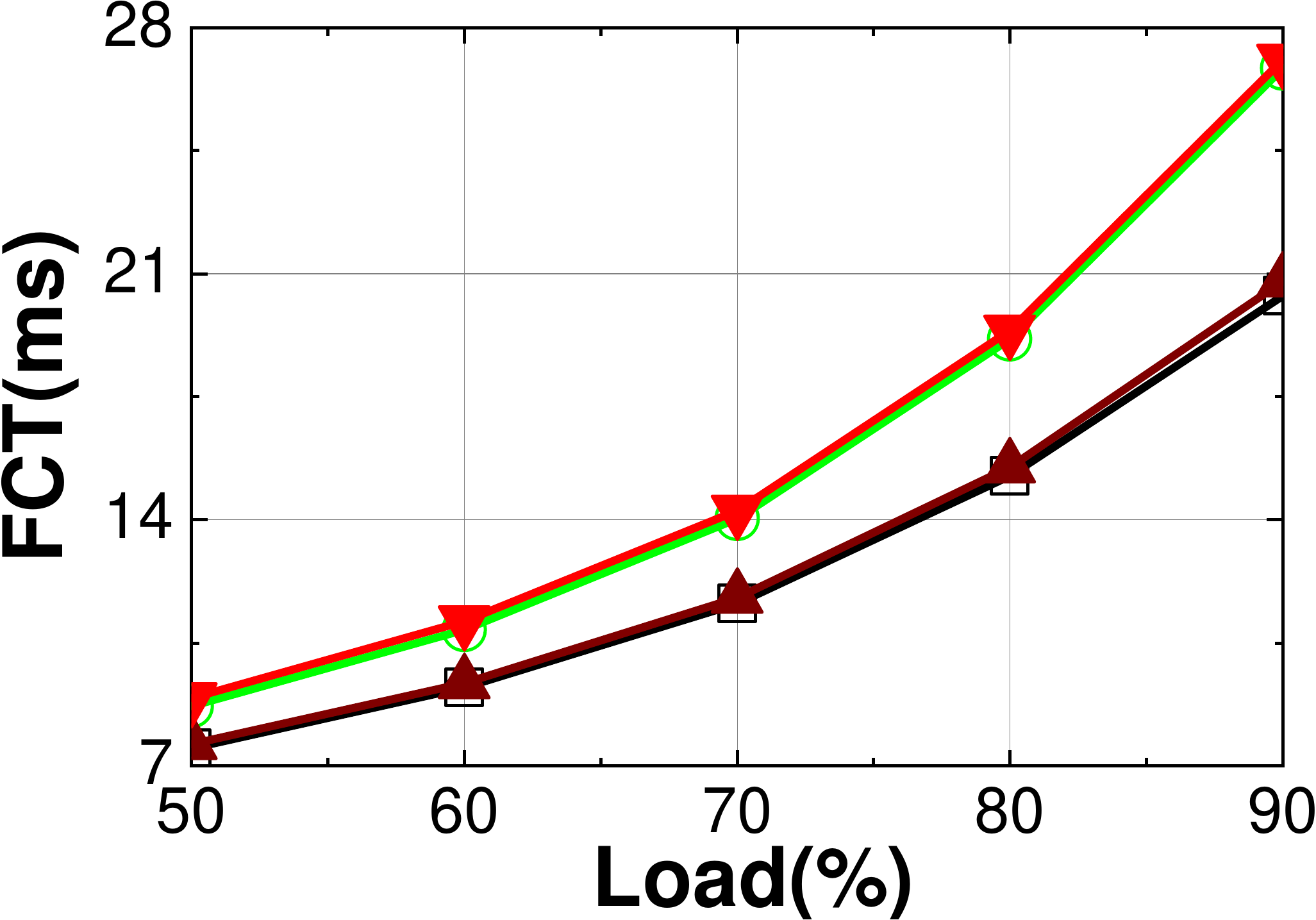} 
\postsubfig
\label{fig:PDA:FCT}
\end{minipage}
}
\postsubfig
\caption{Impact of Packet Disorder Avoidance on W6 for SRPT and LAS.}
\label{fig:PDA}
\end{figure}

\begin{figure*}[htbp]
\setlength\abovecaptionskip{-0.1cm}
\setlength\belowcaptionskip{-0.4cm}
\setlength\subfigcapskip{-0.5cm}
\centering
\subfigure[(0,1KB): Avg]{
\begin{minipage}[b]{0.212\textwidth}
\centering
\includegraphics[width=\textwidth]{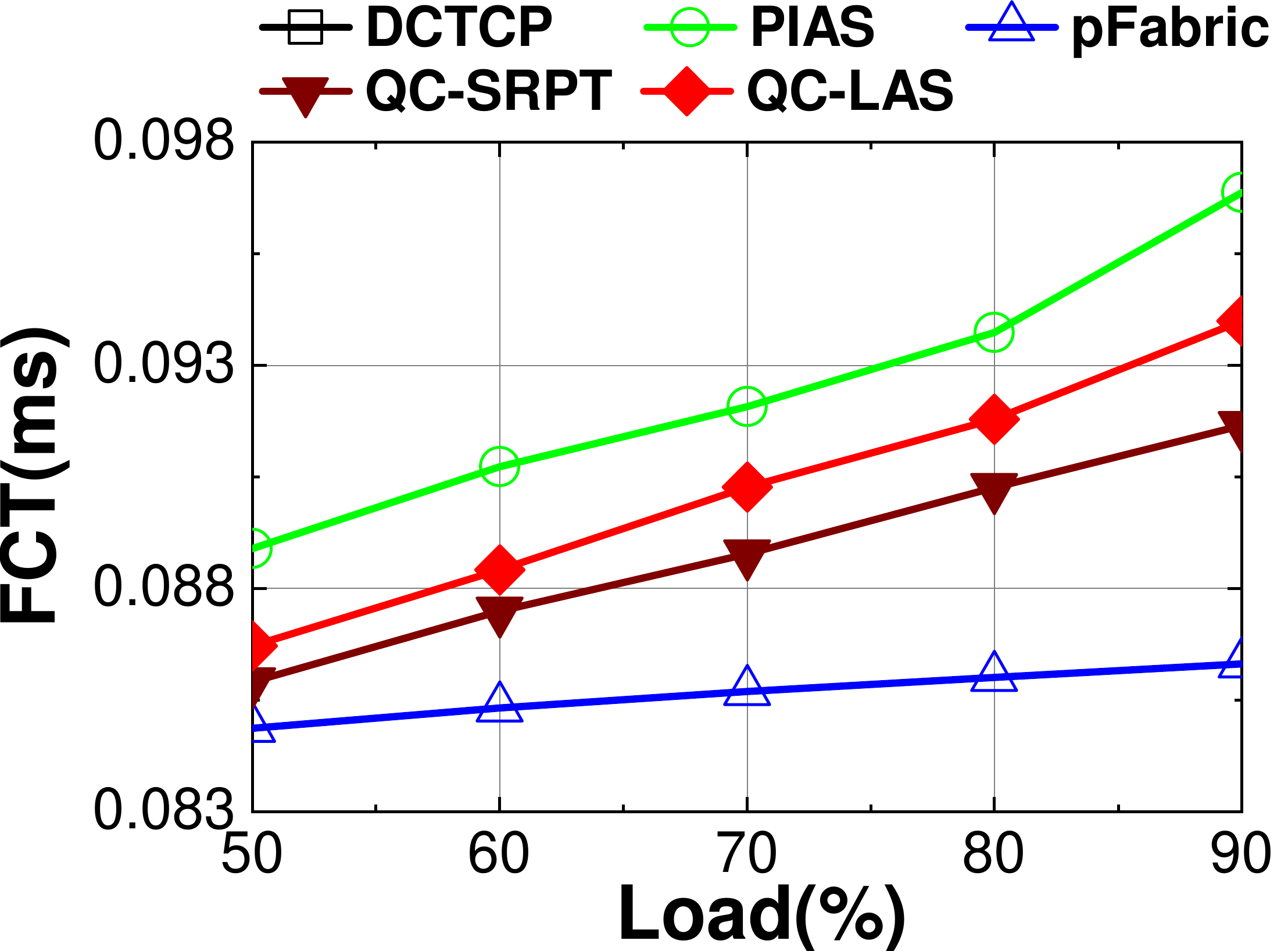}
\postsubfig
\label{fig:W4:1K}
\end{minipage}
}
\subfigure[(0,1KB): 99th Percentile]{
\begin{minipage}[b]{0.192\textwidth}
\centering
\includegraphics[width=\textwidth]{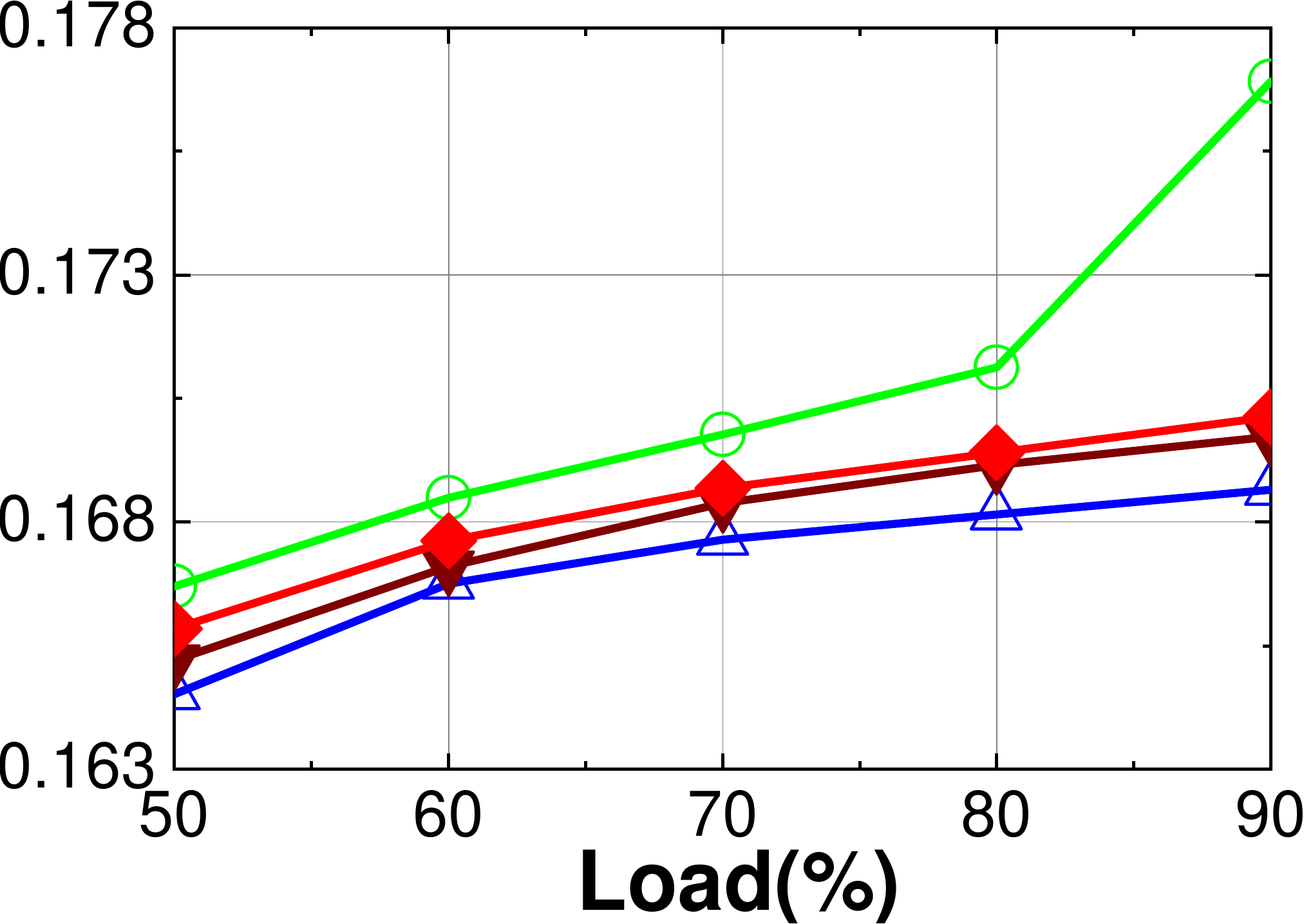}
\postsubfig
\label{fig:W4:99th}
\end{minipage}
}
\subfigure[(1KB,10KB): Avg]{
\begin{minipage}[b]{0.192\textwidth}
\centering
\includegraphics[width=\textwidth]{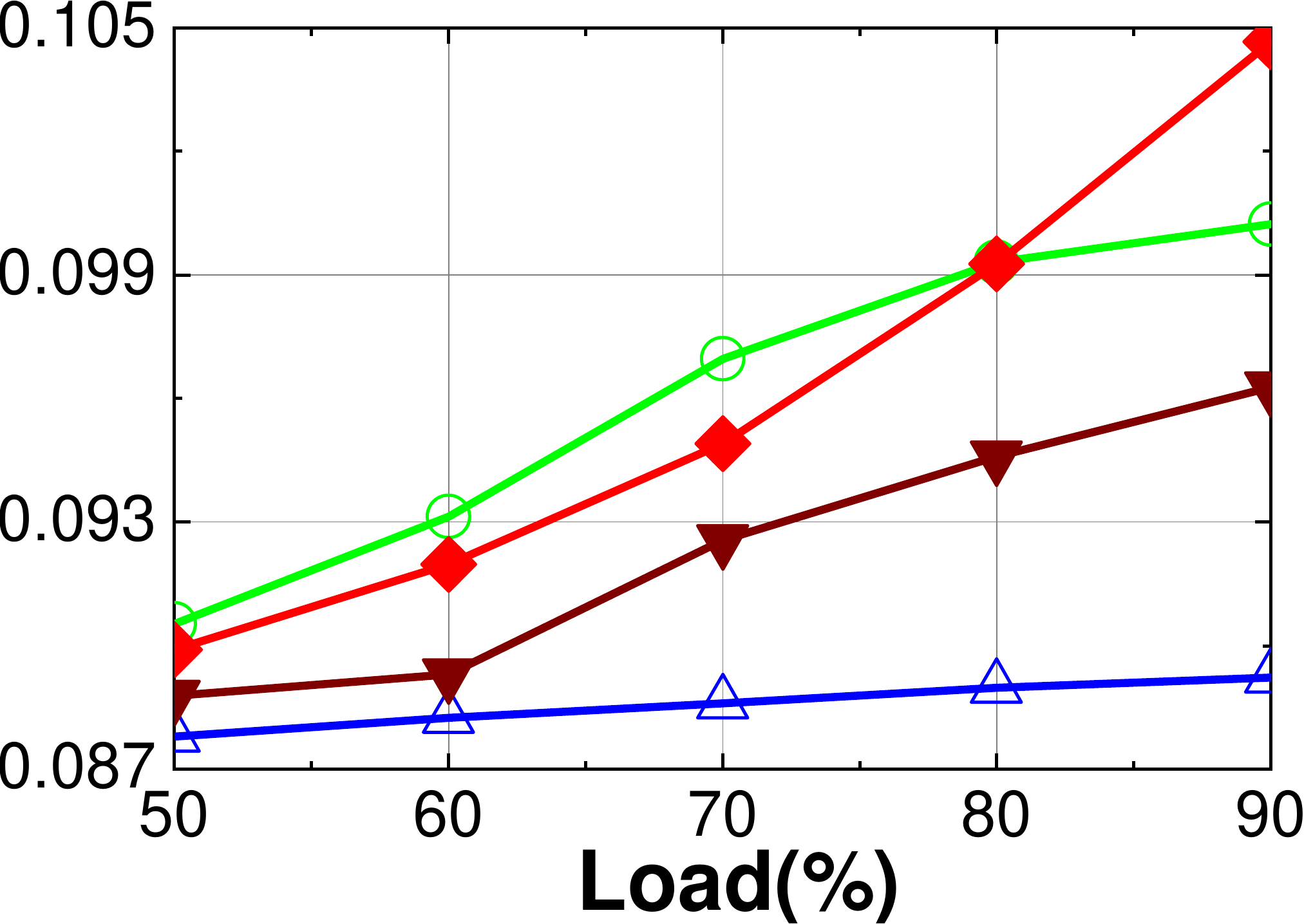}
\postsubfig
\label{fig:W4:10K}
\end{minipage}
}
\subfigure[(10KB,$\infty$): Avg]{
\begin{minipage}[b]{0.18\textwidth}
\centering
\includegraphics[width=\textwidth]{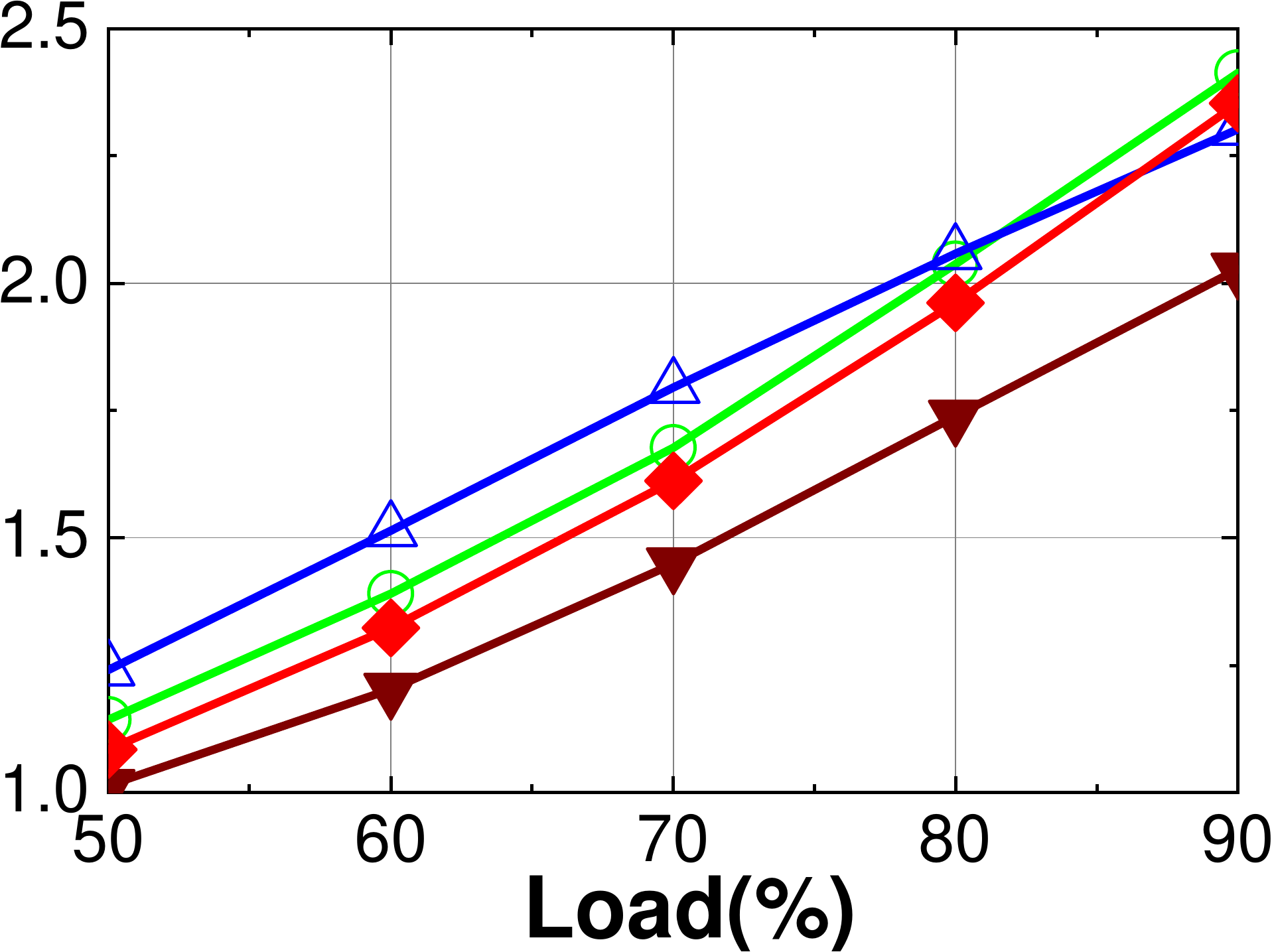}
\postsubfig
\label{fig:W4:infty}
\end{minipage}
}
\postsubfig
\caption{FCT across different flow sizes on W4 for SRPT and LAS.}
\label{fig:W4}
\end{figure*}
\begin{figure*}[htbp]
\setlength\abovecaptionskip{-0.1cm}
\setlength\belowcaptionskip{-0.4cm}
\setlength\subfigcapskip{-0.5cm}
\subfigure[(0KB, 10KB): Avg]{
\begin{minipage}[b]{0.208\textwidth}
\centering
\includegraphics[width=\textwidth]{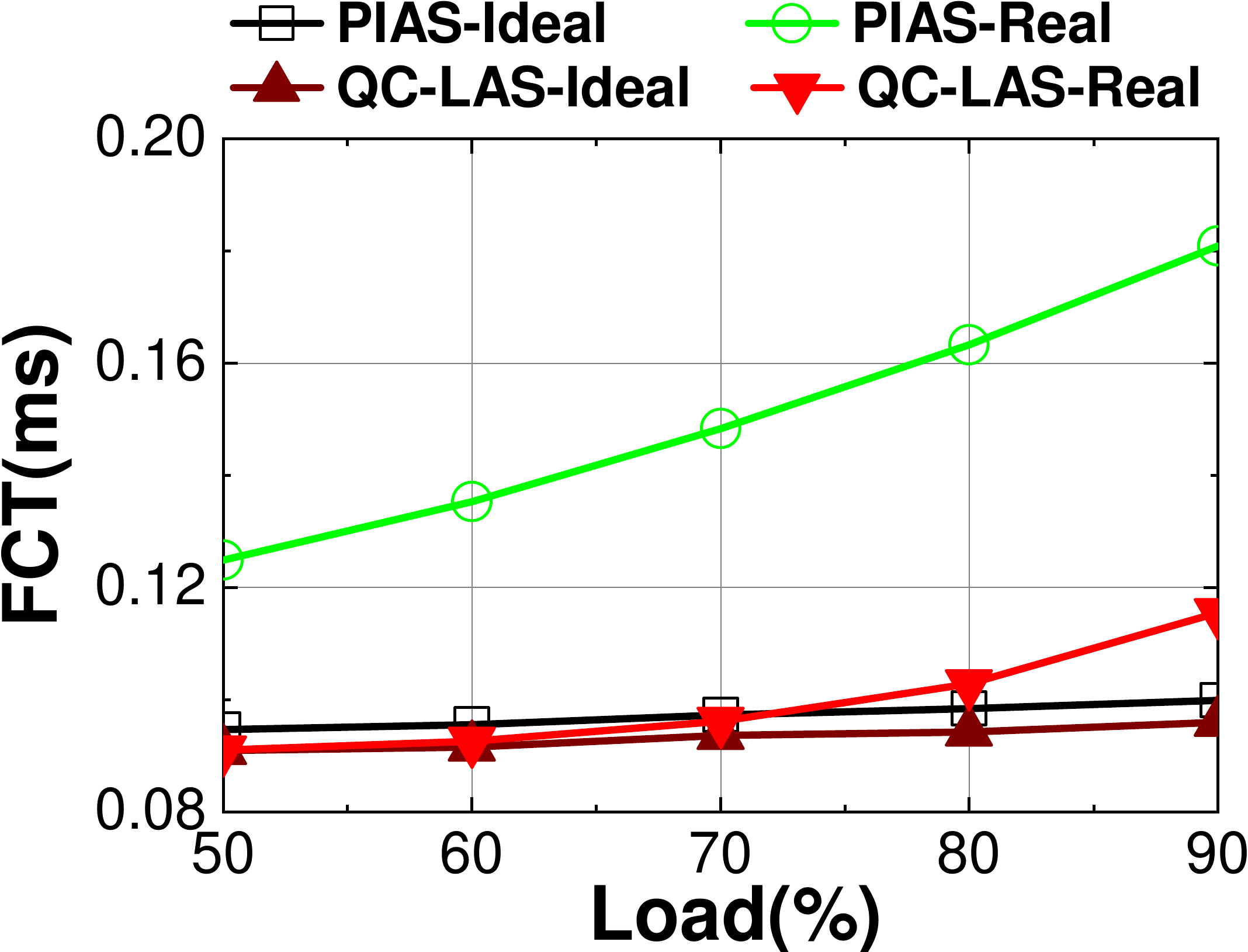}
\postsubfig
\label{fig:mess:10K}
\end{minipage}
}
\subfigure[(0, 10KB): 95th Percentile]{
\begin{minipage}[b]{0.19\textwidth}
\centering
\includegraphics[width=\textwidth]{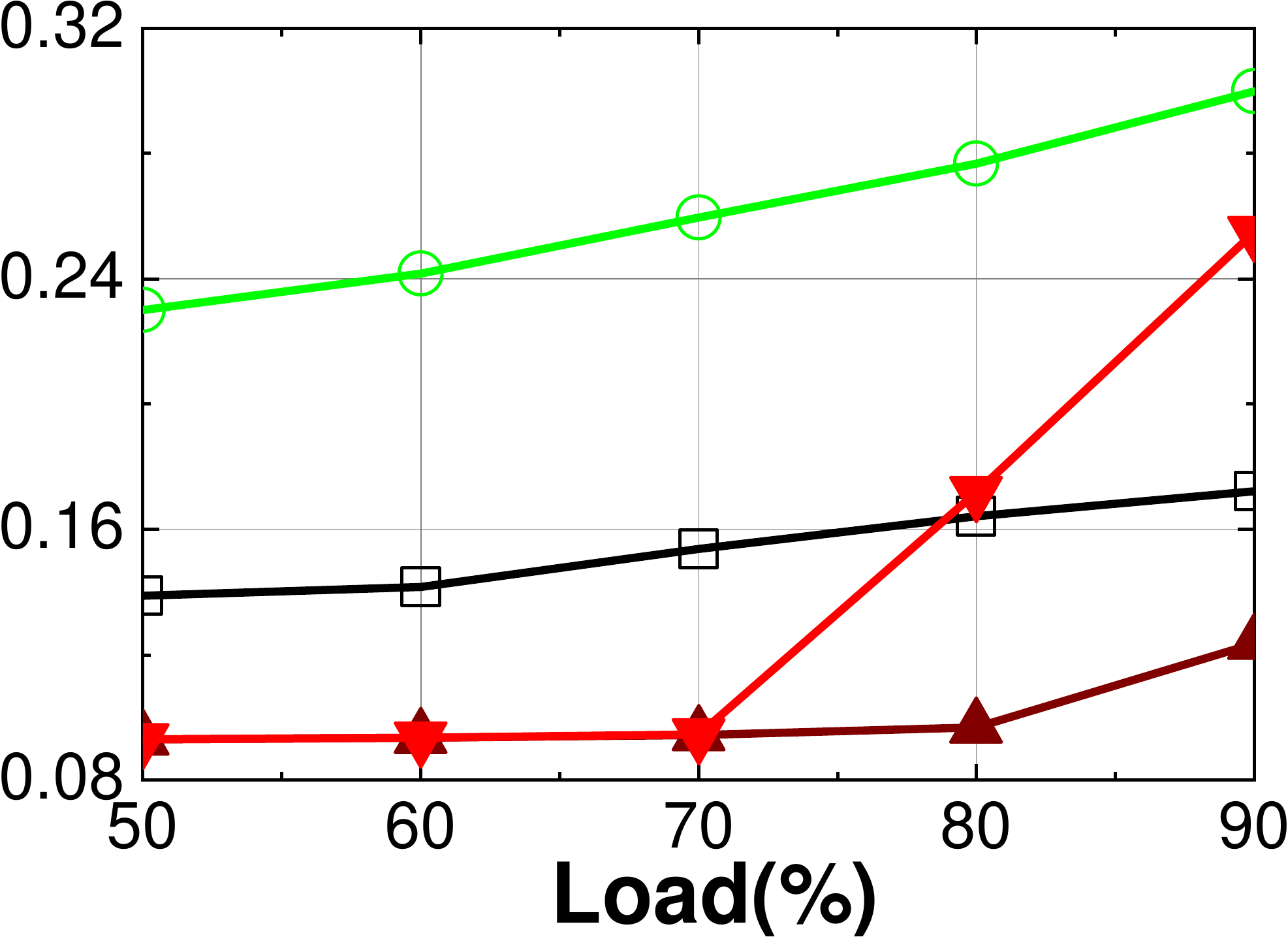}
\postsubfig
\label{fig:mess:95th}
\end{minipage}
}
\subfigure[(10KB,100KB): Avg]{
\begin{minipage}[b]{0.19\textwidth}
\centering
\includegraphics[width=\textwidth]{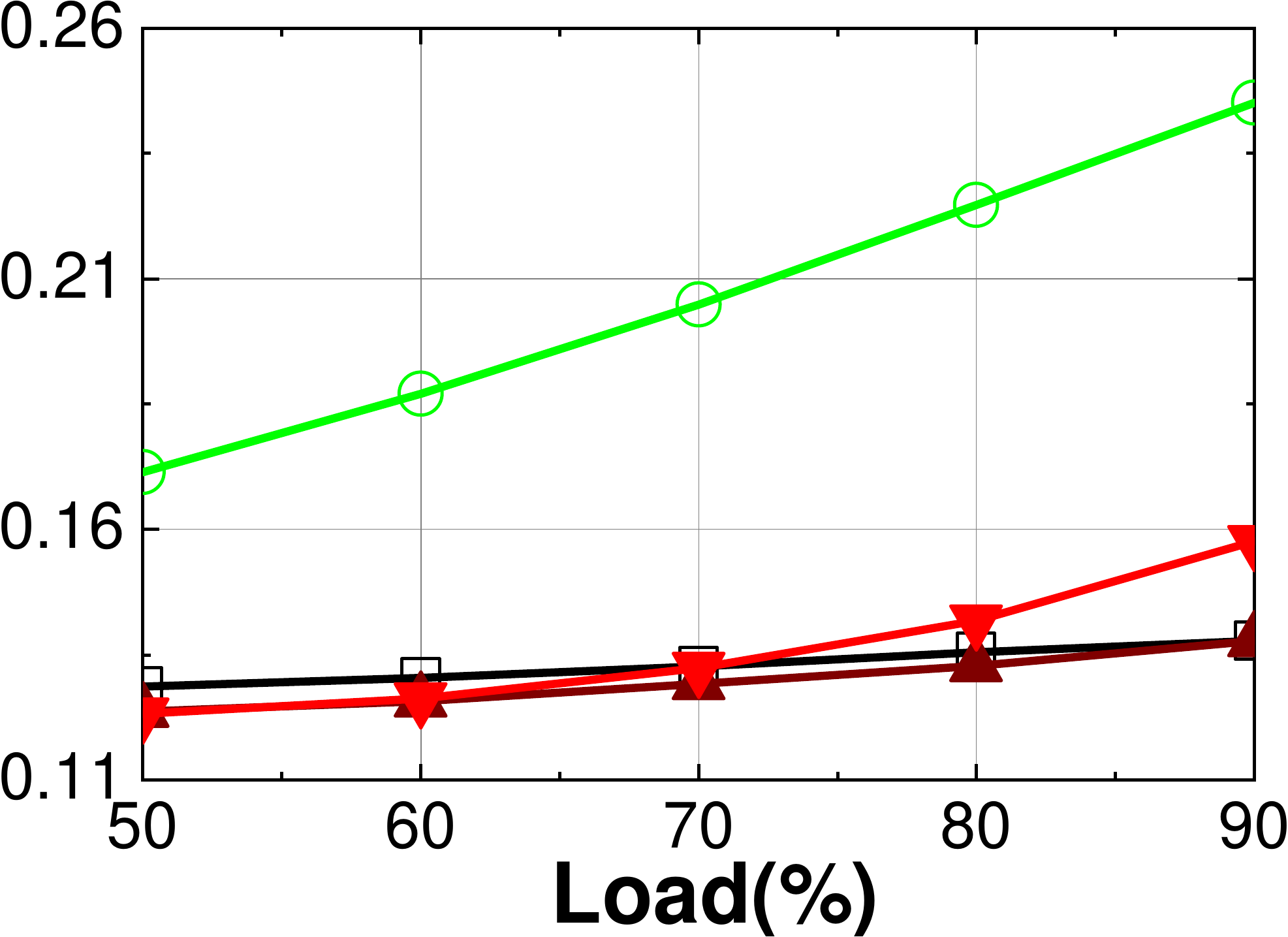}
\postsubfig
\label{fig:mess:100K}
\end{minipage}
}
\subfigure[(100KB,$\infty$): Avg]{
\begin{minipage}[b]{0.18\textwidth}
\centering
\includegraphics[width=\textwidth]{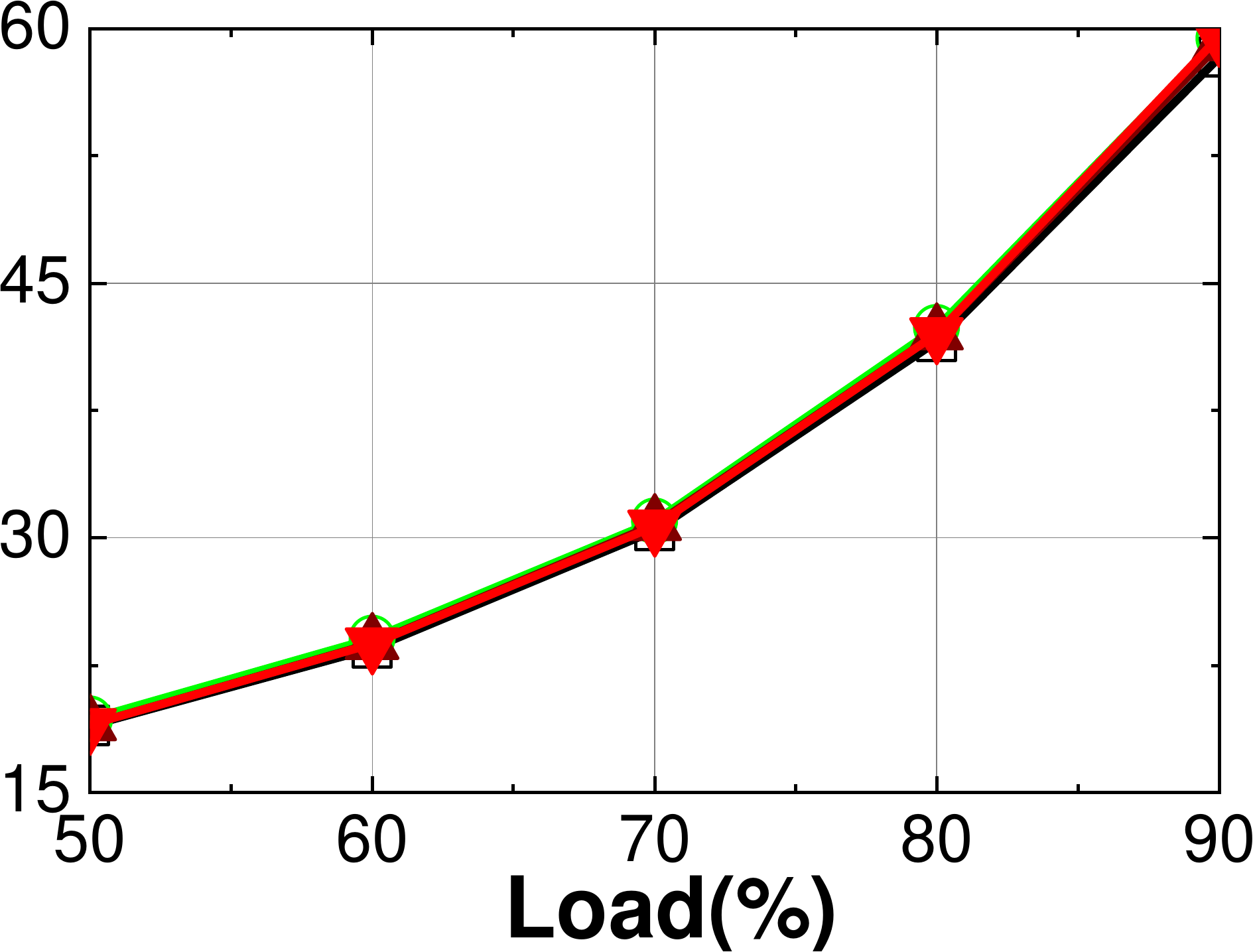}
\postsubfig
\label{fig:mess:infty}
\end{minipage}
}
\postsubfig
\caption{Impact of distinguishing the messages on W6 for LAS.}
\label{fig:mess}
\end{figure*}

\bbb{Performance on W4 (Figure~\ref{fig:W4}): }
Experimental results show that for small flows in (0,1KB), the average FCT of QC-LAS is about 3\% lower than that of PIAS.
As shown in Figure \ref{fig:AVG:W4}, the average FCT of QC-SRPT is about 11.3\% lower than the average FCT of pFabric in W4.
It is because that QC-SRPT decreases the FCT of large flows in (10KB, $\infty$) by about 20\%.
Because the FCT of DCTCP is beyond the range that pictures can represent, we do not show it.

\bbb{Impact of distinguishing the messages (Figure~\ref{fig:mess}):} 
%
%
In the above experiments, QC-LAS uses SCM sketch to distinguish the messages, while PIAS in ns-2 distinguishes the messages totally accurately, which cannot be achieved in practice. 
Therefore, we show how the methods of distinguishing messages influence the FCT. 
We use W6 to show the impact of distinguishing the messages.
%
We add the suffix ``-Ideal'' and ``-Real'' to the algorithms which can accurately distinguish messages and the algorithms which use time interval to distinguish messages, respectively.
As shown in Figure \ref{fig:mess}, the FCT of QC-LAS-Real is about 56.6\% lower than that of PIAS-Real for flows in (0, 10KB). 
Besides, for flows in (0, 10KB) at 90\% load, PIAS-Real increases the FCT by about 80\%, while QC-LAS-Real increases the FCT by about 20\%.
The method of distinguishing messages has a great influence on FCT.
%
Besides, the higher the load, the greater the impact of the method of distinguishing messages. 
%

\bbb{Different methods for Proportional-Cluster-Size: (Figure~\ref{fig:Mean}):}
To achieve Proportional-Cluster-Size, we evaluate the performance of QC-LAS using four methods: adaptive-threshold, geometric mean, harmonic mean, and arithmetic mean.
We observe that the average FCT using adaptive-threshold is about 12.5\%, 1.2\%, and 25.3\% lower than that using geometric mean, harmonic mean, and arithmetic mean, respectively.
Therefore, we recommend using adaptive-threshold.

\bbb{Impact of Packet Disorder Avoidance (PDA) (Figure~\ref{fig:PDA}):}
We observe that using PDA on QC-LAS can reduce the packet disorder by about 2.2\%, and the cost is to increase the average FCT by about 0.9\%.
This means that most of packet disorder in QC-LAS is caused by packet loss, which can not be avoided by PDA.
Using PDA on QC-SRPT can reduce the disorder by about 35.2\%, and the cost is to increase the average FCT by about 2\%.
This is because in SRPT, late-arriving packets have higher priority, which is more likely to cause packet disorder. This kind of disorder can be avoided by PDA.

\bbb{Summary:} 1) QC-LAS outperforms PIAS in general, and improve the average FCT of PIAS by about 4.73\%. 
Taking the impact of distinguishing the messages into consideration, QC-LAS significantly improves the FCT for short flows by about 56.6\%.
2) QC-SPRT performs better in most workloads. The average FCT of QC-SPRT is about 4.3\% lower than that of pFabric.

\presub
\subsubsection{Evaluation on Fair Queueing} 
\postsub~

Because QC-FQ uses the number of packets as a unit, we use W6 in the evaluation of Fair Queueing.

\begin{figure}[htbp]
\setlength\abovecaptionskip{-0.0cm}
\setlength\belowcaptionskip{-0.6cm}
\centering
\includegraphics[width=0.5\linewidth]{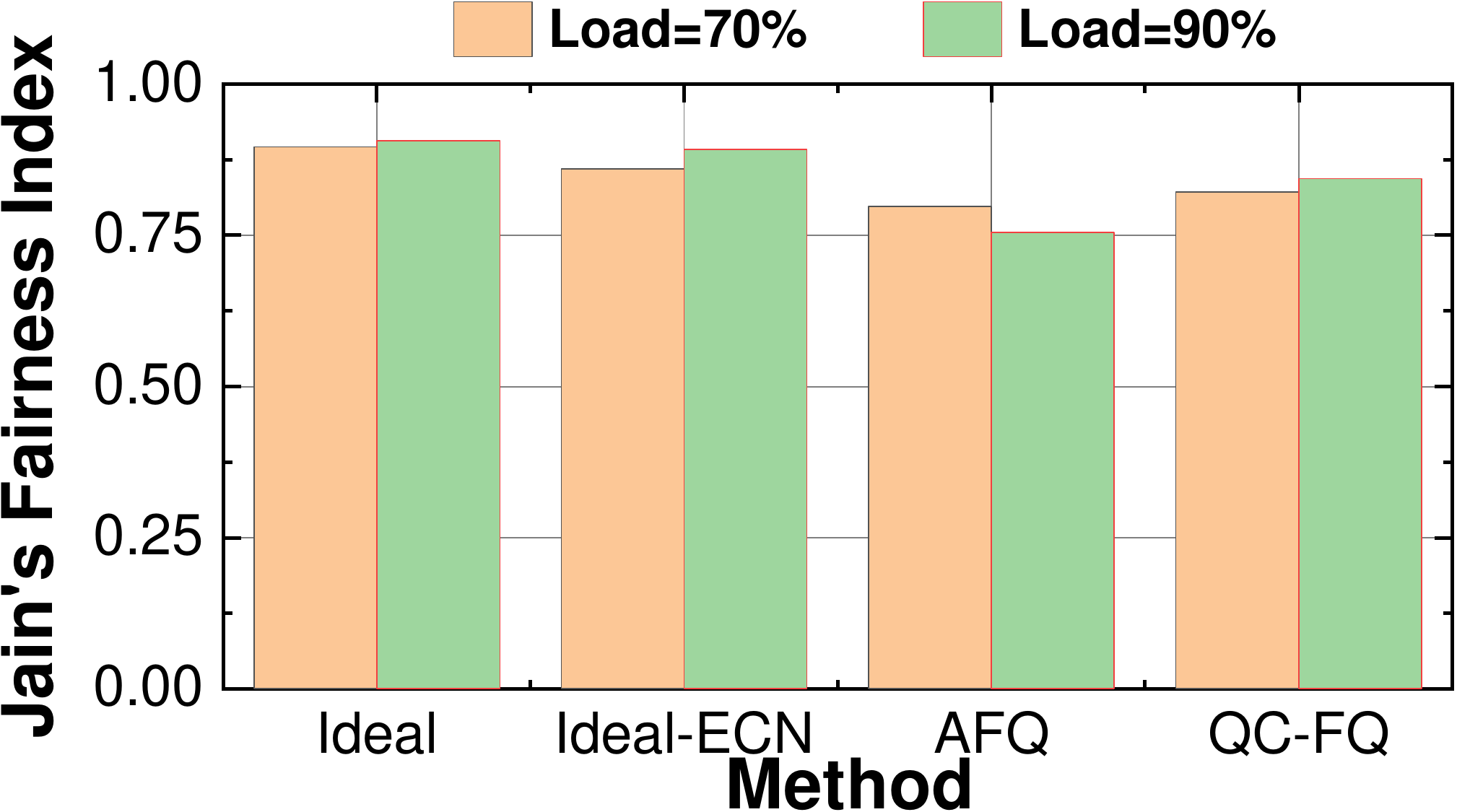}
\vvv \vvv
\caption{Jain's Fairness Index of different Fair Queueing Algorithms on W6.
}
\label{fig:JF}
\end{figure}

\bbb{Jain's Fairness Index (Figure~\ref{fig:JF}):}
As shown in Figure \ref{fig:JF}, The Jain's Fairness Index of QC-FQ is about 5.5\% lower than that of ideal algorithm with ECN and about 11.7\% higher than that of AFQ.
This result shows that QC-FQ can \textbf{halve the gap with the optimal value} compared to the state-of-the-art.

\begin{figure}[htbp]
\setlength\abovecaptionskip{-0.1cm}
\setlength\belowcaptionskip{-0.5cm}
\setlength\subfigcapskip{-0.5cm}
\subfigure[FairQueue: Load 70\%]{
\begin{minipage}[b]{0.19\textwidth}
\includegraphics[width=\textwidth]{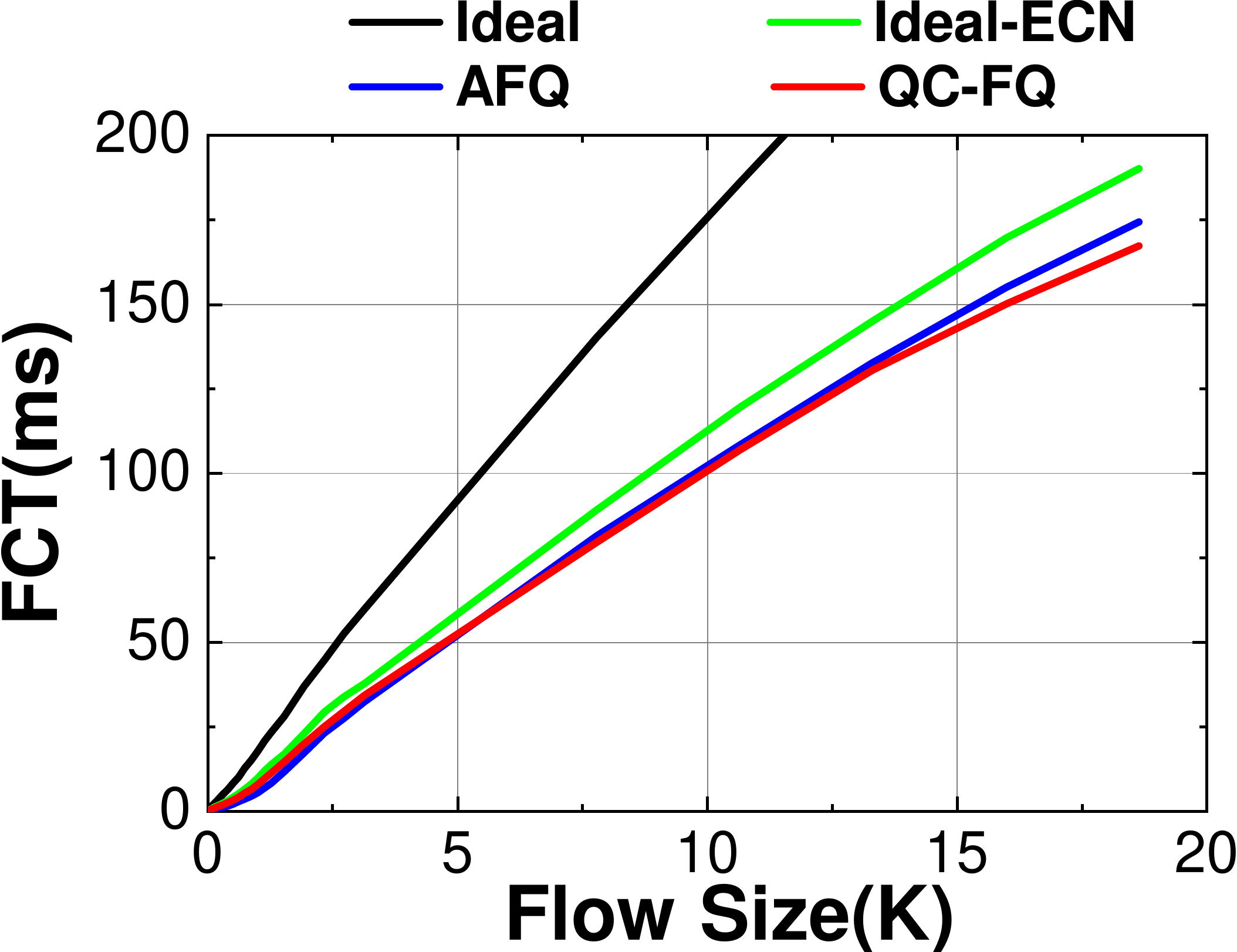}
\postsubfig
\label{fig:FQ1:70}
\end{minipage}
}
\subfigure[FairQueue: Load 90\%]{
\begin{minipage}[b]{0.19\textwidth}
\includegraphics[width=\textwidth]{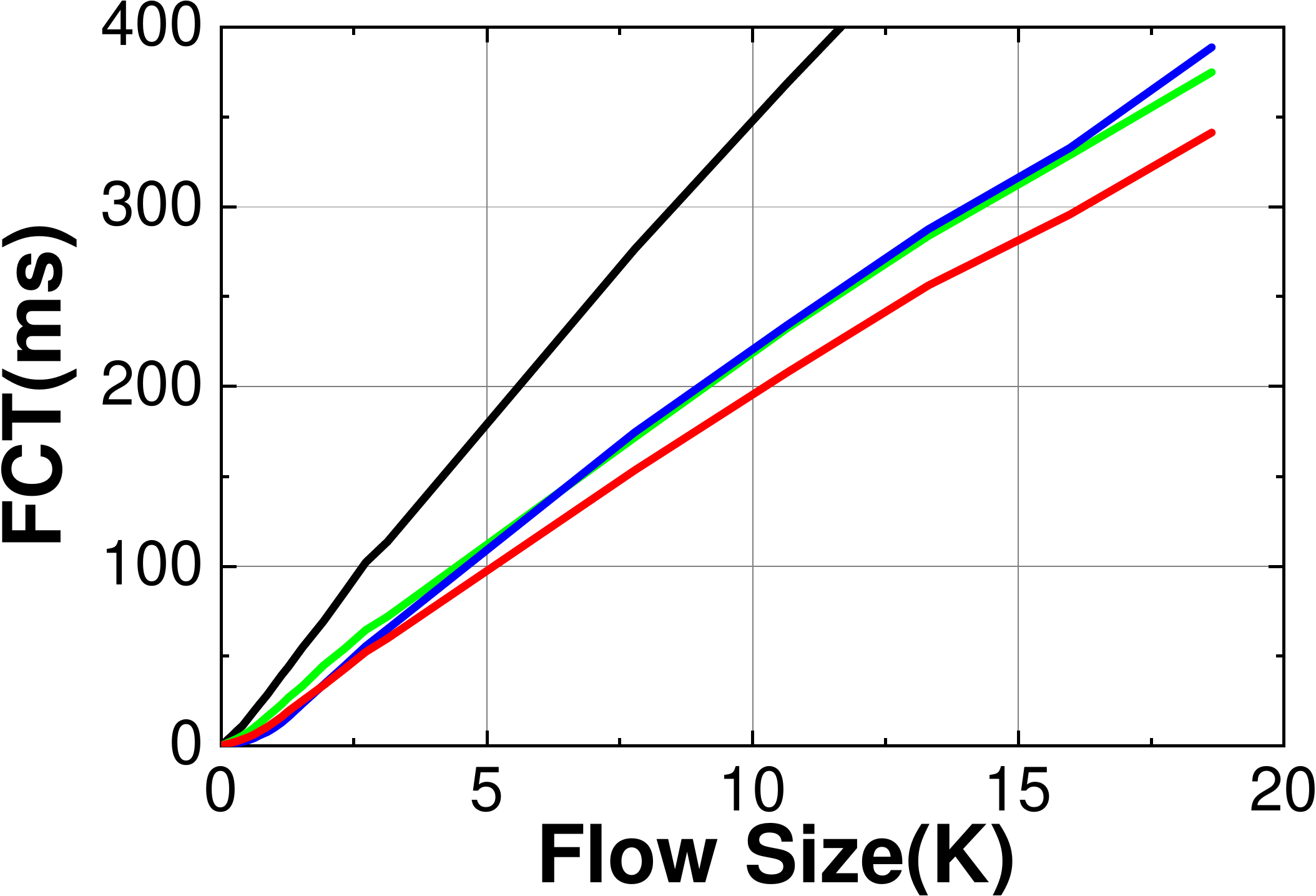}
\postsubfig
\label{fig:FQ1:90}
\end{minipage}
}
\postsubfig
\caption{Average FCT comparisons under different flow size on W6 for Fair Queueing.}
\label{fig:FQ1}
\end{figure}

\bbb{FCT (Figure~\ref{fig:FQ1}):}
The average FCT of QC-FQ is about 13.2\% lower than that of ideal fair queueing with ECN, and about 8.4\% lower than that of AFQ.
We measure the FCT with the Average FCT - Flow Size diagram.
An ideal fair queueing should be a direct proportion function. As we can see from Figure \ref{fig:FQ1},  though the performance of our algorithm is not exactly a straight line compared to that of ideal fair queueing, it nearly satisfies the Fair Queueing requirement. 
Meanwhile, it improves the average FCT.

\bbb{Summary:} 1) Though QC-FQ does not achieve as high fairness as the ideal fair queueing, the overall FCT of QC-FQ is about 13.2\% lower. 
2) Compared with AFQ, QC-FQ achieves both higher fairness and lower latency. 
The Jain's Fairness Index of QC-FQ is about 11.7\% higher and the overall FCT of QC-FQ is about 8.4\% lower.


\subsubsection{Evaluation on Deadline-Aware Scheduling} \postsub~


In this experiment, we only assign deadlines for flows that are smaller than 100KB in W4. 
The deadlines are assumed to be exponentially distributed similar to prior work \cite{pfabric}.

\begin{figure}[htbp]
\setlength\abovecaptionskip{-0.0cm}
\setlength\belowcaptionskip{-0.4cm}
\setlength\subfigcapskip{-0.5cm}
\subfigure[Application Throughput]{
\begin{minipage}[b]{0.19\textwidth}
\includegraphics[width=\textwidth]{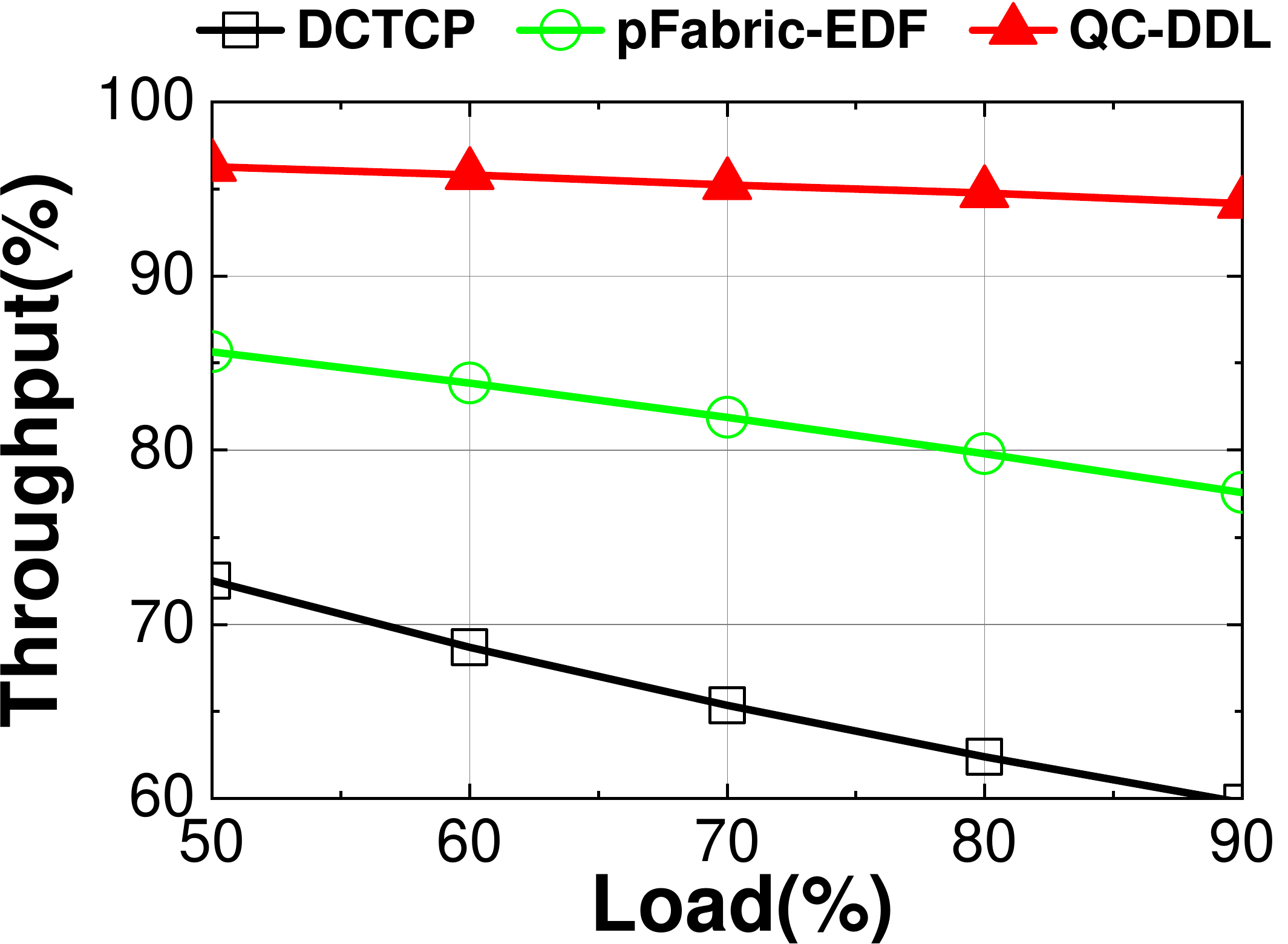}
\postsubfig
\label{fig:DDL:thp}
\end{minipage}
}
\subfigure[FCT for non-deadline flows]{
\begin{minipage}[b]{0.175\textwidth}
\includegraphics[width=\textwidth]{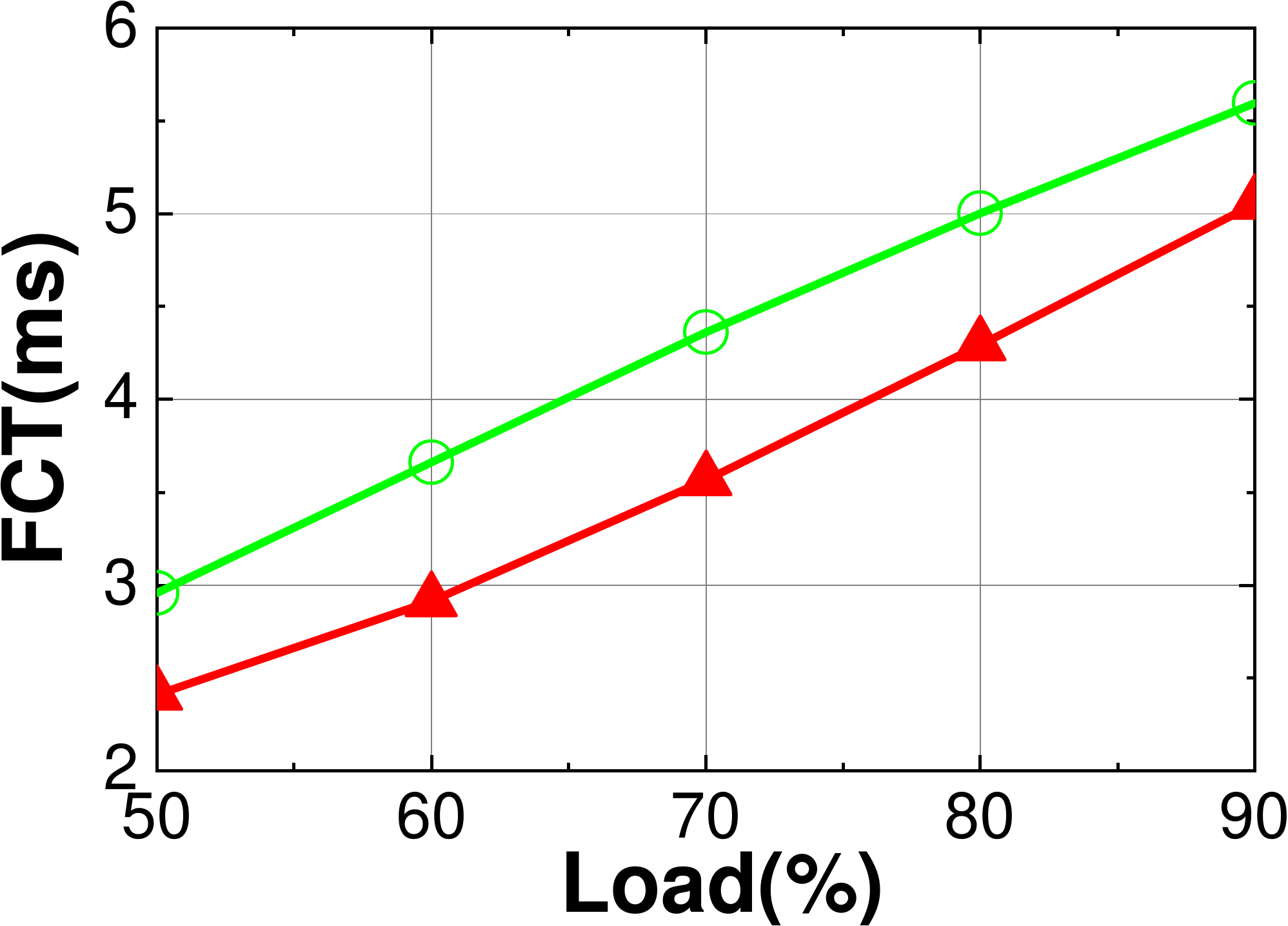}
\postsubfig
\label{fig:DDL:fct}
\end{minipage}
}
\vspace{-0.05in}
\caption{Deadline-Aware Scheduling on W4.}
\label{fig:DDL}
\end{figure}

\bbb{Application Throughput (Figure~\ref{fig:DDL:thp}):}
Compared to pFabric-EDF, QC-DDL can increase the application throughput by about 15\%. 
And the application throughput of QC-DDL is about 29\% higher than that of DCTCP.
Because when switches begin to send the large deadline flow, pFabric-EDF could be too late to catch up with its deadline. 
Besides, when the deadline is a large number, deadline flows may be blocked by non-deadline flows regardless of the remaining time to deadline in pFabric-EDF.

\bbb{FCT (Figure~\ref{fig:DDL:fct}):}
the FCT of non-deadline flows in QC-DDL is about 15\% lower than that of pFabric-EDF.
Because pFabric-EDF does not take the size of flows into consideration, it will send the deadline flows aggressively, which will hurt the FCT of non-deadline flows.
The FCT of DCTCP is beyond the range that pictures can represent, so we do not show it in this figure.

\bbb{Summary:}
1) The application throughput of QC-DDL is about 29\% higher than that of DCTCP and about 15\% higher than that of DCTCP.
2) The FCT of non-deadline flows of  QC-DDL is about 15\% lower than that of pFabric-EDF.

\section{Conclusion} 
\label{sec:conclusion}

This paper proposes a framework, QCluster, to adapt existing flow scheduling solutions (SRPT, LAS, Fair Queueing, and Deadline-Aware Scheduling) for limited number of queues without measuring traffic in advance.
The key idea of QCluster is to cluster packets with similar weights/properties into the same queue.
We also propose the PDA algorithm to avoid packet disorder incurred by scheduling.  
We apply QCluster to four typical scheduling problems, and also show how to apply QCluster to other scheduling problems.
We implement QCluster with PDA in Tofino switches, achieving a clustering speed of 3.2 Tbps.
We also implement QCluster in large-scale ns-2 simulations for four kinds of scheduling problems.
Experimental results in testbed and ns-2 show that QCluster achieves better or comparable performance compared to the state-of-the-art algorithms for four typical scheduling policies.
All the source codes in ns-2 are available in Github without identity information \cite{opensource}.

	\clearpage	
	\balance
	\bibliographystyle{ACM-Reference-Format}
	\bibliography{reference}
	
	\clearpage
	\nobalance
	\appendixpage
\appendix

\section{Packet Disorder Avoidance}
\label{appendix:pda}

\noindent For an incoming packet $a_{now}$, we use the SCM sketch to get the queue where the previous packet stays.
The update and query of queue ID are different between LAS and Fair Queueing.

\bbb{Update of queue ID:}
We check the timestamp of each mapped bucket in the SCM sketch.
Suppose that the time now is $t_{now}$, and the chosen queue for $a_{now}$ is the $i^{th}$ queue.
If the timestamp $t_{bucket}$ is smaller/older than $t_{now}-\Delta T_{Flowlet}$, we directly set the queue ID of this bucket to $i$.
Otherwise, for LAS, we update the queue ID to $i$ only if the $i^{th}$ queue has lower-priority than the queue recorded in the bucket, and for fair queueing, we do not update the queue ID.

\bbb{Query of queue ID:}
We only query the queue ID if $a_{now}$ is not the first packet of a new Flowlet.
For LAS, we choose the queue with the highest priority in all mapped buckets.
For Fair Queueing, we choose the queue with the smallest/oldest timestamp in all mapped buckets.
Both methods can minimize the minus effect of hash collisions.

The hash collisions can still cause the SCM sketch giving the wrong queue IDs.
For LAS, this mistake will slightly downgrade the performance, but it will not incur packet disorder, because the SCM sketch only gives the overestimation on queue IDs. In other words, the SCM sketch may give a lower-priority queue than the queue of the previous packet, and it will never give a higher-priority queue, so the incoming packet will not go to the higher-priority queue and packet disorder will not happen. For Fair Queueing, all queues have the same priority but different weights, so there are still chances that packet disorder happens.


\section{Other Experiment results in testbed}
\label{app:testbed}

\begin{figure}[h]
\setlength\abovecaptionskip{-0.0cm}
\setlength\belowcaptionskip{-0.0cm}
\setlength\subfigcapskip{-0.2cm}
\centering
\subfigure[(0,10KB): Avg]{
\begin{minipage}[b]{0.23\textwidth}
\centering
\includegraphics[width=\textwidth]{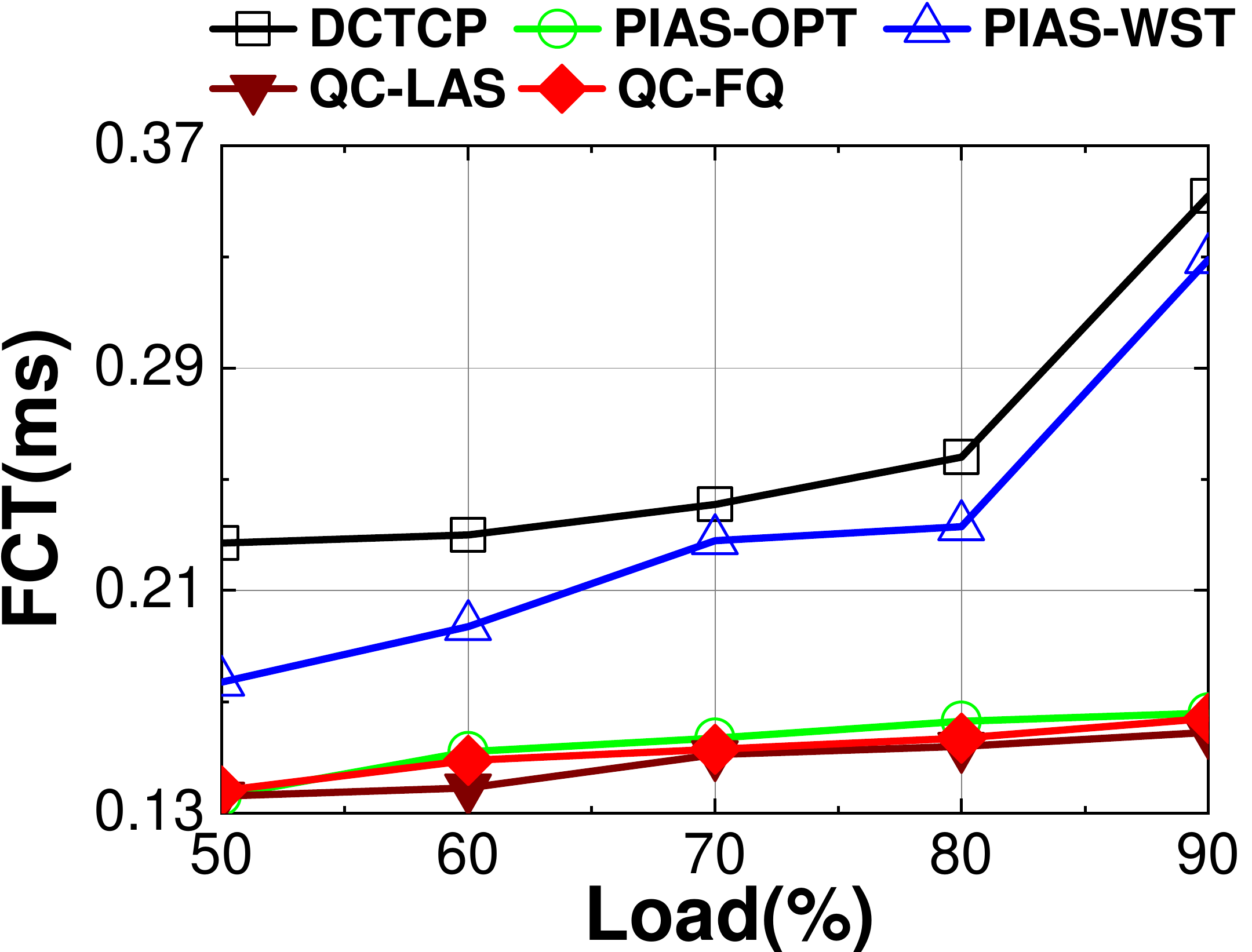}
\label{fig:tb_end:W6:1K}
\end{minipage}
}
\subfigure[(0,10KB): 99th Percentile]{
\begin{minipage}[b]{0.21\textwidth}
\centering
\includegraphics[width=\textwidth]{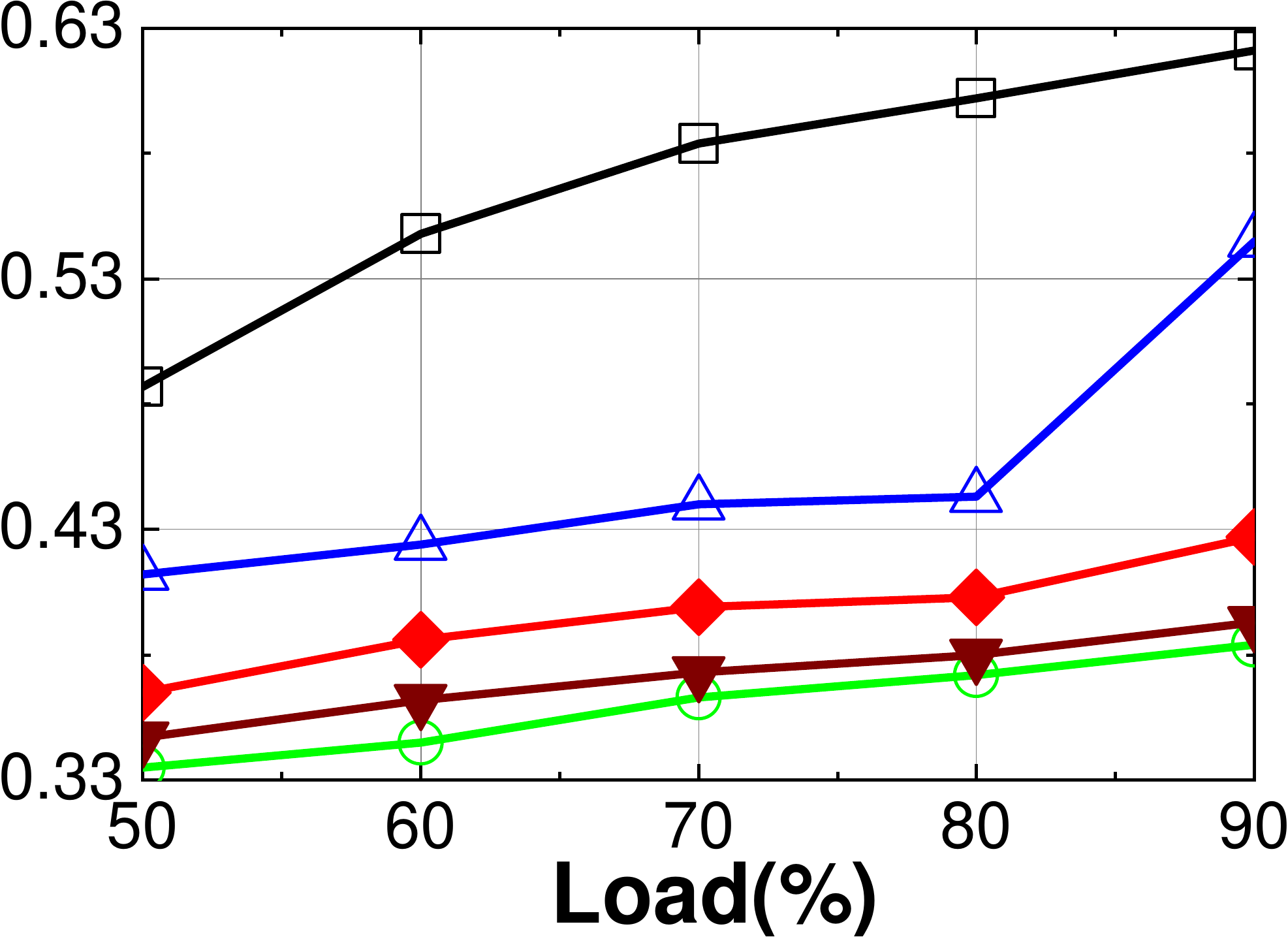}
\label{fig:tb_end:W6:99th}
\end{minipage}
}
\subfigure[(10KB,100KB): Avg]{
\hspace{0.25cm}
\begin{minipage}[b]{0.21\textwidth}
\centering
\includegraphics[width=\textwidth]{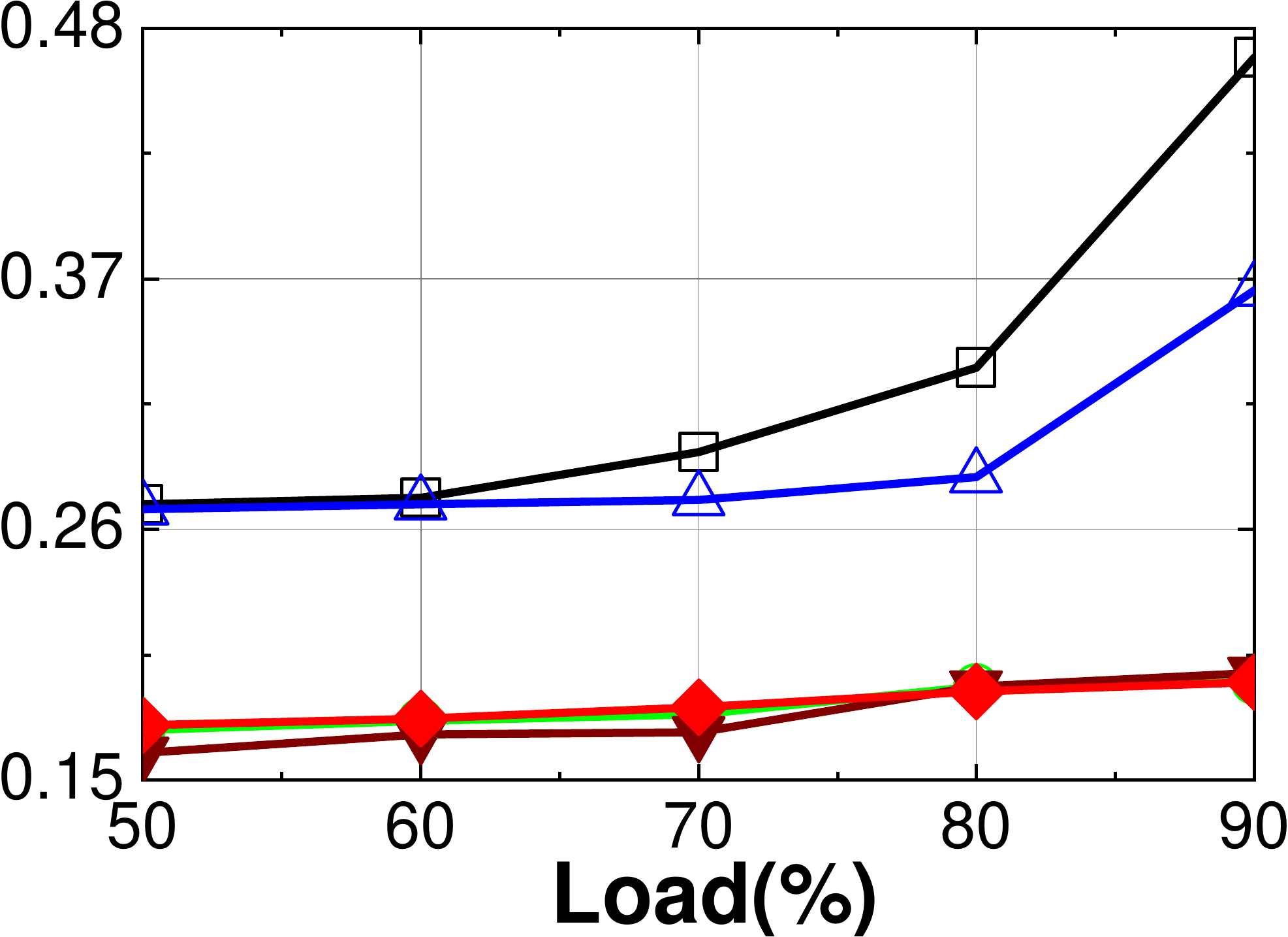}
\label{fig:tb_end:W6:10K}
\end{minipage}
}
\subfigure[(100KB,$\infty$): Avg]{
\hspace{0.1cm}
\begin{minipage}[b]{0.205\textwidth}
\centering
\includegraphics[width=\textwidth]{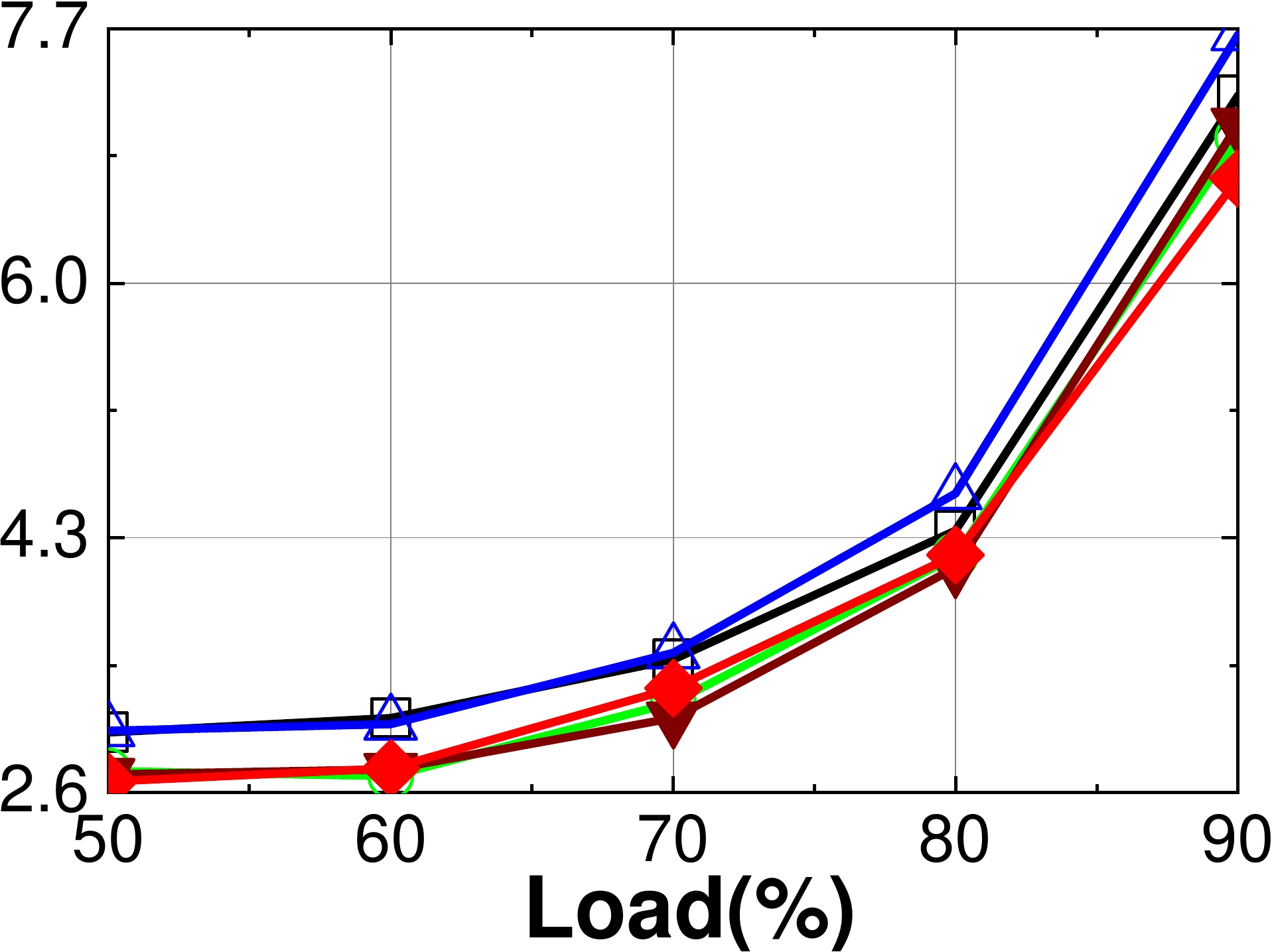}
\label{fig:tb_end:W6:infty}
\end{minipage}
}
\caption{FCT across different flow sizes on W6 in testbed.}
\label{fig:tb_end_all_W6}
\end{figure}

\bbb{Performance on W6
(Figure~\ref{fig:tb_end_all_W6}): }
For small flows in (0, 10KB), the FCT of QC-FQ is about 50.2\% lower than that of PIAS-WST and about 53.4\% lower than that of DCTCP.
For the 99th percentile flow of small flows, the FCT of QC-FQ is about 21.7\% lower than that of PIAS-WST.
For middle flows in (10KB, 100KB), QC-FQ reduces the FCT by about 47.1\% compared to PIAS-WST. 
For large flows in (100KB, $\infty$), our QC-FQ reduces the FCT by about 12.5\% compared to PIAS-WST.

\section{Other Experiment results on LAS and SRPT in ns2}
\label{app:pic}

\begin{figure}[h]
\setlength\abovecaptionskip{-0.0cm}
\setlength\belowcaptionskip{-0.0cm}
\setlength\subfigcapskip{-0.2cm}
\subfigure[(0, 1KB): Avg]{
\begin{minipage}[b]{0.23\textwidth}
\centering
\includegraphics[width=\textwidth]{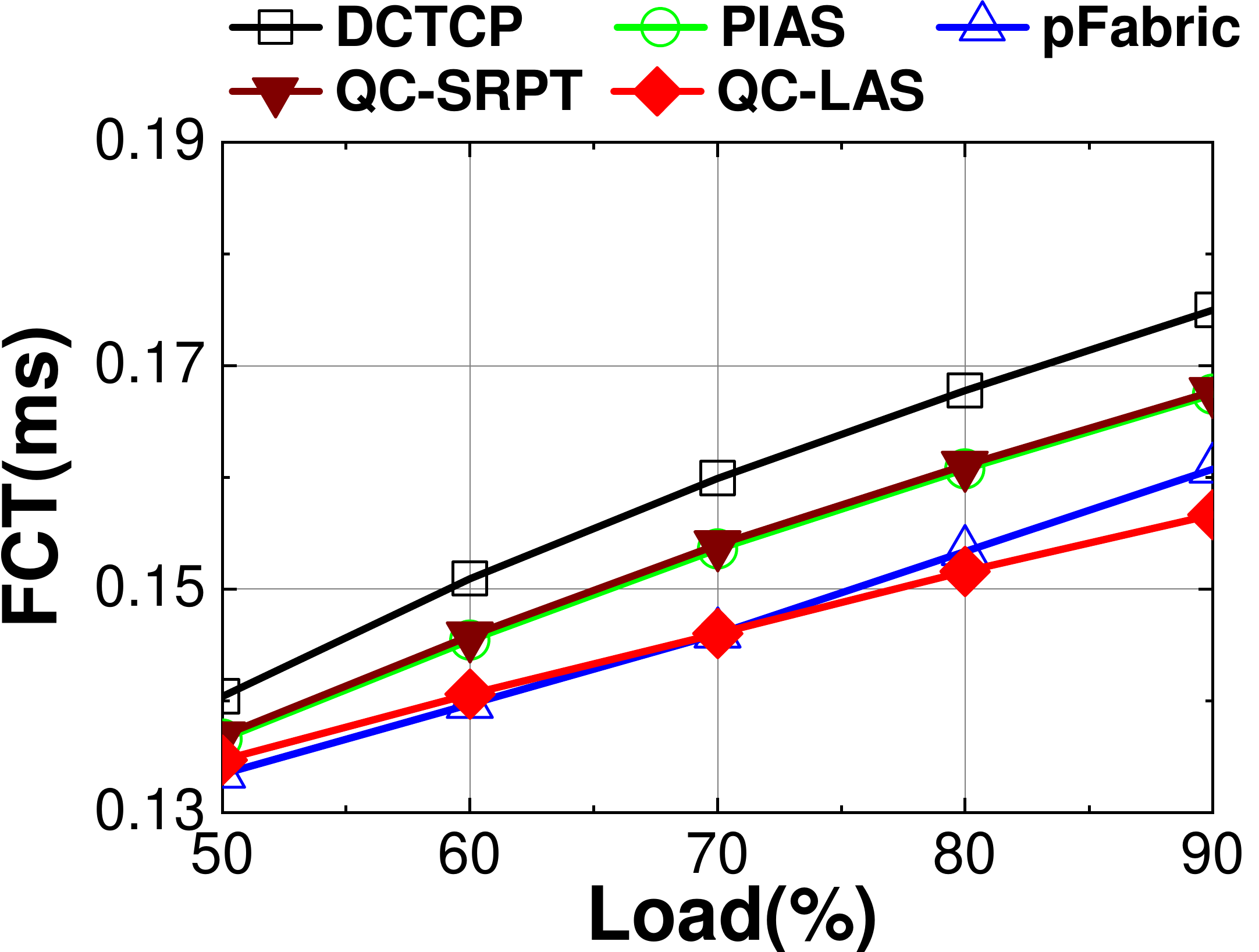}
\postsubfig
\label{fig:W1:1K}
\end{minipage}
}
\subfigure[(0, 1KB): 99th Percentile]{
\begin{minipage}[b]{0.21\textwidth}
\centering
\includegraphics[width=\textwidth]{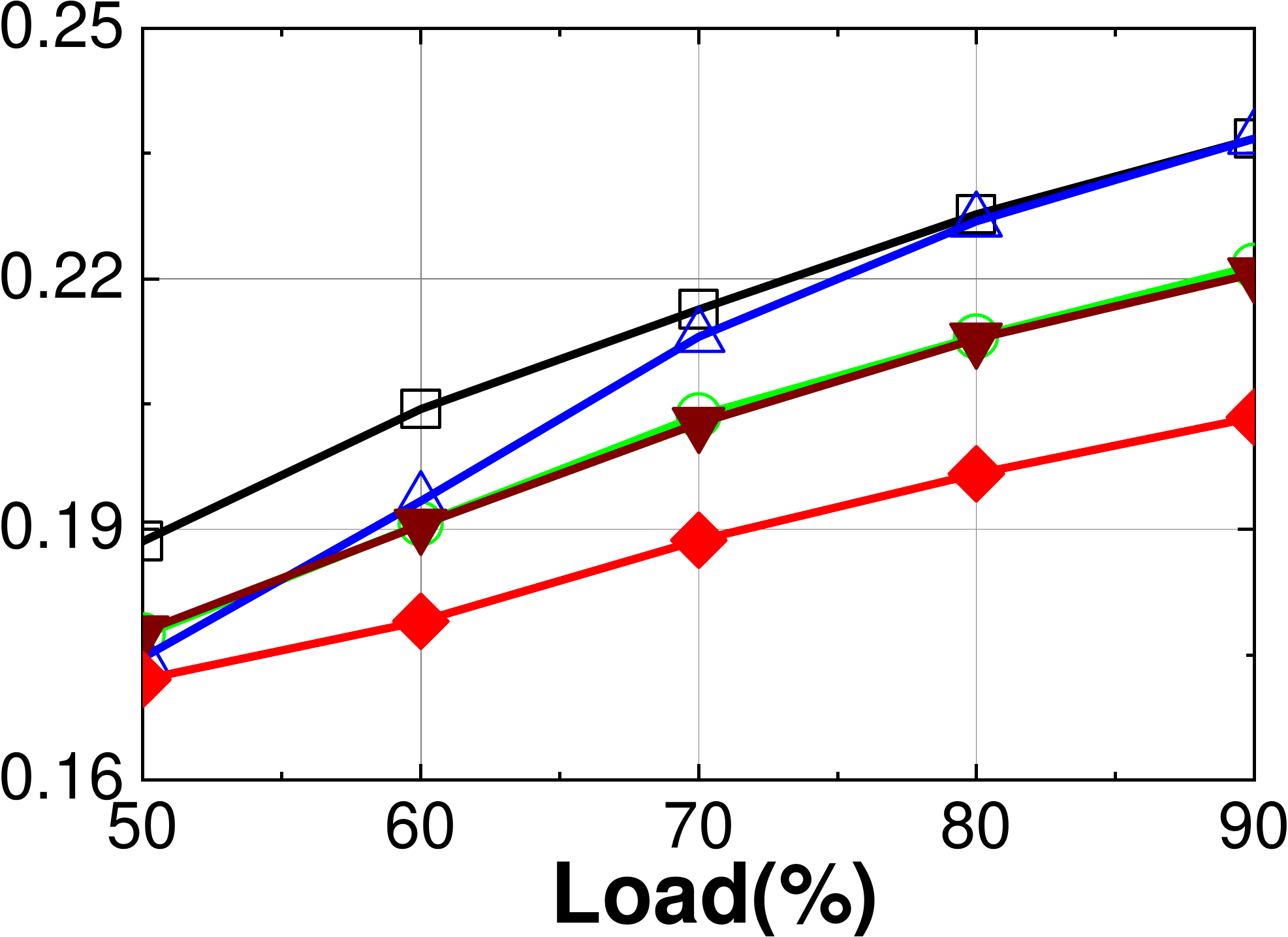}
\postsubfig
\label{fig:W1:99th}
\end{minipage}
}
\subfigure[(1KB,10KB): Avg]{
\hspace{0.35cm}
\begin{minipage}[b]{0.21\textwidth}
\centering
\includegraphics[width=\textwidth]{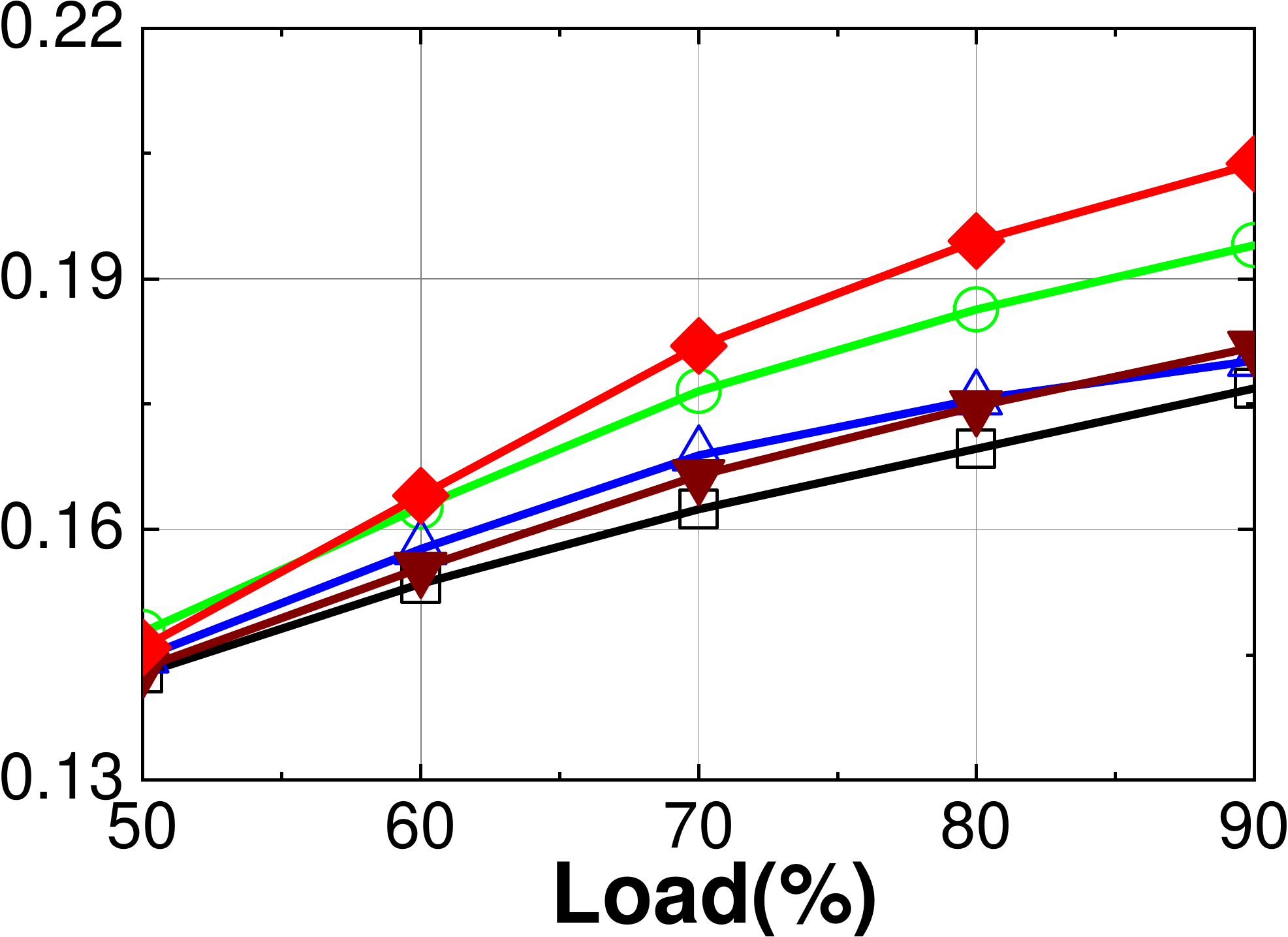}
\postsubfig
\label{fig:W1:10K}
\end{minipage}
}
\subfigure[(10KB,$\infty$): Avg]{
\begin{minipage}[b]{0.21\textwidth}
\centering
\includegraphics[width=\textwidth]{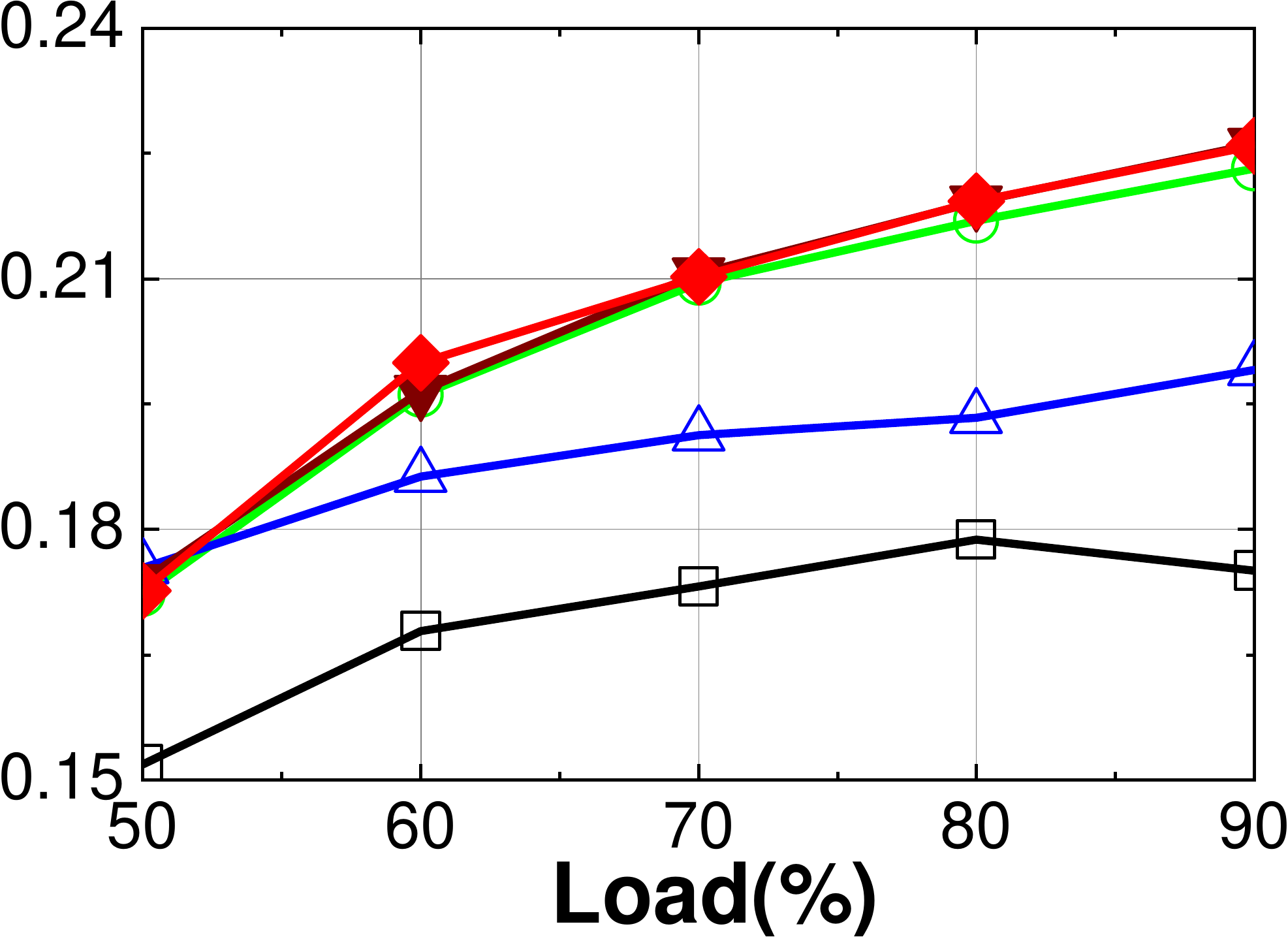}
\postsubfig
\label{fig:W1:infty}
\end{minipage}
}
\postsubfig
\caption{FCT across different flow sizes on W1 for SRPT and LAS.}
\label{fig:W1}
\end{figure}
\begin{figure}[h]
\setlength\abovecaptionskip{-0.0cm}
\setlength\belowcaptionskip{-0.0cm}
\setlength\subfigcapskip{-0.2cm}
\subfigure[(0,1KB): Avg]{
\begin{minipage}[b]{0.23\textwidth}
\centering
\includegraphics[width=\textwidth]{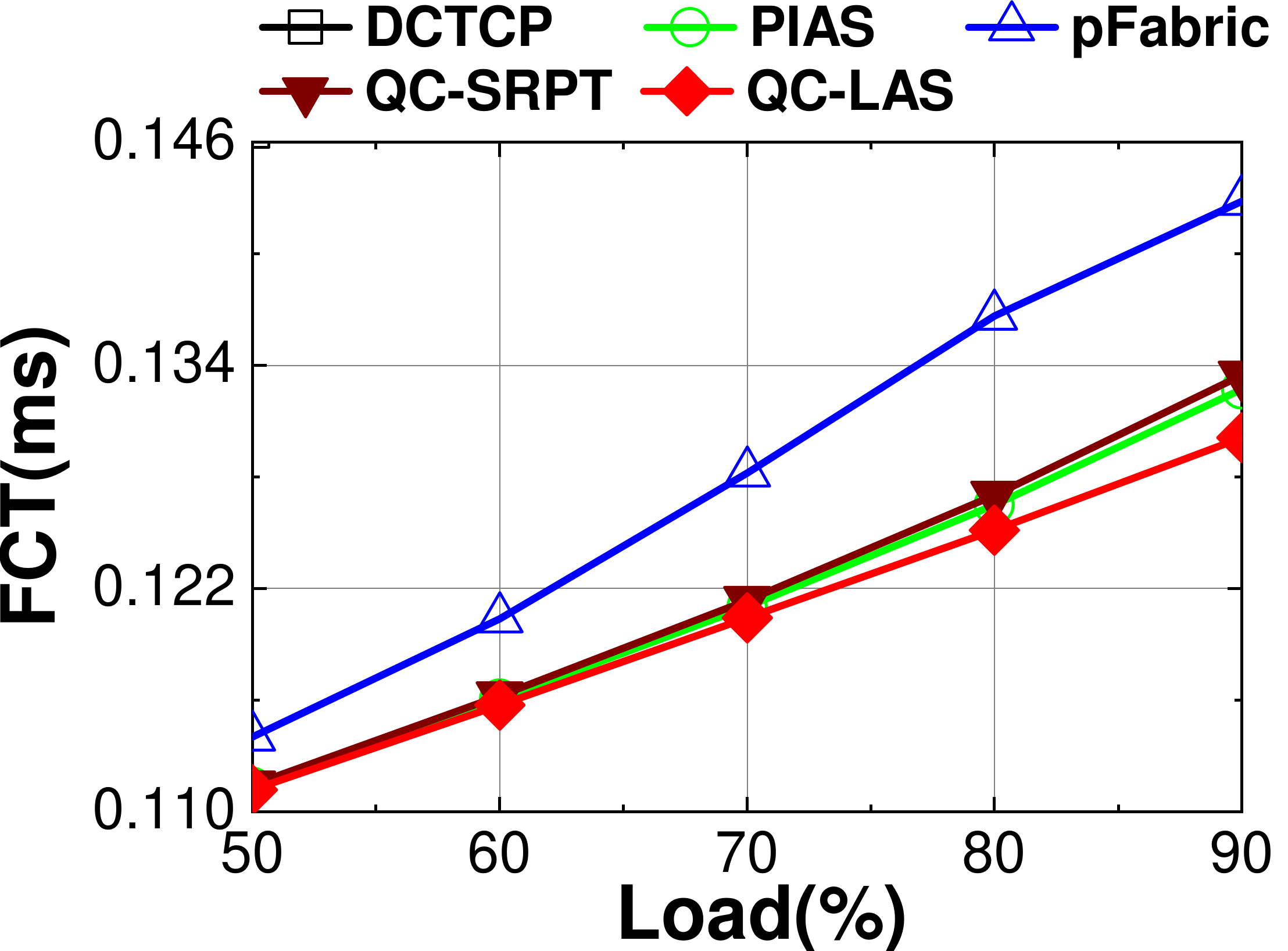}
\postsubfig
\label{fig:W2:1K}
\end{minipage}
}
\subfigure[(0,1KB): 99th Percentile]{
\begin{minipage}[b]{0.21\textwidth}
\centering
\includegraphics[width=\textwidth]{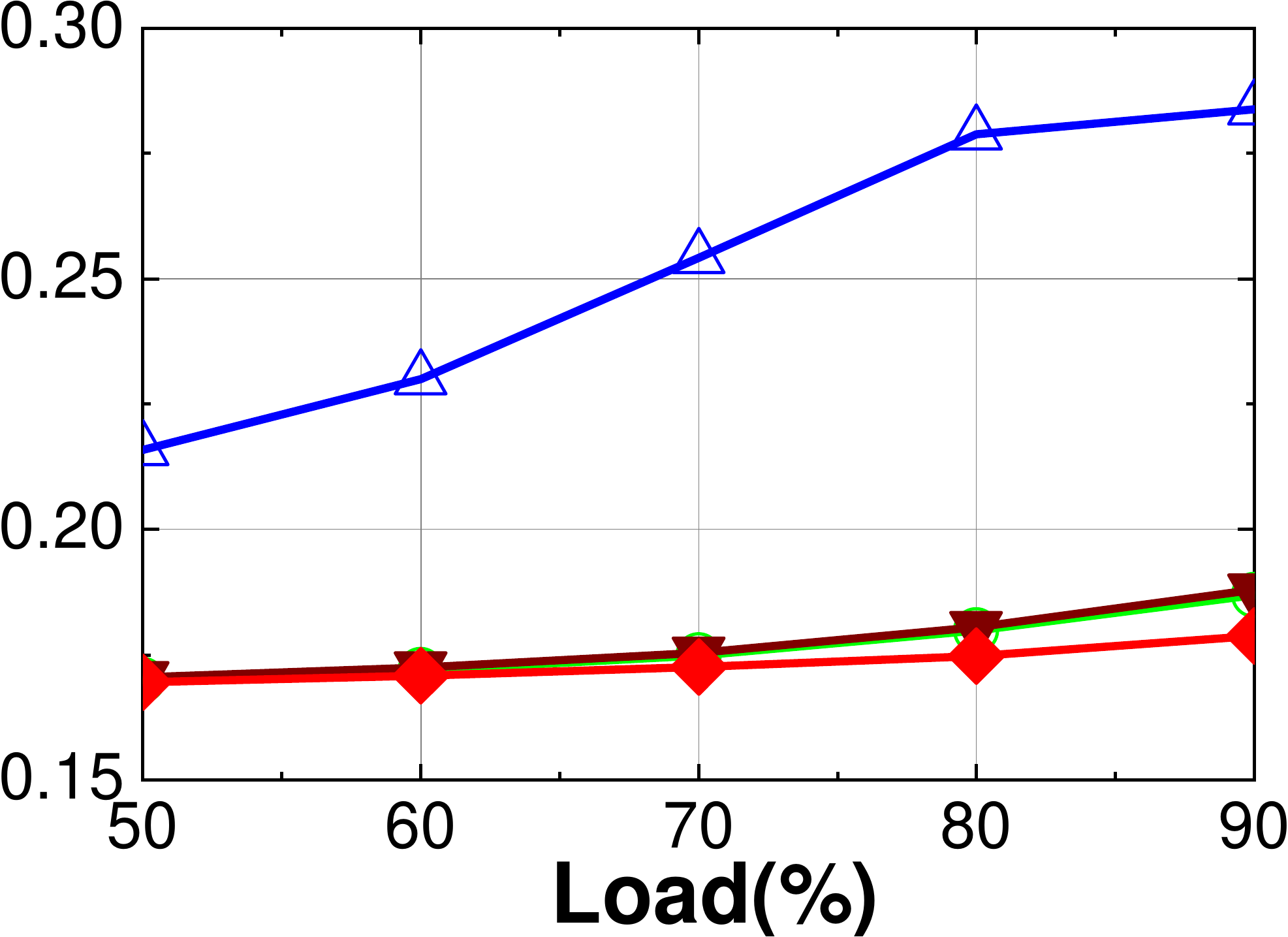}
\label{fig:W2:99th}
\end{minipage}
}
\subfigure[(1KB,10KB): Avg]{
\hspace{0.35cm}
\begin{minipage}[b]{0.21\textwidth}
\centering
\includegraphics[width=\textwidth]{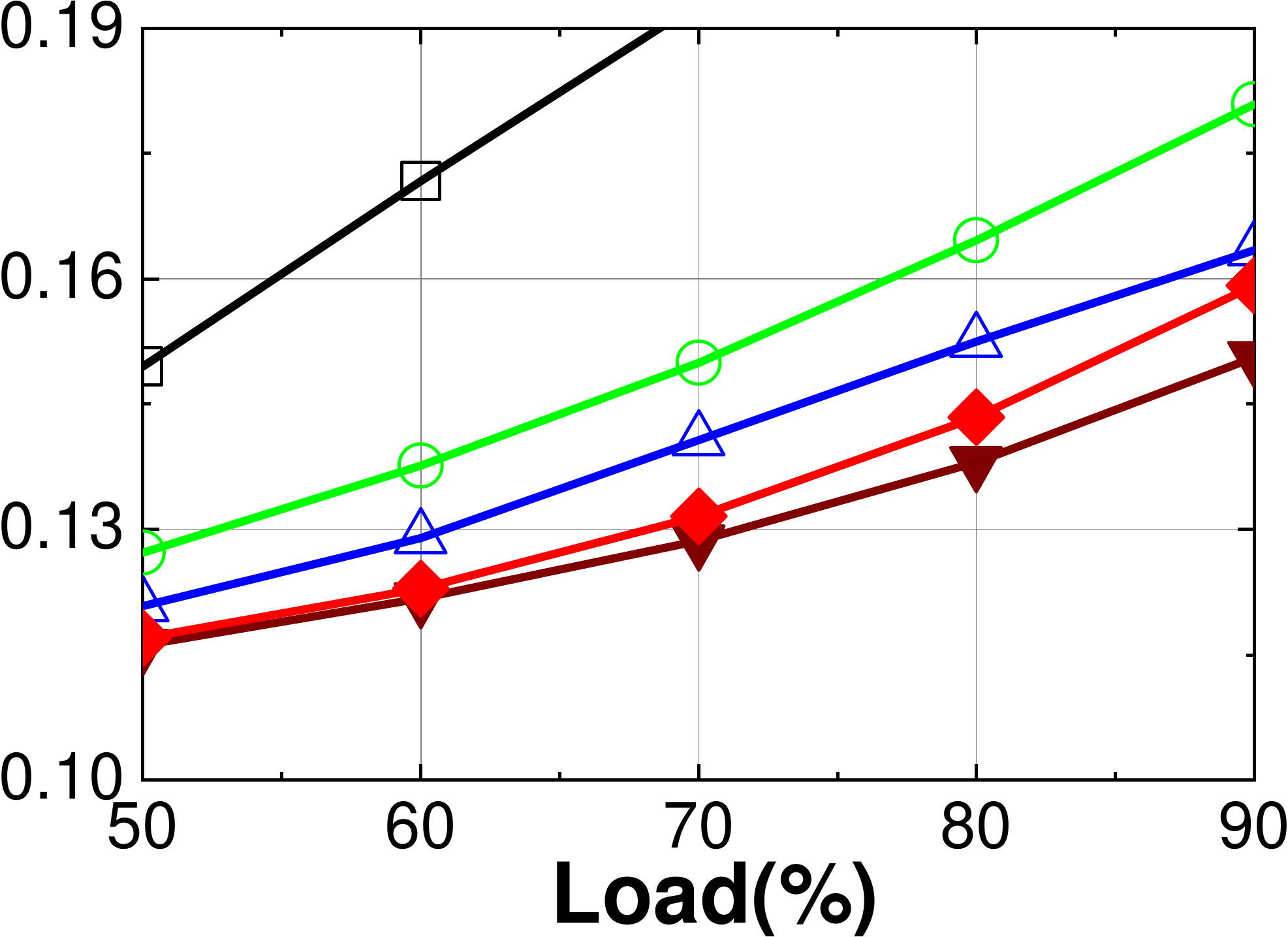}
\label{fig:W2:10KB}
\end{minipage}
}
\subfigure[(10KB,$\infty$): Avg]{
\begin{minipage}[b]{0.21\textwidth}
\centering
\includegraphics[width=\textwidth]{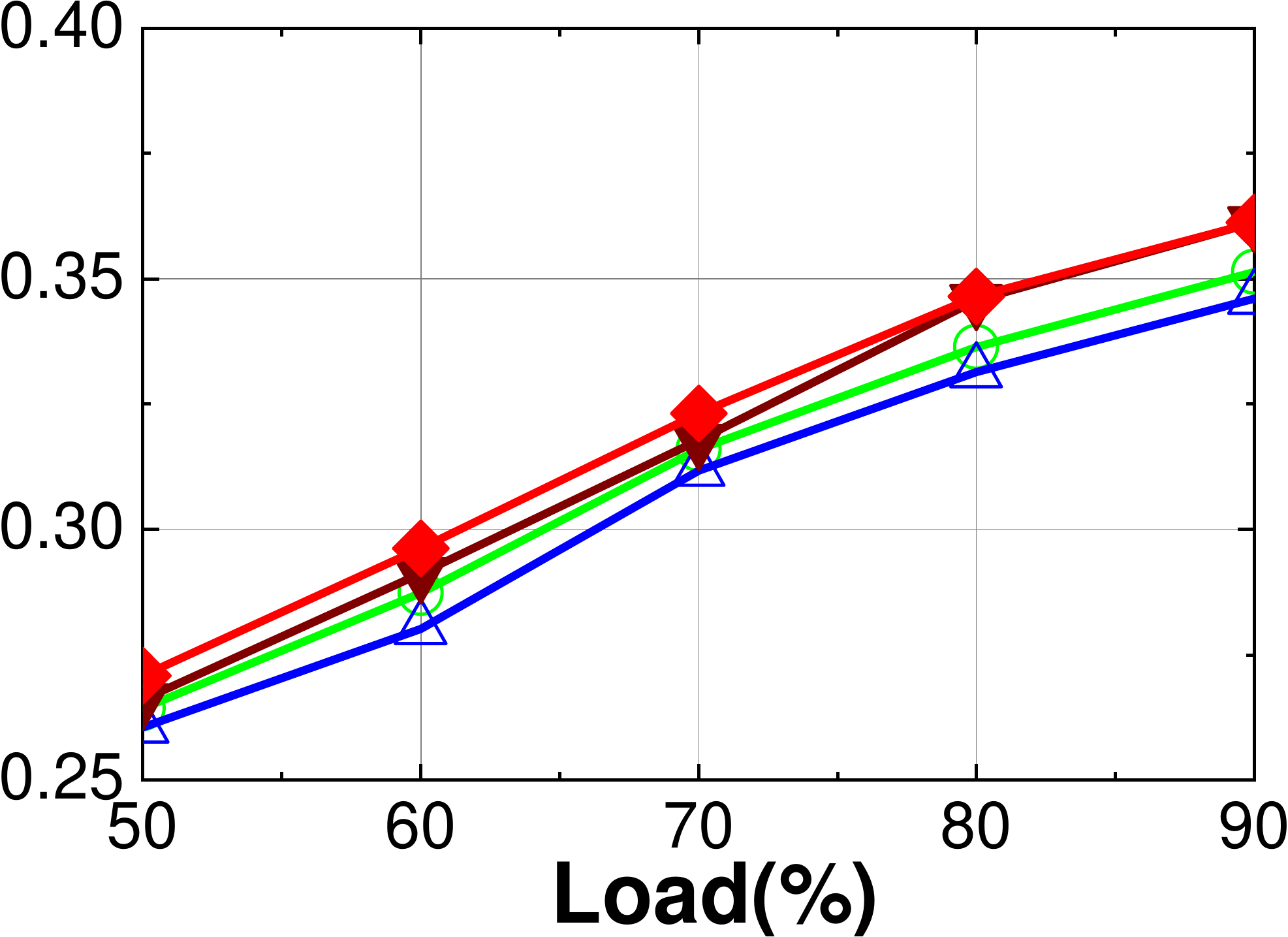}
\label{fig:W2:infty}
\end{minipage}
}
\caption{FCT across different flow sizes on W2 for SRPT and LAS.}
\label{fig:W2}
\end{figure}

\bbb{Performance on W1 (Figure~\ref{fig:W1}): }
In W1, more than 95\% of flows are smaller than 1KB. That is to say, more than 95\% of flows can be transmitted in one TCP packet. In this situation, the advantage of SRPT over LAS is not obvious. Therefore, our QC-LAS may outperform pFabric on W1 in some loads. For short flows in (0, 1KB), the FCT of our QC-LAS is about 6.9\% lower than that of PIAS. Without ECN for congestion control, congestion is more likely to occur in switches. Therefore, the 99th percentile FCT for small flows of pFabric is even higher than that of DCTCP at 90\% load.

\bbb{Performance on W2 (Figure~\ref{fig:W2}): }
For small flows in (0,1KB), the average FCT of the QC-SRPT is about 7\% lower than that of the pFabric.
Moreover, QC-SRPT also reduces the 99th percentile FCT by about 50.1\% at 90\% load. 
Similar to W1, pFabric does not perform so well when data centers are dominated by very small flows. 
The FCT of QC-LAS is about 1.2\% lower than the FCT of PIAS for small flows in (0, 1KB). And QC-LAS reduces the FCT for middle flows in (1KB, 10KB) by about 13\%.

\begin{figure}[h]
\setlength\abovecaptionskip{-0.0cm}
\setlength\belowcaptionskip{-0.0cm}
\setlength\subfigcapskip{-0.2cm}
\subfigure[(0, 1KB): Avg]{
\begin{minipage}[b]{0.23\textwidth}
\centering
\includegraphics[width=\textwidth]{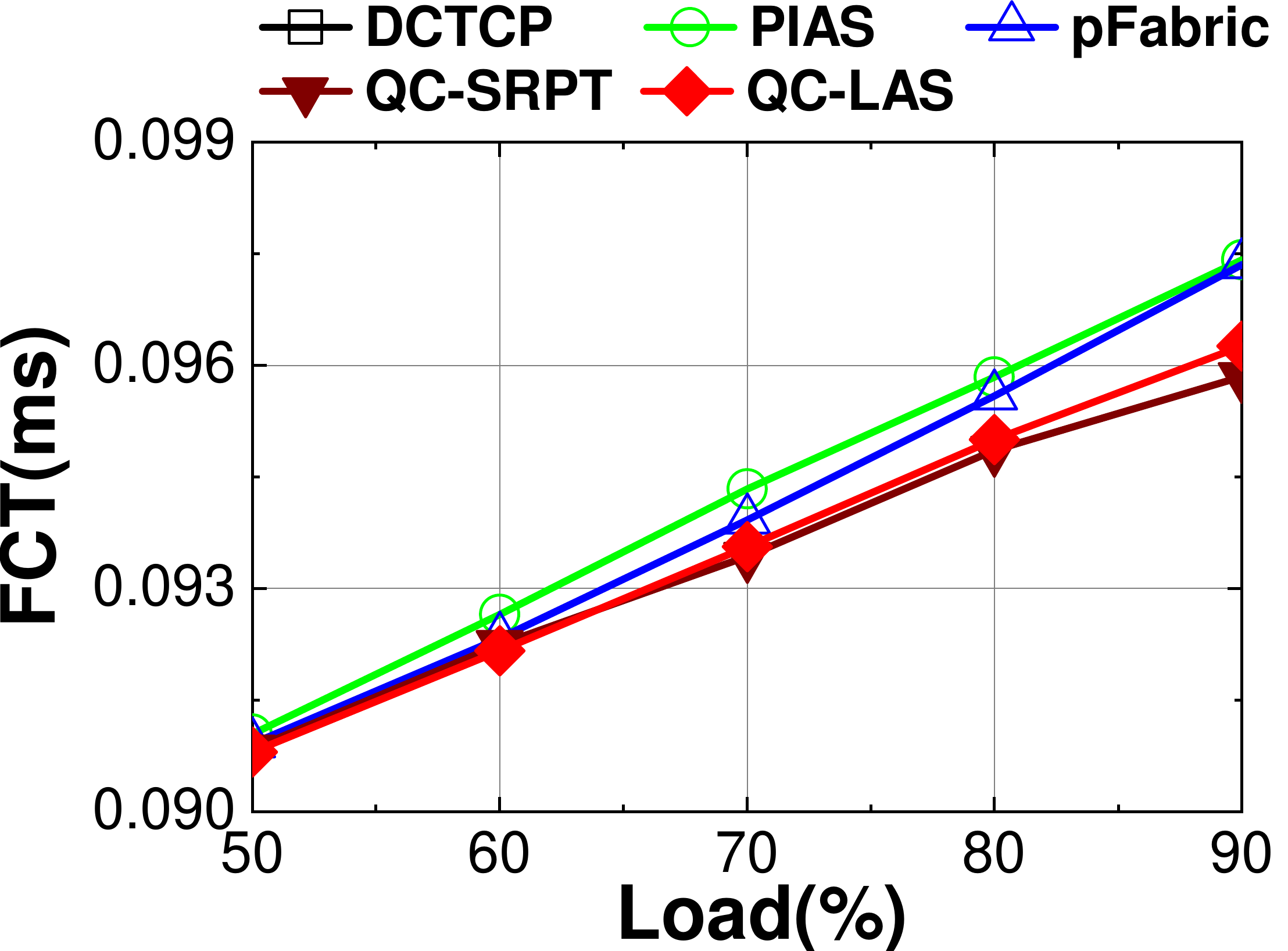}
\postsubfig
\label{fig:W3:1K}
\end{minipage}
}
\subfigure[(0, 1KB): 99th Percentile]{
\begin{minipage}[b]{0.21\textwidth}
\centering
\includegraphics[width=\textwidth]{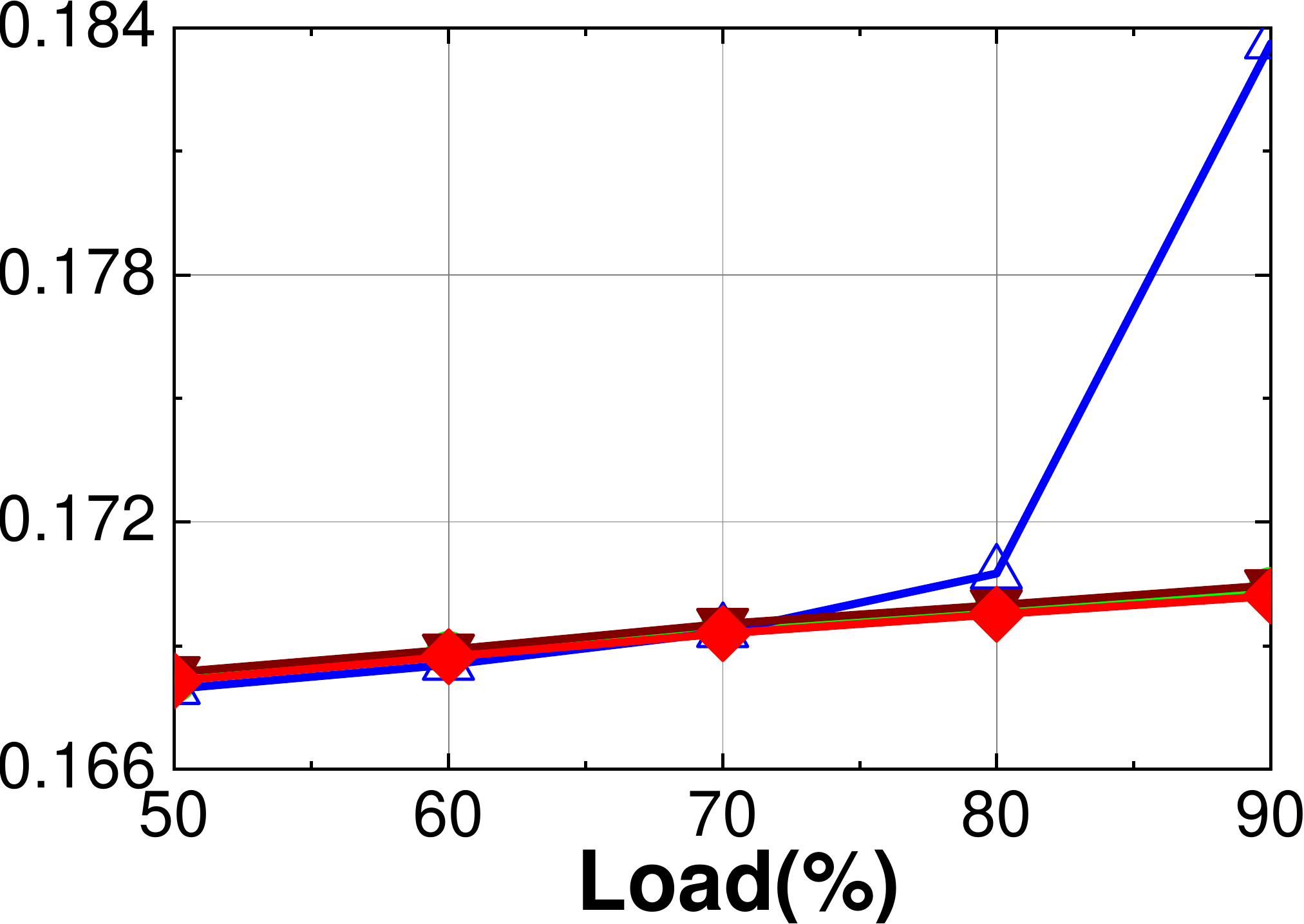}
\postsubfig
\label{fig:W3:99th}
\end{minipage}
}
\subfigure[(1KB,10KB): Avg]{
\hspace{0.35cm}
\begin{minipage}[b]{0.21\textwidth}
\centering
\includegraphics[width=\textwidth]{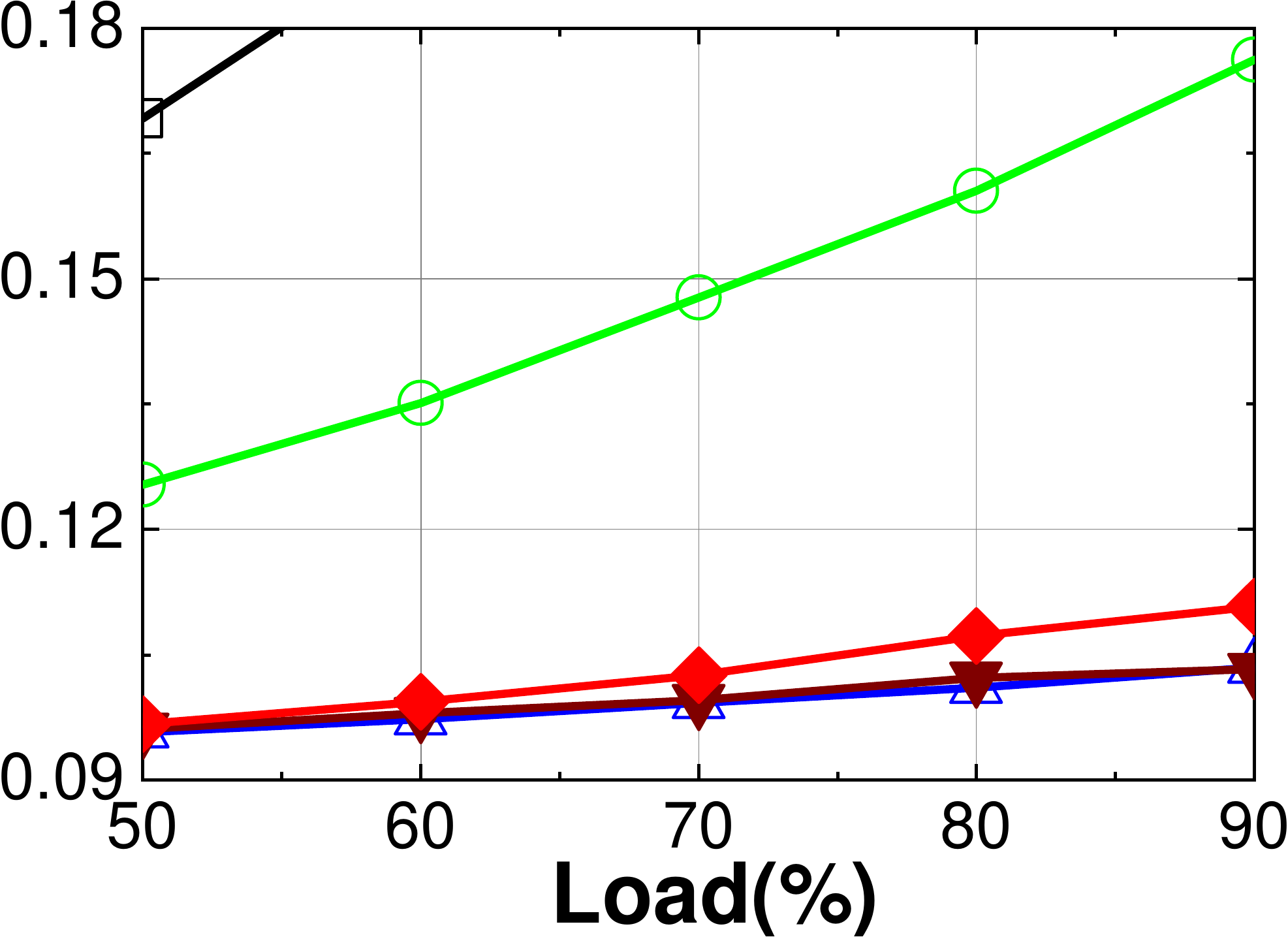}
\postsubfig
\label{fig:W3:10K}
\end{minipage}
}
\subfigure[(10KB,$\infty$): Avg]{
\begin{minipage}[b]{0.21\textwidth}
\centering
\includegraphics[width=\textwidth]{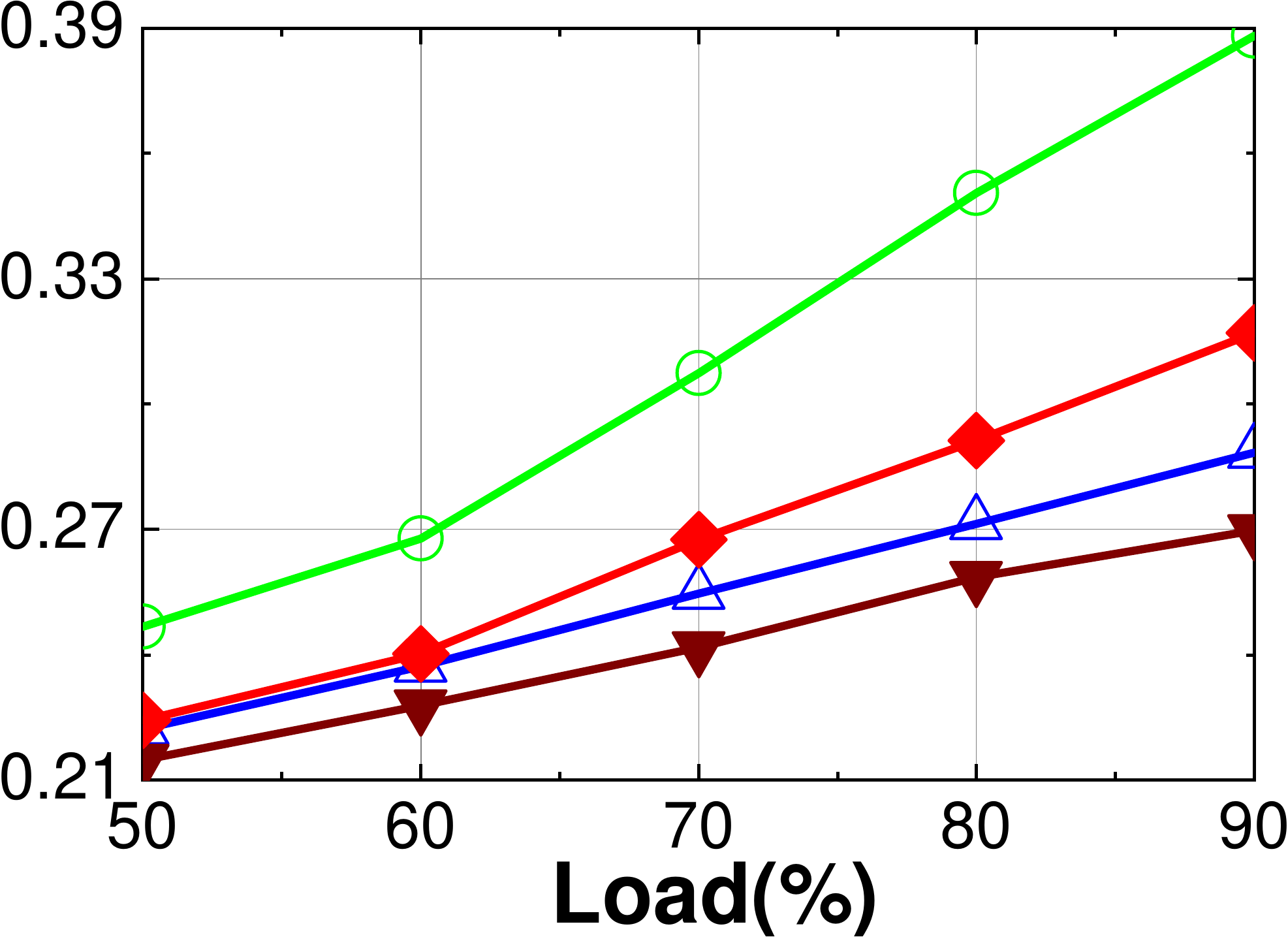}
\postsubfig
\label{fig:W3:infty}
\end{minipage}
}
\postsubfig
\caption{FCT across different flow sizes on W3 for SRPT and LAS.}
\label{fig:W3}
\end{figure}
\begin{figure}[h]
\setlength\abovecaptionskip{-0.0cm}
\setlength\belowcaptionskip{-0.0cm}
\setlength\subfigcapskip{-0.2cm}
\subfigure[(0, 1KB): Avg]{
\begin{minipage}[b]{0.23\textwidth}
\centering
\includegraphics[width=\textwidth]{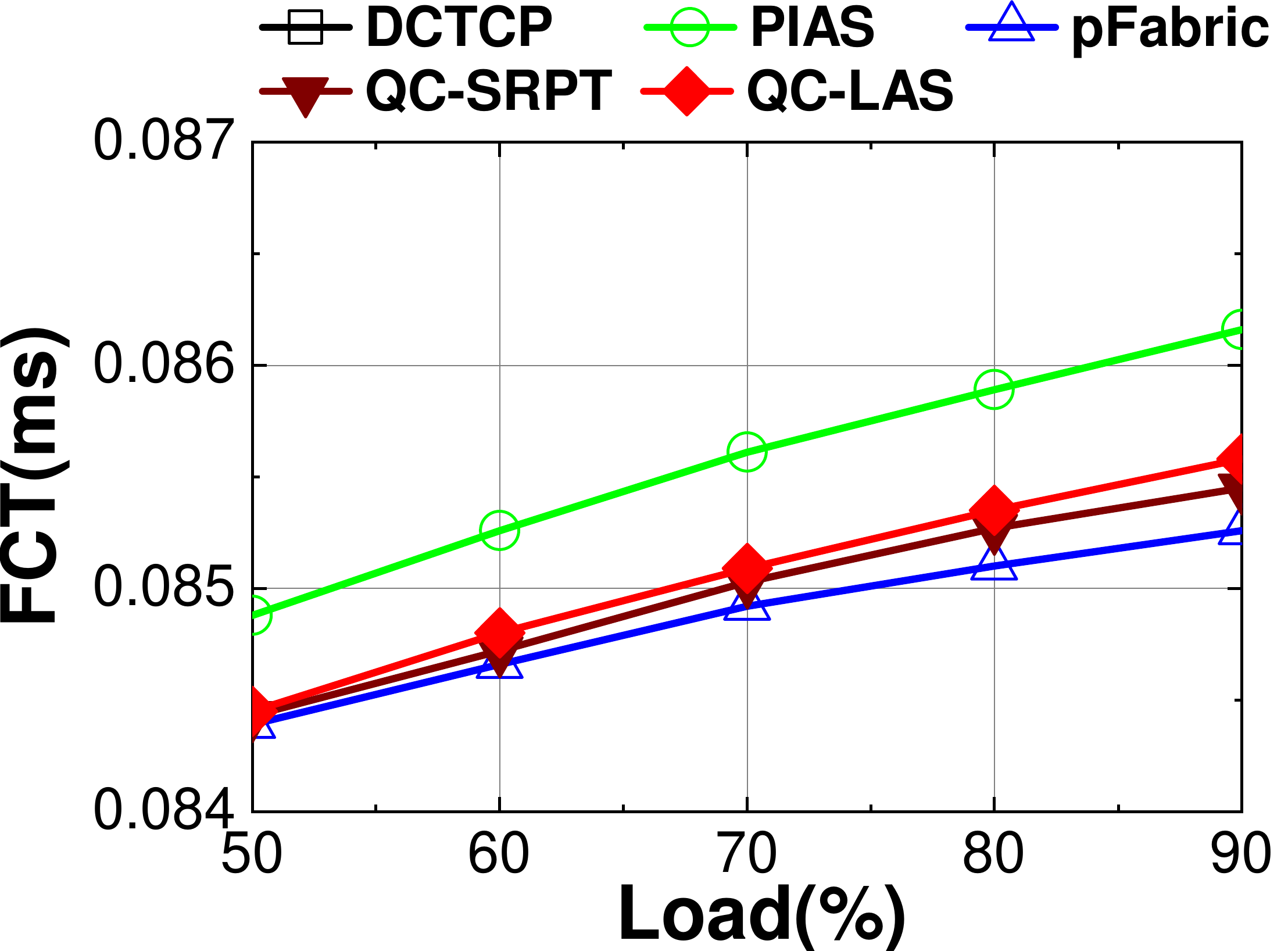}
\postsubfig
\label{fig:W5:1K}
\end{minipage}
}
\subfigure[(0, 1KB): 99th Percentile]{
\begin{minipage}[b]{0.21\textwidth}
\centering
\includegraphics[width=\textwidth]{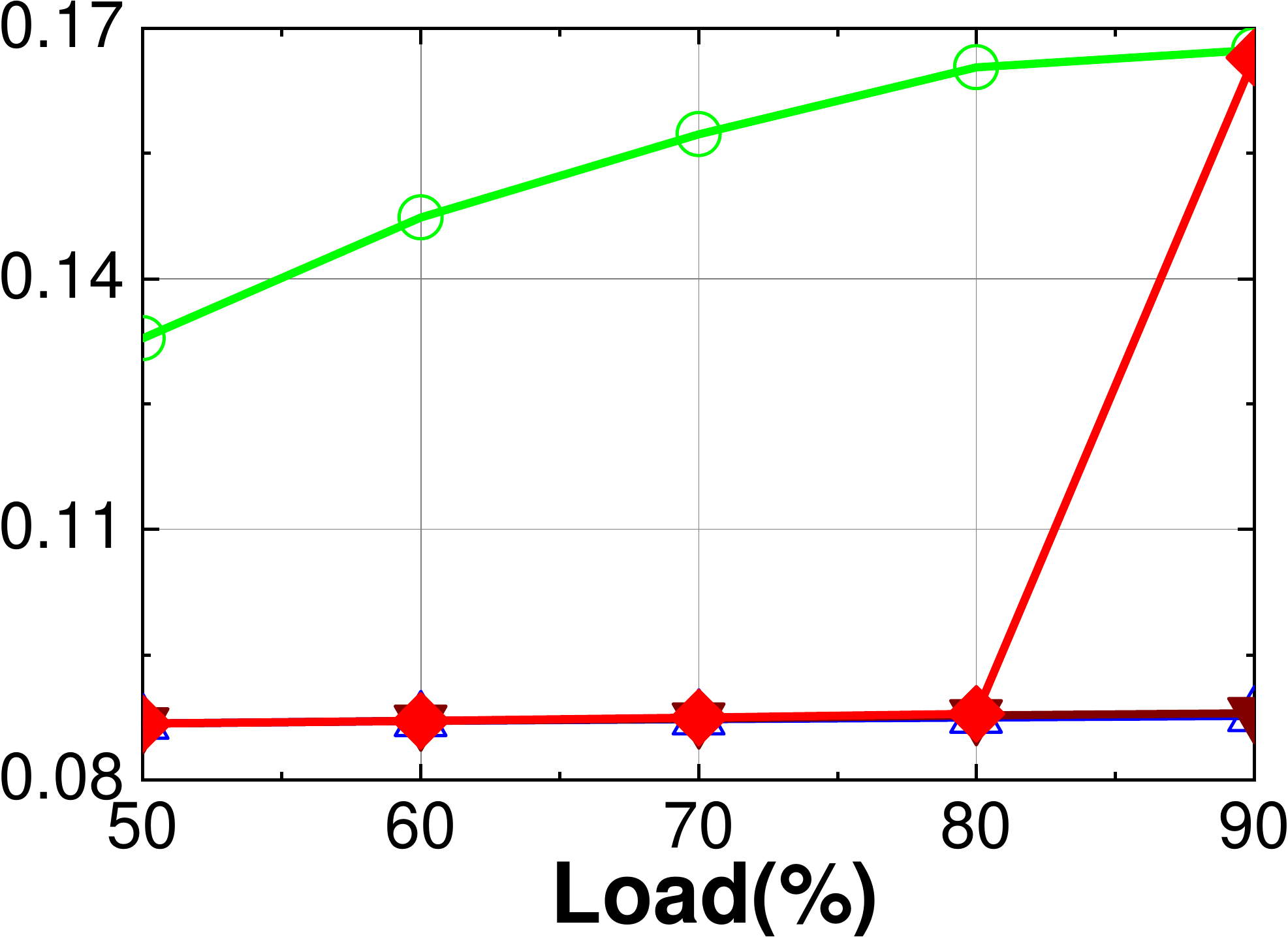}
\postsubfig
\label{fig:W5:99th}
\end{minipage}
}
\subfigure[(1KB,10KB): Avg]{
\hspace{0.35cm}
\begin{minipage}[b]{0.21\textwidth}
\centering
\includegraphics[width=\textwidth]{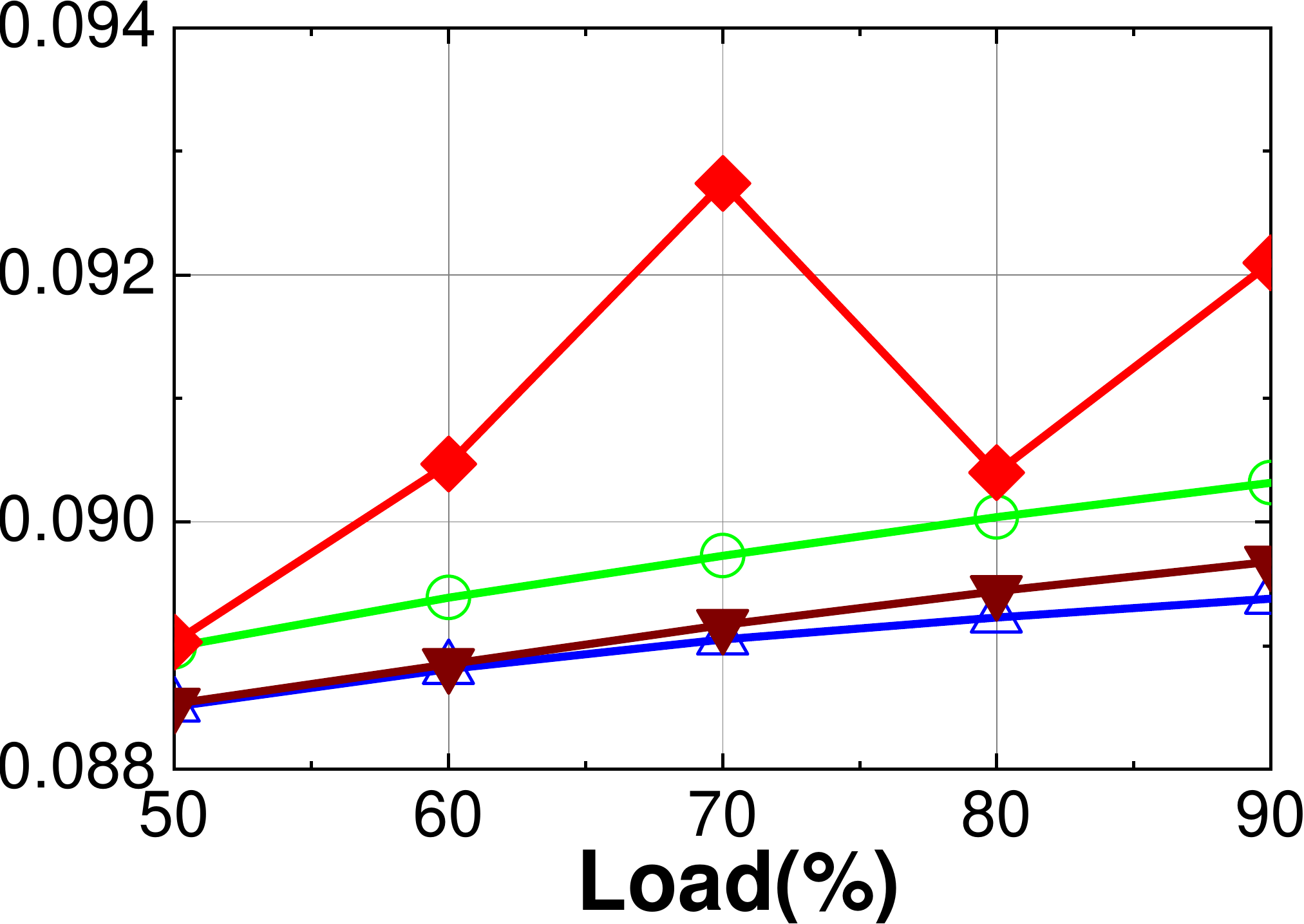}
\postsubfig
\label{fig:W5:10K}
\end{minipage}
}
\subfigure[(10KB,$\infty$): Avg]{
\begin{minipage}[b]{0.21\textwidth}
\centering
\includegraphics[width=\textwidth]{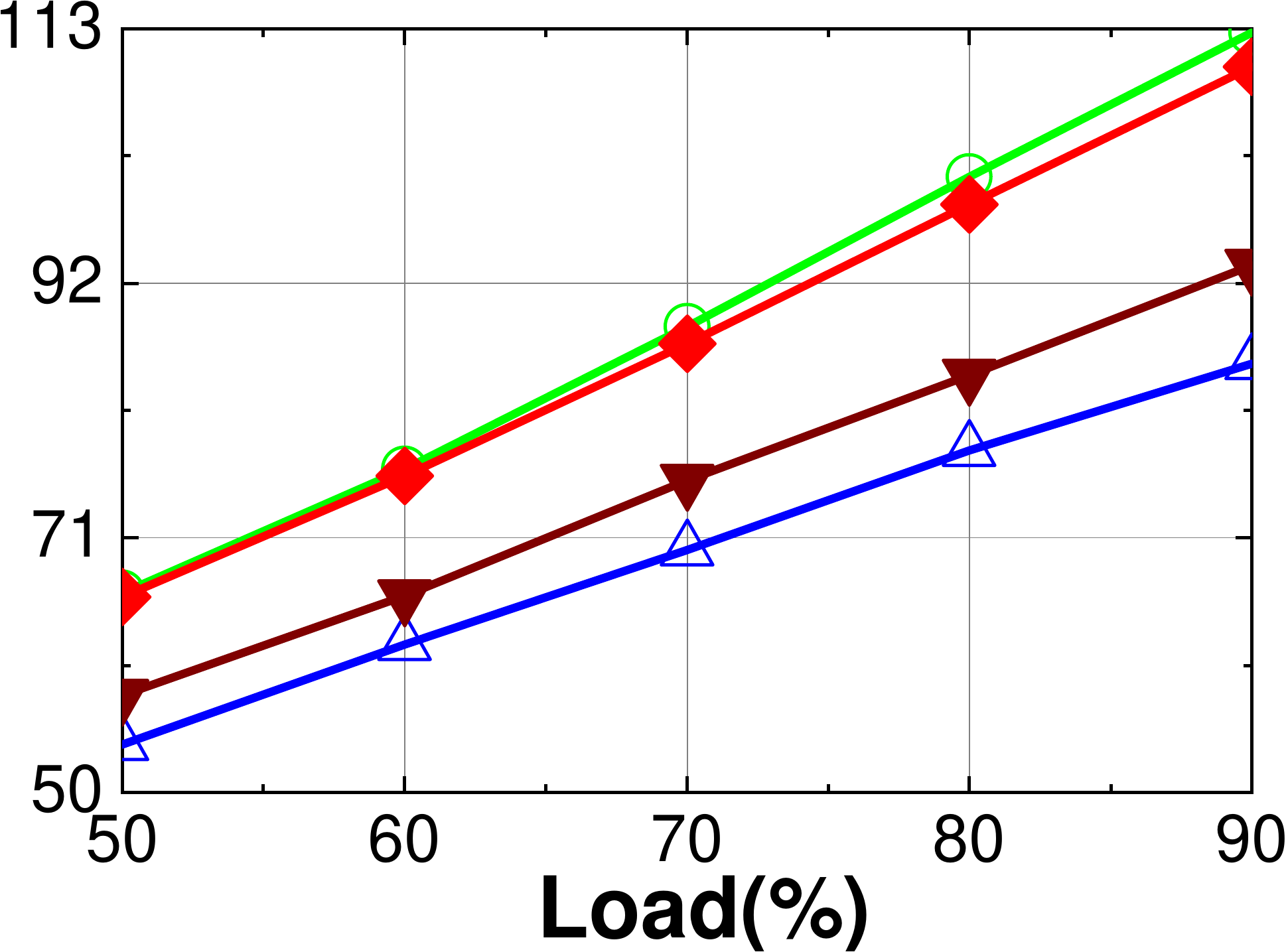}
\postsubfig
\label{fig:W5:infty}
\end{minipage}
}
\postsubfig
\caption{FCT across different flow sizes on W5 for SRPT and LAS.}
\label{fig:W5}
\end{figure}
\begin{figure}[h]
\setlength\abovecaptionskip{-0.0cm}
\setlength\belowcaptionskip{-0.0cm}
\setlength\subfigcapskip{-0.2cm}
\subfigure[(0KB, 10KB): Avg]{
\begin{minipage}[b]{0.235\textwidth}
\centering
\includegraphics[width=\textwidth]{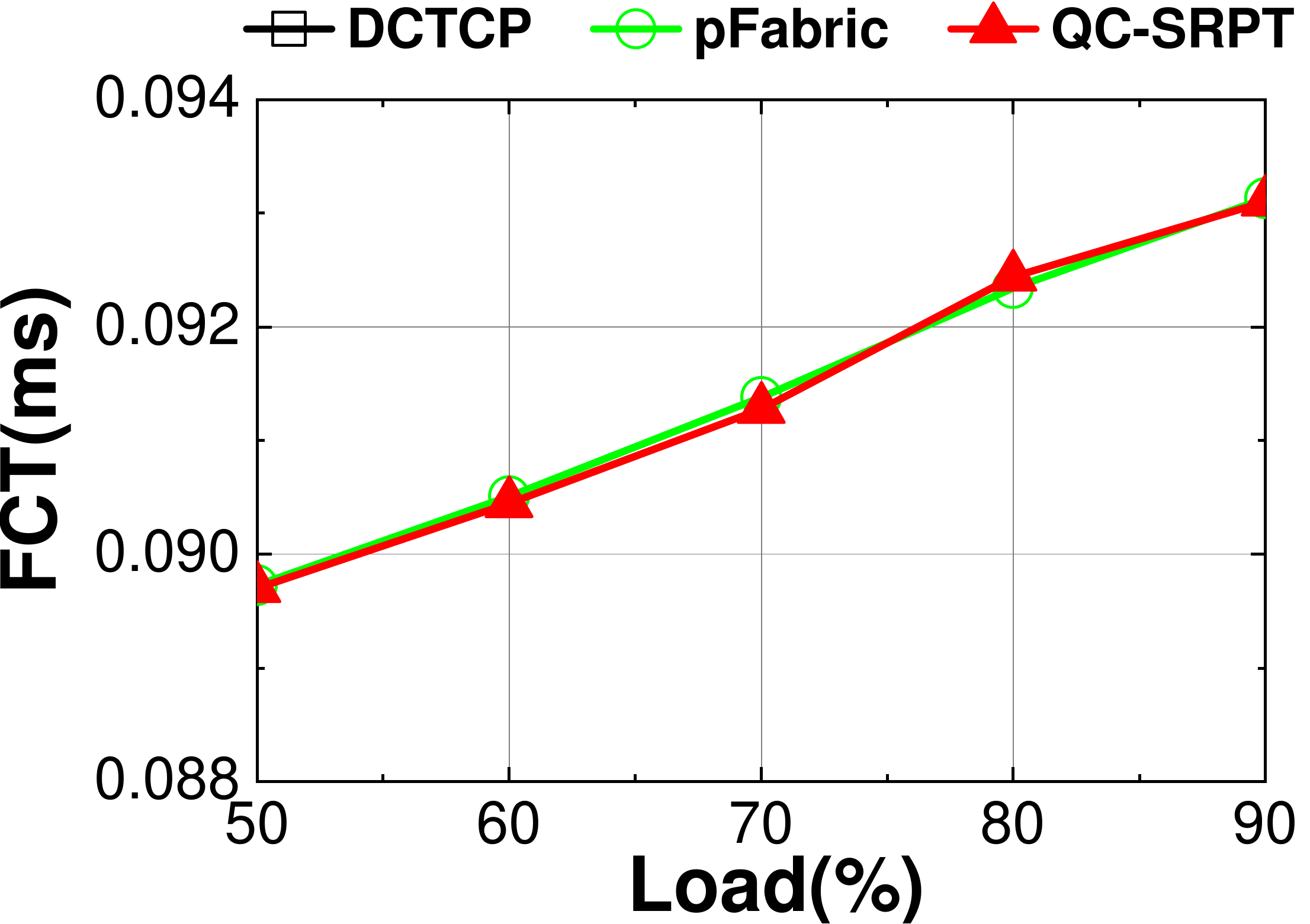}
\postsubfig
\label{fig:W6:10K}
\end{minipage}
}
\subfigure[(0, 10KB): 99th Percentile]{
\begin{minipage}[b]{0.215\textwidth}
\centering
\includegraphics[width=\textwidth]{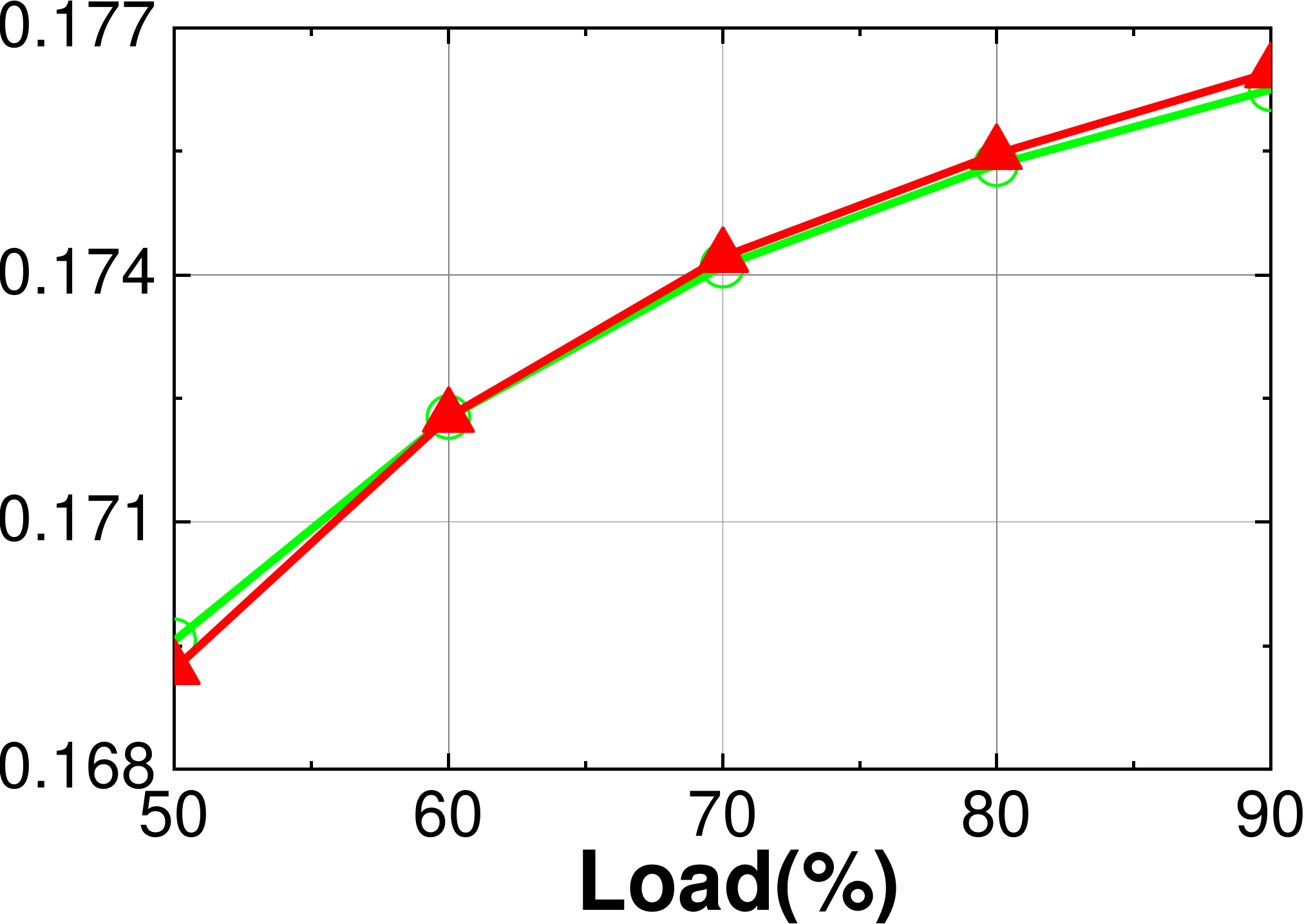}
\postsubfig
\label{fig:W6:99th}
\end{minipage}
}
\subfigure[(10KB,100KB): Avg]{
\hspace{0.2cm}
\begin{minipage}[b]{0.215\textwidth}
\centering
\includegraphics[width=\textwidth]{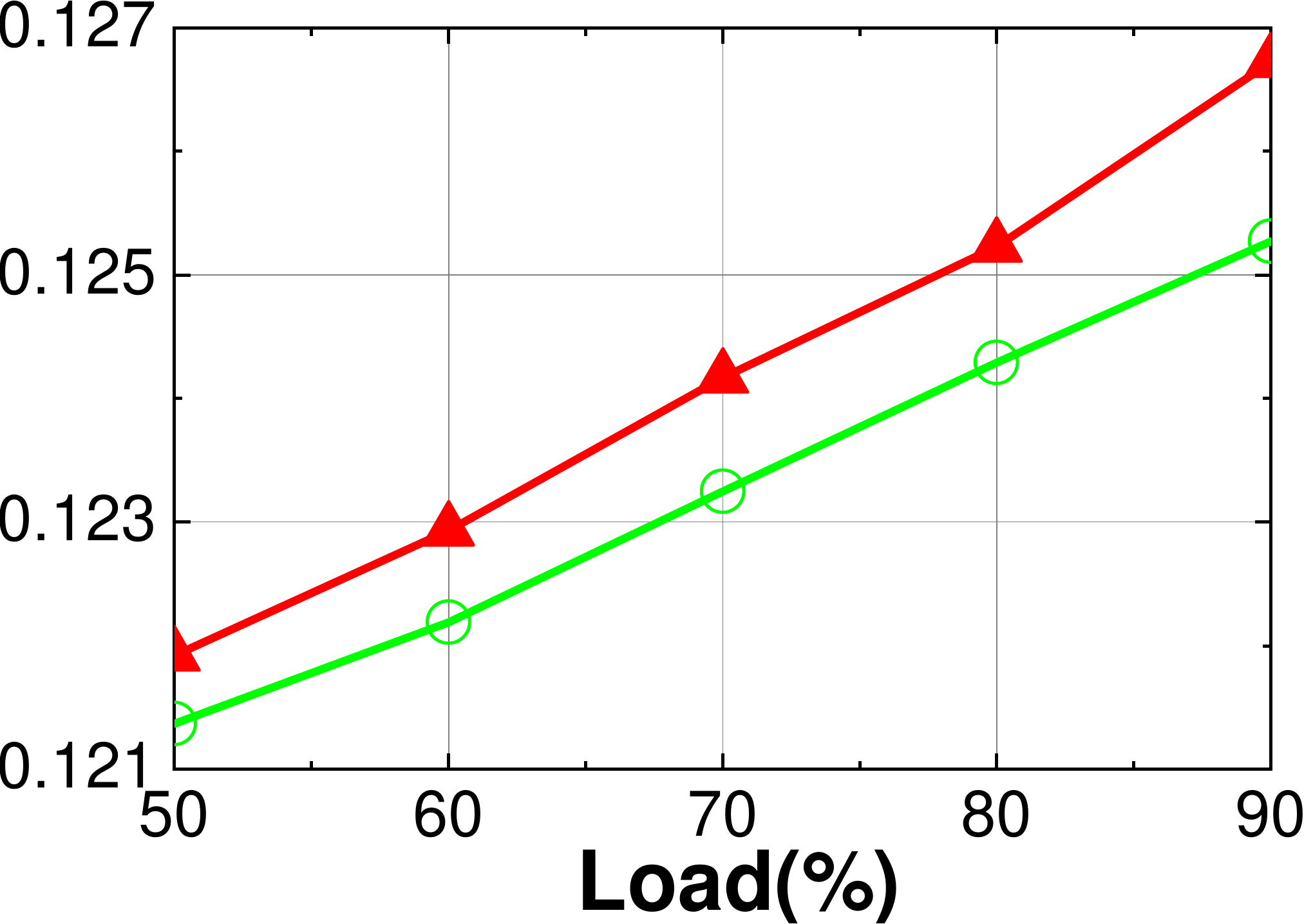}
\postsubfig
\label{fig:W6:100K}
\end{minipage}
}
\subfigure[(100KB,$\infty$): Avg]{
\hspace{0.3cm}
\begin{minipage}[b]{0.2\textwidth}
\centering
\includegraphics[width=\textwidth]{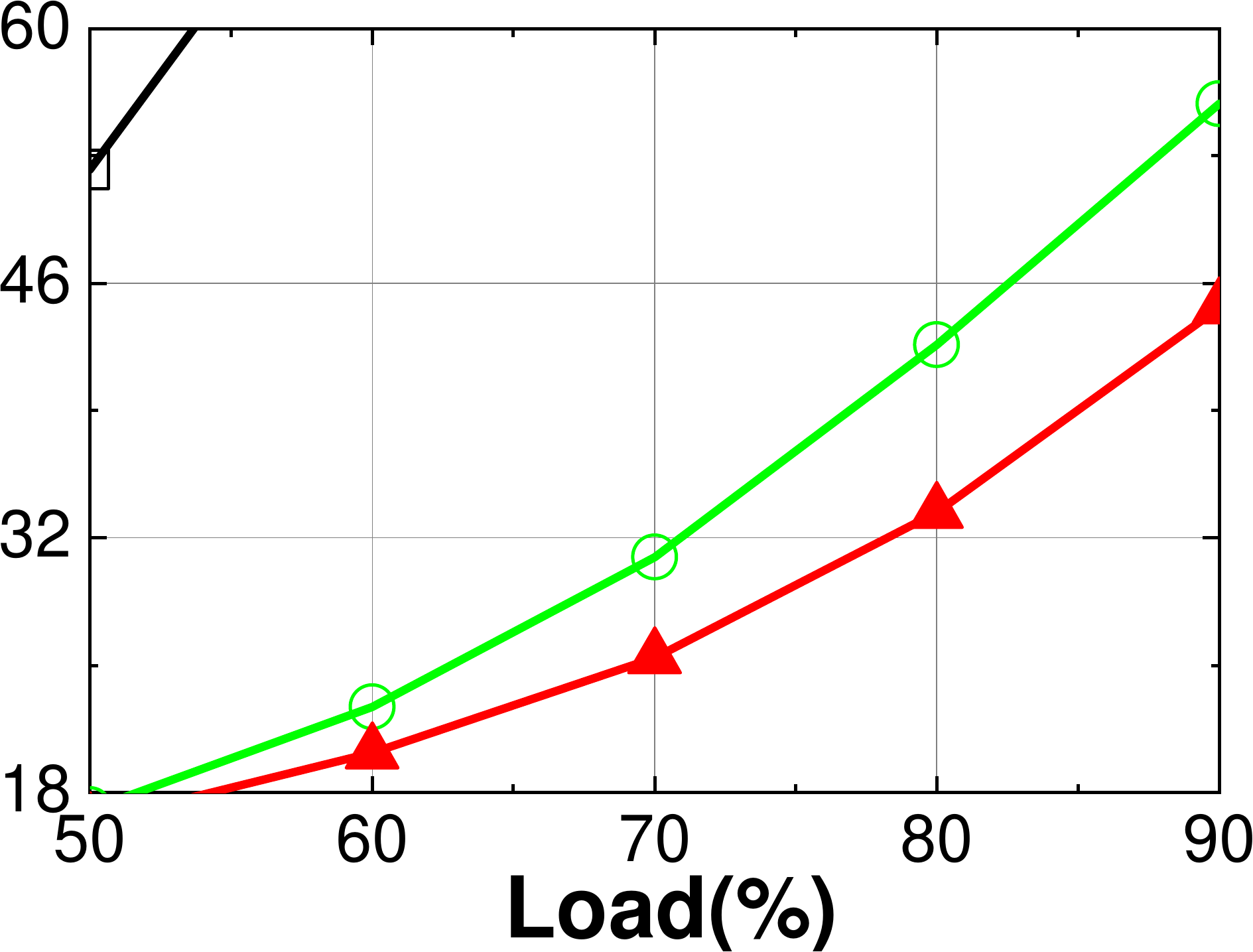}
\postsubfig
\label{fig:W6:infty}
\end{minipage}
}
\postsubfig
\caption{FCT across different flow sizes on W6 for SRPT.}
\label{fig:W6}
\end{figure}

\bbb{Performance on W3 (Figure~\ref{fig:W3}): }
For small flows in (0,1KB), our QC-LAS, QC-SRPT, PIAS and pFabric achieve similar FCT.
However, the 99th percentile FCT for small flows of pFabric is about 8\% higher than that of QC-SRPT.
For middle flows in (1KB,10KB), compared to PIAS, our QC-LAS reduces the FCT by about 60.8\%. Besides, our QC-LAS reduces the FCT of large flows by about 22.4\% compared to PIAS.

\bbb{Performance on W5 (Figure~\ref{fig:W5}): }
Our QC-SRPT and pFabric achieve similar FCT for small flows. And the FCT for small flows of PIAS is about 1\% higher than that of QC-LAS. The 99th percentile FCT for small flows of PIAS is about 87.9\% higher than that of QC-LAS.

\bbb{Performance on W6 (Figure~\ref{fig:W6}): }
For small flows in (0,10KB), the average FCT of the QC-SRPT is nearly the same as that of pFabric.
For flows in (10KB, 100KB), the FCT of QC-SRPT is higher than that of pFabric.
However, similar to W4, the average FCT of QC-SRPT is about 21.7\% lower than the average FCT of pFabric in W6. It is because that QC-SRPT decreases the FCT of large flows in (10KB, $\infty$) by about 25\%.

\end{document}